\documentclass[12pt,sort&compress]{article}
\usepackage{xcolor}

\usepackage{amsmath}
\usepackage{graphicx}
\usepackage{tabularx}
\usepackage{subcaption}
\usepackage{cite}
\usepackage{amssymb}
\usepackage{mathtools}
\usepackage{bbm}
\usepackage{lmodern}
\usepackage[sort&compress,numbers]{natbib}
\usepackage[toc,page]{appendix}
\usepackage{multicol}
\usepackage{multirow}
\usepackage[toc,page]{appendix}

\usepackage{tabularx,ragged2e}
\newcolumntype{C}{>{\Centering\arraybackslash}X} % centered "X" column
\usepackage[outercaption]{sidecap}    
\usepackage{hyperref}

\newcommand{\qinst}{\tilde{q}}

\begin{document} 
%\title{Local resetting in non-conserving zero-range processes with extensive rates}
%\date{}
%\author{Pascal Grange\\
%{\emph{Division of Natural and Applied Sciences}}\\
%{\emph{and Zu Chongzhi Center for Mathematics and Computational Science}}\\
% Duke Kunshan University\\
%8 Duke Avenue, Kunshan, 215316 Jiangsu, China\\
%\normalsize{{\ttfamily{pascal.grange@dukekunshan.edu.cn}}}}
%\maketitle
\title{First passage of a run-and-tumble particle with exponentially-distributed tumble duration in the presence of a drift}
\date{}
\author{Pascal Grange\\
{\emph{Division of Natural and Applied Sciences and Zu Chongzhi Center}}\\
 Duke Kunshan University\\
8 Duke Avenue, Kunshan, 215316 Jiangsu, China\\
\normalsize{\ttfamily{pascal.grange@dukekunshan.edu.cn}}
\vspace{5mm}\\
Linglong Yuan\\
{\emph{Department of Mathematical Sciences}}\\
University of Liverpool\\
Liverpool, United Kingdom\\
\normalsize{\ttfamily{linglong.yuan@liverpool.ac.uk}}}
\maketitle

%\tableofcontents

\abstract{We consider a run-and-tumble particle on a finite interval $[a,b]$ with two absorbing end points.  The particle has an internal velocity state that switches between three values $v,0,-v$ at exponential times, thus incorporating positive tumble times. Moreover, a constant drift is added to the run-and-tumble motion at all times. The combination of these two features constitutes the main novelty of our model.  The densities of the first-passage time through $a$ (given the initial position and velocity states) satisfy certain forward Fokker--Planck equations. 
 The Laplace transforms of these equations induce evolution equations for the exit probabilities and  mean first-passage times of the particle. We solve these equations explicitly for all possible initial states. We consider the limiting regimes of instantaneous tumble and/or the limit of large $b$  to confirm consistency with existing results in the literature. In particular, in the limit of a half-line (large $b$), the mean first-passage time conditioned on the exit through $a$ is an affine function of the initial position if the drift is positive, as in the case of instantaneous tumble.} %If the drift is negative, we find corrections that vanish exponentially with the distance to the end of the half line.}

\pagebreak

\tableofcontents

\section{Introduction and main results}

The movement of a bacterium (we will use bacterium and particle interchangeably) of certain species (such as \textit{E.\ coli}) in a fluid environment is typically modeled as a run-and-tumble motion \cite{berg2008coli}.  The state of the particle alternates between running and tumbling.  At the end of each tumble,  the direction of the next run is chosen. The tumble time is relatively short compared to the run time, so the former has often been assumed to be zero in mathematical models of the run-and-tumble particle (or RTP), 
see \cite{berg1972chemotaxis,berg2008coli,ramaswamy2010mechanics,cates2015motility}.  Exact results on the first-passage properties of the RTP are still sparse, compared to the classic results on diffusive particles (see \cite{redner2001guide} for a review). The survival probability of a free RTP on a half-line has been calculated for an RTP with instantaneous tumble, which yielded exact results on the non-crossing probability of two free RTPs \cite{le2019noncrossing}. The model of an RTP in higher dimension with an absorbing hyperplane has been solved in the presence of a constant drift in \cite{de2021survival}.\\

%\cite{berg2008coli}{\bf{\textcolor{red}{insert references}}}.\\ %For instance, a tumble of \textit{E.\ coli} lasts about 0.1 seconds, compared to  1 second for a run \cite{berg2004coli}. 

 Two directions of generalization have emerged in recent developments. Firstly, one may consider that the velocity of the bacterium results not only from its own run-and-tumble movement, but also from the flow in the fluid environment, see \cite{bearon2000modelling, lopez2014asymmetric, de2021survival, gueneau2024relating, grange2024mean}. In particular, a drift term is added to the run-and-tumble motion to model the effect of a uniform flow of the fluid  in which the particle is swimming \cite{de2021survival} (for a model coupling diffusion to the run-and-tumble dynamics, see \cite{malakar2018steady}). Secondly, the tumble time may be assumed to be positive, see \cite{sun2025exact, romo2025run}. This is a  more realistic setting. Indeed, the typical duration of tumbles in \textit{E.\ coli}  is $0.1$ second, compared to $1$ second for the typical duration of runs; thus in this %the \textit{E.\ coli} 
case, the typical tumble duration is approximately 10\% of a run time. The survival probability of an RTP (both in the case of instantaneous and exponentially-distributed tumble times) has been calculated in dimension $d$, with an absorbing hyperplane and zero force field, if the particle starts on the hyperplane \cite{mori2020universal}. Quite remarkably, the result does not depend on the dimension.
Assuming positive tumble times, \cite{sun2025exact} considers that the particle is constantly under an external harmonic potential. In \cite{romo2025run}, $\sigma(t)$ can change from any value to another in $\{1,0,-1\}$ in a way that breaks detailed balance, while the drift term consists of a Gaussian white noise. \\

% {\bf{\textcolor{red}{Issues of dimensionality.
% We restrict ourselves to a one-dimensional model. The results of \cite{de2021survival} include higher-dimensional cases (in which the projection of the internal velocity state onto the direction of the drift has a density $W$ which depends on  the dimension.}}}\textcolor{blue}{It seems that \cite{de2021survival} only discussed the higher-dimension issue in Summary and Conclusion as a possible generalization.}\\
%cite: thesis Touzo Gueneau for reviews, Basu et al. in higher dimension, Bressloff, universality in higher dimensions.

The purpose of this paper is to combine the above two generalizations in a one-dimensional model of an RTP, i.e. adding a drift to the motion and assuming positive (exponentially-distributed) tumble times. More precisely, we consider a model in which the velocity of the RTP is the sum of a drift and the internal velocity. We therefore consider the following equation:
\begin{align}\label{mainmodel}
\frac{dx(t)}{dt}=\mu+v\sigma(t), \quad x\in[a,b], \quad t\geq 0,
\end{align}
where $-\infty<a<b<\infty$ are two absorbing points,  $x(t)$ is the position of the particle, $\mu\in\mathbb R$ is the drift, $v>0$ is the absolute value of the internal velocity when the particle is in the running state, $v\sigma(t)$ is the internal velocity, and $\sigma(t)$
is the telegraphic noise that switches between three values: $-1,0,1$ according to the following process: 
\begin{itemize}
\item If $\sigma(t)=0$,  the particle is tumbling, and the remaining tumble time follows the exponential distribution with parameter $\varphi>0$; once the tumble is finished, at time $t'$ say, then $\sigma(t')$ is drawn uniformly from $\{-1,+1\}$;
\item If $\sigma(t)=1$ or $-1$, the particle is running, and the remaining run time follows the exponential distribution with parameter $\gamma>0$; once the run is finished, at time $t'$ say, then $\sigma(t')=0$;
\item All the run times and tumble times are independent. 
\end{itemize}
Note that in this model, the drift term $\mu$ is always present, in both running and tumbling states, so that the possible values of the total velocity of the particle are $\mu-v,\mu$ and $\mu+v$. We will assume that the drift is subcritical, meaning
\begin{equation}
    |\mu|<v,
\end{equation}
 so that the particle can run in both directions. We also exclude the free case $\mu=0$, for which the  the mean-first-passage time at $a$ is infinite, and the survival probability has been calculated exactly in any dimension \cite{mori2020universal}.\\

 If $\sigma(t)$ switches only between two values $1,-1$ (i.e. it always changes from $1$ to $-1$ and from $-1$ to $1$) and $b=\infty$, then the tumbling is instantaneous and this is the model considered in \cite{lopez2014asymmetric}. In \cite{gueneau2024relating}, the authors considered the movement in a finite interval with two absorbing end points, with $\sigma(t)$ switching between $1,-1$ and the drift term replaced by $F(x(t))$ which is a function of the location $x(t)$ of the particle. In the model \eqref{mainmodel}, during two consecutive runs, note that the internal velocity states of the particle are identical with probability $1/2$ (whereas in the process with instantaneous tumbling, they are identical with probability zero). Hence, in the limit of zero mean duration of tumbles (meaning $\varphi\to\infty$), the process is equivalent to the one with two values $1,-1$ for $\sigma(t)$  with rate $\gamma / 2$ (this results from the so-called Poisson thinning). The initial state of the system consists of the initial position $x(0)$ of the particle in the interval and the internal velocity state $\sigma(0)$  at the beginning of the process. The coordinate $x(0)$ is in an interval $I$, which can be either a closed interval denoted by $[a,b]$, or the half-line denoted by $[a,\infty[$. 
 The two main quantities of interest are:\\
\begin{enumerate}
\item the exit probability through $a$, denoted by $\mathcal{E}_a(x(0),\sigma(0);I,\varphi)$,\\
\item the  mean exit time through $a$, given that the particle exits the system through $a$, denoted by $\mathcal{T}_a(x(0),\sigma(0);I,\varphi)$.
\end{enumerate}

\vspace{3mm}
 
  Studying Eq.\ \eqref{mainmodel} will allow to check consistency with the results of  \cite{lopez2014asymmetric, de2021survival, gueneau2024relating} by letting $\varphi\to\infty$ and/or $b\to\infty$. Indeed, 
 The exit probability from a segment $[a,b]$ through $a$ was obtained in \cite{gueneau2024relating} for any background force field taking only subcritical values. Specializing to a constant drift and  taking Poisson thinning into account, the result reads as follows (in a system of units where $v$ is the unit velocity and $\gamma$ is the unit of frequency\footnote{the unit of time is therefore the mean duration $\gamma^{-1}$ of runs, and the unit of length is the mean length $\gamma^{-1}v$ of runs.}):
\begin{equation}\label{TGE}
\mathcal{E}_a(x,\pm;[a,b],\infty) = 
\frac{\left(\mu - 1\right)\exp\left( -\frac{\mu(b-a)}{1 -\mu^2}  \right) + \left( \mp \mu +1\right)\exp\left( -\frac{\mu(x-a)}{1 -\mu^2}  \right) }{ \mu + 1 + \left( \mu- 1 \right)\exp\left( -\frac{\mu(b-a)}{1 -\mu^2}  \right)}.
\end{equation}
%\begin{equation}\label{TGE}
%\mathcal{E}_a(x,\pm;[a,b],\infty) = 
%\frac{\left(1 - \frac{v}{\mu}\right)\exp\left( -\frac{\gamma\mu(b-a)}{v^2 -\mu^2}  \right) + \left( \mp 1 +\frac{v}{\mu}\right)\exp\left( -\frac{\gamma\mu(x-a)}{v^2 -\mu^2}  \right) }{ 1 + \frac{v}{\mu} + \left( 1-\frac{v}{\mu}\right)\exp\left( -\frac{\gamma\mu(b-a)}{v^2 -\mu^2}  \right)}.
%\end{equation}

  The exact expression of the exit probability of the RTP on the semi-infinite  interval $[a,\infty[$ with instantaneous tumble was obtained directly in  \cite{de2021survival}, and also follows from the large-$b$ limit of the above equation. It is given for subcritical values of the drift as
%\begin{equation}
%\begin{split}
%\mathcal{E}_a(x,+;[a,\infty[,\infty)=& \mathbbm{1}( \mu<0) +\frac{v-\mu}{v+\mu} 
% \exp\left( -\frac{\gamma \mu}{v^2 - \mu^2}  (x-a)\right)\mathbbm{1}( \mu>0),\\
% \mathcal{E}_a(x,-;[a,\infty[,\infty)=& \mathbbm{1}( \mu<0) +
% \exp\left( -\frac{\gamma \mu}{v^2 - \mu^2}  (x-a)\right)\mathbbm{1}( \mu>0).\\
% \end{split}
%\end{equation}
\begin{equation}\label{exitOrdinary}
\begin{split}
\mathcal{E}_a(x,+;[a,\infty[,\infty)=& \mathbbm{1}( \mu<0) +\frac{1-\mu}{1+\mu} 
 \exp\left( -\frac{ \mu}{1 - \mu^2}  (x-a)\right)\mathbbm{1}( \mu>0),\\
 \mathcal{E}_a(x,-;[a,\infty[,\infty)=& \mathbbm{1}( \mu<0) +
 \exp\left( -\frac{ \mu}{1 - \mu^2}  (x-a)\right)\mathbbm{1}( \mu>0).\\
 \end{split}
\end{equation}

 Morover, the  conditional average of the first passage-time at the left end of the half-line $[a,\infty[$ (over the trajectories that eventually exit the system through $a$) has been obtained in \cite{de2021survival} for instantaneous tumble as
%\begin{equation}
%\begin{split}
%\mathcal{T}_a(x,+;[a,\infty[,\infty)=&
%\left( \frac{v^2+\mu^2}{v^2-\mu^2}\frac{x-a}{\mu} +2 \frac{v}{\gamma \mu}\right) \mathbbm{1}( \mu>0) +
%\left(\frac{x-a}{|\mu|} +2 \frac{v}{\gamma |\mu|}\right) \mathbbm{1}( \mu<0)\\
%\mathcal{T}_a(x,-;[a,\infty[,\infty)=&\left( \frac{v^2+\mu^2}{v^2-\mu^2}\frac{x-a}{\mu} \right) \mathbbm{1}( \mu>0) 
%+\frac{x-a}{|\mu|} \mathbbm{1}( \mu<0)  \\
%\end{split}    
%\end{equation}
 \begin{equation}\label{MFPTOrdinary}
\begin{split}
\mathcal{T}_a(x,+;[a,\infty[,\infty)=&
\left( \frac{1+\mu^2}{1-\mu^2}\frac{x-a}{\mu} + \frac{2}{\mu}\right) \mathbbm{1}( \mu>0) +
\left(\frac{x-a}{|\mu|} + \frac{2}{ |\mu|}\right) \mathbbm{1}( \mu<0),\\
\mathcal{T}_a(x,-;[a,\infty[,\infty)=&\left( \frac{1+\mu^2}{1-\mu^2}\frac{x-a}{\mu} \right) \mathbbm{1}( \mu>0) 
+\frac{x-a}{|\mu|} \mathbbm{1}( \mu<0).\\
\end{split}    
\end{equation}

This paper aims to investigate the model of Eq. \eqref{mainmodel}, focusing  on the exit probability  of a particle starting within the open segment $]a,b[$ and its mean exit time (conditional on exit). Exiting from $b$ can be studied by symmetry and is omitted in this paper.  We stress that models like \eqref{mainmodel} have been studied in the literature of fluid flow models and Markov additive processes, see \cite{asmussen1995stationary, kyprianou2003fluctuations, badescu2005risk, jiang2008perpetual, ivanovs2012occupation}. But to our best knowledge, the explicit results for the exit probability and its mean exit time have not been stated for the model \eqref{mainmodel}.\\   
%The case where the absolute value of the drift $\mu$ is strictly lower than the absolute value of the internal velocity $v$ of the RTP is especially intriguing, because it allows the sign of the velocity of the particle to take a positive and a negative value. %In particular, the exit probability of the particle and the mean exit time (conditional on the exit) have been calculated exactly in \cite{}. 
%Following \cite{de2021survival}, we are going to take the absolute interval velocity $v$ as the unit of velocity and parameter $\gamma$ of the run time as the unit of frequency. 

We extend Eq. (\ref{TGE}) to the case of a finite rate $\varphi$ (see Eq. \eqref{resExitNegativeDrift}). Taking the limit of large-$b$ yields the following generalization of Eq. \eqref{exitOrdinary} for the model on a half-line:
\begin{equation}\label{exitFinite}
\begin{split}
\mathcal{E}_a(x,0;[a,\infty[,\varphi)=&  \mathbbm{1}( \mu<0)\\
&+\frac{2+ \varphi( 1 - \mu^2)  \left( 1 - \sqrt{1 + \frac{4\mu^2}{(1-\mu^2)^2 \varphi^2} }\right)}{  1+ \mu^2 + 2\mu\left( 1+\frac{1}{\varphi}\right) +(1-\mu^2 ) \sqrt{1 + \frac{4\mu^2}{(1-\mu^2)^2 \varphi^2} }} \times \exp\left(  
\lambda_- (x-a) 
\right) \mathbbm{1}( \mu>0),\\
\mathcal{E}_a(x,+;[a,\infty[,\varphi)=&  \mathbbm{1}( \mu<0)\\
&+\frac{ 1+ \mu^2 - 2\mu\left( 1+\frac{1}{\varphi}\right) +(1-\mu^2 ) \sqrt{1 + \frac{4\mu^2}{(1-\mu^2)^2 \varphi^2} }}{ 1+ \mu^2 + 2\mu\left( 1+\frac{1}{\varphi}\right) +(1-\mu^2 ) \sqrt{1 + \frac{4\mu^2}{(1-\mu^2)^2 \varphi^2} }}\times 
\exp\left(  \lambda_-(x-a) 
\right) \mathbbm{1}( \mu>0),\\
\mathcal{E}_a(x,-;[a,\infty[,\varphi)=&  \mathbbm{1}( \mu<0) +\exp\left(  \lambda_-(x-a) 
\right) \mathbbm{1}( \mu>0),\\
{\mathrm{with}} \qquad &\lambda_\pm = -\frac{\mu}{1-\mu^2} +\frac{\varphi}{2\mu} \left( 1 \pm \sqrt{1 + \frac{4\mu^2}{(1-\mu^2)^2 \varphi^2} }\right).
\end{split}
\end{equation}
On the other hand,  we obtain expressions generalizing  Eq. (\ref{MFPTOrdinary}) to finite values of the rate $\varphi$. The  dependence of   $\mathcal{T}_a(x,\sigma(0);[a,\infty[,\varphi)$ on the coordinate $x$ is affine (but dependent on $\varphi$) if the drift is positive. If the drift is negative, there are exponential corrections:   
\begin{equation}\label{resMeanAnnounced}
\begin{split}
     \mathcal{T}_a(x,0;[a,\infty[,\varphi)=& \left(-\frac{1}{\mu}(x-a) + Q_0 + F_0 e^{\lambda_+(x-a)} \right)\mathbbm{1}( \mu < 0) \\
     & + \left(  
 \frac{\varphi(1-\mu^2) + 2\mu^2 + \mu^2\sqrt{\varphi^2(1-\mu^2)^2 + 4\mu^2}}{\mu(1-\mu^2)\sqrt{\varphi^2(1-\mu^2)^2 + 4\mu^2}}(x-a) + {\mathcal{T}}_a( a, 0; [a,\infty[, \varphi ) \right) \mathbbm{1}( \mu > 0), \\
     \mathcal{T}_a(x,+;[a,\infty[,\varphi)=&  \left(-\frac{1}{\mu}(x-a) + Q_+ + F_+ e^{\lambda_+(x-a)} \right)\mathbbm{1}( \mu < 0) \\
     & + \left(  
 \frac{\varphi(1-\mu^2) + 2\mu^2 + \mu^2\sqrt{\varphi^2(1-\mu^2)^2 + 4\mu^2}}{\mu(1-\mu^2)\sqrt{\varphi^2(1-\mu^2)^2 + 4\mu^2}}(x-a) + {\mathcal{T}}_a( a, +; [a,\infty[, \varphi ) \right) \mathbbm{1}( \mu > 0),\\
     \mathcal{T}_a(x,-;[a,\infty[,\varphi)=&   \left(-\frac{1}{\mu}(x-a) + Q_- + F_- e^{\lambda_+(x-a)} \right)\mathbbm{1}( \mu < 0) \\
     & + \left( 
 \frac{\varphi(1-\mu^2) + 2\mu^2 + \mu^2\sqrt{\varphi^2(1-\mu^2)^2 + 4\mu^2}}{\mu(1-\mu^2)\sqrt{\varphi^2(1-\mu^2)^2 + 4\mu^2}}(x-a) \right) \mathbbm{1}( \mu > 0),\\
\end{split}
\end{equation}
 where the constant $\lambda_+$ is defined in the previous equation, and   the constants $Q_0,Q_+,Q_-,F_0,F_+, F_-$ are expressed in closed form in Eq. (\ref{closedFormMinus}). The constants 
 ${\mathcal{T}}_a( a, 0; [a,\infty[, \varphi )$ and 
 ${\mathcal{T}}_a( a, +; [a,\infty[, \varphi )$ are derived and expressed in closed form in Eq. (\ref{closedFormPlus}).\\

The paper is organized as follows. In Section \ref{sec:notation}, we set notations for the densities of the exit time at coordinate $a$ for a particle starting within $]a,b[$, and we obtain a system of differential equations satisfied by their Laplace transforms. In Section \ref{sec:evolution}, we use this system to establish the differential equations satisfied by the exit probabilities and mean exit times (or mean first-passage times, conditional on exit through $a$). We proceed to solving the equations in Section \ref{sec:solutionproba} and Section \ref{sec:solutiontime}, respectively for exit probabilities and mean first-passage times. Positive and negative drifts induce different boundary conditions, which are addressed in separate subsections. In particular, we obtain the limits of the solutions when $b$ goes to infinity, and check them against direct numerical simulations of the system. As the particle is active, the mean lifetime of a particle starting its motion may be positive. For a negative drift, this is the case if the initial internal velocity state is positive. For a positive drift, this is the case if the initial internal velocity state is positive or zero. The formula giving the mean lifetime as a function of the initial coordinate has to be extrapolated to a negative coordinate, known as the Milne length by analogy with neutron scattering. In Section \ref{sec:Milne} we 
  express the Milne length in closed form.
In Appendices, we provide technical details for some of the computations\footnote{A Matlab \href{https://www.dropbox.com/scl/fi/ua9dpojeld3l7wl7blf57/MFPT_tumble_algebra.m?rlkey=6jjgw9viw4jokp7ewz5ncv83x&st=n69zw5i8&dl=0}{script} is available for symbolic calculation}. %\textcolor{red}{Comments on $\mu=0$?}

\section{Notations and preliminaries% quantities of interest
}\label{sec:notation}

 Let us recall the model \eqref{mainmodel}. A run-and-tumble particle (RTP) moves on a segment $[a,b]$ of the real line with coordinate $x$, with two absorbing points at $x=a$ and $x = b$.
  The duration of each run is exponentially distributed with parameter $\gamma$. After each run, the particle tumbles: its internal velocity becomes $0$ (and its total velocity becomes $\mu$).  The duration of the tumble is exponentially distributed, with parameter $\varphi$. At the end of each tumble, the internal velocity state of the particle is drawn uniformly\footnote{In this section we shall keep the velocity $v$ and the rate $\gamma$ for the sake of generality. In Section \ref{sec:evolution} we will set $v$ and $\gamma$ to $1$, choosing them as the unit of velocity and frequency respectively.} from $\{ -v,+v \}$ (and its total velocity becomes $\mu\pm v$). We require that $|\mu|<v$ so that the particle can move in both directions.\\

Let us denote by $\rho_a(x,\pm,t)$ (resp. $\tau_a(x,t)$) the density of first-passage time at coordinate $a$  of a particle that started its motion at position $x$ in $]a,b[$, in the internal velocity state  $\pm v$ (resp. in a tumbling state):
\begin{equation}
\begin{split}
\rho_a(x,\pm,t) \,dt =&P( {\mathrm{first}}\;{\mathrm{passage}}\;{\mathrm{through}}\;a\;{\mathrm{in}}\;[t,t+dt]|
  {\mathrm{internal}}\;{\mathrm{velocity}}\; \pm v \;{\mathrm{at}}\; {\mathrm{position}} \;x\; {\mathrm{at}} \;t=0),\\
 \tau_a(x,t) \, dt =& P( {\mathrm{first}}\;{\mathrm{passage}}\;{\mathrm{through}}\;a\;{\mathrm{in}}\;[t,t+dt]|
  {\mathrm{tumbling}}\;{\mathrm{at}}\; {\mathrm{position}} \;x\; {\mathrm{at}} \;t=0).\\
  \end{split}
\end{equation}

Consider a particle that starts its motion at coordinate $x$ (at time $0$), in a tumbling state. Consider the interval of time $[0,dt]$, where $dt$ is an infinitesimal time (this reasoning was used to derive the evolution equation of the survival probability of the RTP with instantaneous tumbles, see \cite{bray2013persistence,dhar2019run}). If the tumble does not end in $[0,dt]$ (which is the case with probability $1-\varphi dt$), the particle is at coordinate $x+\mu dt$ at time $dt$, in a tumbling state (and the residual duration until time $t$ is $t-dt$). If the tumble ends in $[0,dt]$ (which is the case with probability $\varphi dt$), the particle picks an internal velocity state uniformly. Hence,
\begin{equation}
\tau_a(x,t)dt = ( 1-\varphi dt) \tau_a(x+\mu dt,t-dt) +
\varphi dt \left(\frac{1}{2} \rho_a(x,+,t)+
\frac{1}{2}  \rho_a(x,-,t) \right), \qquad( x\in ]a,b[).
\end{equation}
If the particle starts its motion at coordinate $x$ (at time $0$), in a given internal velocity state $\epsilon v$, and if it does not tumble before $dt$ (which is the case with probability $1-\gamma dt$), the particle is at coordinate $x+(\mu+\epsilon v) dt$ at time $dt$, with the internal velocity state $\epsilon$, and has time $t-dt$ to reach $a$ for the first time. If a tumble event starts before $dt$  (which is the case with probability $\gamma dt$), the particle enters a tumble state. Hence:
\begin{equation}
\rho_a(x,\epsilon,t)dt = ( 1-\gamma dt) \rho_a(x+(\mu +\epsilon v) dt,t-dt) + \gamma dt \tau_a(x,t),\;\;\;\;\epsilon \in \{-,+\}, \qquad( x\in ]a,b[).
\end{equation}
 Taylor expansion  (in the time $dt$) yields the system of evolution equations
\begin{equation}\label{sysEv}
\begin{split}
 \frac{\partial \tau_a(x,t)}{\partial t} =& \mu \frac{\partial \tau_a(x,t)}{\partial x} -\varphi \tau_a(x,t)+\frac{1}{2}\varphi \rho_a( x,+,t) + \frac{1}{2}\varphi \rho_a( x,-,t)  ,\\
\frac{\partial \rho_a(x,+,t)}{\partial t} =& (\mu+v) \frac{\partial \rho_a(x,+,t)}{\partial x}   -\gamma \rho_a(x,+,t) + \gamma \tau_a(x,t) ,\\
\frac{\partial \rho_a(x,-,t)}{\partial t} =&  (\mu -v) \frac{\partial \rho_a(x,-,t)}{\partial x}  -\gamma \rho_a(x,-,t)  +\gamma \tau_a(x,t),\quad( x\in ]a,b[). 
\end{split}
\end{equation}

 Let us denote by $\tilde{f}$ the Laplace transform of any quantity $f$ depending on time (and possibly other variables):
\begin{equation}
\tilde{f}(s) := \int_0^\infty e^{-st} f(t) dt. 
\end{equation}
 Integrating by parts, the Laplace transforms of the densities $\rho_a$ and $\tau_a$ satisfy 
 \begin{equation}
\begin{split}
\int_0^\infty  \frac{\partial\rho_a(x,\epsilon,t)}{\partial t} e^{-st} dt=& [ \rho_a(x,\epsilon, t) e^{-s t}]_0^\infty + s\int_0^\infty   \rho_a(x,\epsilon,t) e^{-st} dt = s \widetilde{\rho_a}(x,\epsilon, s),\\
\int_0^\infty  \frac{\partial\tau_a(x,t)}{\partial t} e^{-st} dt=& [ \tau_a(x, t) e^{-s t}]_0^\infty + s\int_0^\infty   \tau_a(x,t) e^{-st} dt = s \widetilde{\tau_a}(x, s),\qquad( x\in ]a,b[),
\end{split}     
 \end{equation}
 where we have used the fact that there is no flow of particle through the ends of the segment at time $0$ if the particle is at $x$ in $]a,b[$ a time $0$, and we have  assumed that $\lim_{t\to\infty}\rho_a(x,\epsilon, t) e^{-s t}=0$ and $\lim_{t\to\infty}\tau_a(x, t) e^{-s t}=0$. The main quantities of interest are related to the behavior of the Laplace transforms of the densities as follows. The exit probabilities through  $a$ are obtained by integrating the densities over time:\\
\begin{equation}
\begin{split}
\mathcal{E}_a(x,0; [a,b],\varphi) =& \int_0^\infty \tau_a(x,t) dt = \widetilde{\tau_a}(x,0) ,\\
\mathcal{E}_a(x,+; [a,b],\varphi) =& \int_0^\infty \tau_a(x,t) dt = \widetilde{\rho_a}(x,+,t)dt= \widetilde{\rho_a}(x,+,0),\\
\mathcal{E}_a(x,-; [a,b],\varphi) =& \int_0^\infty \tau_a(x,t) dt = \widetilde{\rho_a}(x,-,t)dt= \widetilde{\rho_a}(x,-,0).
\end{split}    
\end{equation}
Moreover, the conditional\footnote{In this context, we use the word conditional as a shorthand for {\emph{conditional on the exit of the particle through $a$}} (we condition on trajectories that reach $a$ before reaching $b$).} mean the first-passage time at $a$ given the initial position and velocity state is expressed as
\begin{equation}\label{timesOfInterestDef}
\begin{split}
\mathcal{T}_a(x,0;[a,b],\varphi) =& \frac{\int_0^\infty t \,\tau_a(x,t) \,dt }{\mathcal{E}_a(x,0; [a,b],\varphi)} = \frac{1}{\widetilde{\tau_a}(x,0) }\frac{\partial}{\partial s }\widetilde{\tau_a}(x,s)|_{s=0},\\
\mathcal{T}_a(x,+;[a,b],\varphi) =&    \frac{\int_0^\infty t \,\rho_a(x,+,t) \,dt }{\mathcal{E}_a(x,+; [a,b],\varphi)}  =\frac{1}{\widetilde{\rho_a}(x,+,0) }\frac{\partial}{\partial s }\widetilde{\rho_a}(x,+,s)|_{s=0},\\
\mathcal{T}_a(x,-;[a,b],\varphi) =&  \frac{\int_0^\infty t \,\rho_a(x,-,t) \,dt }{\mathcal{E}_a(x,-; [a,b],\varphi)}= \frac{1}{\widetilde{\rho_a}(x,-,0) }\frac{\partial}{\partial s }\widetilde{\rho_a}(x,-,s)|_{s=0}.
\end{split}
\end{equation}

 The Laplace transform of Eq. (\ref{sysEv}) reads
 \begin{equation}\label{LapEv}
\begin{split}
 s \widetilde{\tau_a}(x,s) =& \mu \frac{\partial \widetilde{\tau_a}(x,s)}{\partial x} -\varphi \widetilde{\tau_a}(x,s)+\frac{\varphi}{2} \widetilde{\rho_a}( x,+,s) + \frac{\varphi}{2} \widetilde{\rho_a}( x,-,s)  ,\\
s \widetilde{\rho_a}(x,+,s) =& (\mu+v) \frac{\partial \widetilde{\rho_a}(x,+,s)}{\partial x}   -\gamma \widetilde{\rho_a}(x,+,s) + \gamma \widetilde{\tau_a}(x,s) ,\\
s \widetilde{\rho_a}(x,-,s) =&  (\mu -v) \frac{\partial \widetilde{\rho_a}(x,-,s)}{\partial x}  -\gamma \widetilde{\rho_a}(x,-,s)  +\gamma \widetilde{\tau_a}(x,s),\quad( x\in ]a,b[).\\ 
\end{split}
\end{equation}

Let us introduce the combinations of functions
 \begin{equation}\label{SDeff}
 \begin{split}
S_a(x,s):=& \frac{1}{2}\left( \widetilde{\rho_a}(x,+,s) 
 + \widetilde{\rho_a}(x,-,s)\right),\\ 
D_a(x,s):=& \frac{1}{2}\left( \widetilde{\rho_a}(x,+,s) 
 - \widetilde{\rho_a}(x,-,s)\right).
\end{split}
 \end{equation}
 In terms of these unknown functions, Eq.  \eqref{LapEv} becomes
 \begin{equation}
\begin{split}
 0 =& \mu \frac{\partial \widetilde{\tau_a}(x,s)}{\partial x} -(\varphi + s) \widetilde{\tau_a}(x,s)+\varphi S_a(x,s),\\
0 =& (\mu+v) \frac{\partial (S_a + D_a)(x,s)}{\partial x}   -(\gamma +s)(S_a+D_a)(x,s) + \gamma \widetilde{\tau_a}(x,s) ,\\
0 =&  (\mu -v) \frac{\partial( S_a-D_a )(x,s)  }{\partial x}  -(\gamma +s)( S_a-D_a)(x,s)  +\gamma \widetilde{\tau_a}(x,s).\\ 
\end{split}
\end{equation}
Taking the sum and the difference of the last two equations:
\begin{equation}\label{LapSDtau}
\begin{split}
 0 =& \mu \frac{\partial \widetilde{\tau_a}}{\partial x} -(\varphi + s) \widetilde{\tau_a}+\varphi S_a,\\
0 =& \mu \frac{\partial S_a}{\partial x} +v  \frac{\partial D_a}{\partial x}
 -(\gamma +s)S_a + \gamma \widetilde{\tau_a},\\
0 =&  \mu \frac{\partial D_a}{\partial x} +v  \frac{\partial S_a}{\partial x}
 -(\gamma +s)D_a .\\ 
\end{split}
\end{equation}
The Taylor expansion of this system of equations around $s=0$ yields evolution equations for the exit probability through $a$ and for the moments of the first-passage time through $a$.

\section{Evolution equations}\label{sec:evolution}
\subsection{Exit probabilities}\label{subsec:exit}

Let us introduce the following notations:
\begin{equation}\label{notationsE}
\begin{split}
Z_a(x):=& \widetilde{\tau_a}(x,0),\\
E_a(x):=& \frac{1}{2}\left(  \mathcal{E}_a(x,+; [a,b],\varphi) + \mathcal{E}_a(x,-; [a,b],\varphi)  \right) =S_a(x,s=0),\\ 
e_a(x):=& \frac{1}{2}\left( \mathcal{E}_a(x,+; [a,b],\varphi) - \mathcal{E}_a(x,-; [a,b],\varphi)  \right) =D_a( x,s=0).\\
\end{split}
\end{equation}
The notations $E_a(x)$ and $e_a(x)$ for the symmetric and antisymmetric combinations of the exit probabilities (through $a$) with positive and negative initial velocities  were used in \cite{gueneau2024relating}. We supplement them with $Z_a$ to account for the additional value of the internal velocity state in our model.\\

Specializing Eq. (\ref{LapSDtau}) at $s=0$ and 
 taking linear combinations of the equations yields
\begin{equation}\label{combEvol}
\begin{split}
Z_a' =& \frac{\varphi}{\mu} Z_a - \frac{\varphi}{\mu} E_a,\\
E_a' =& \frac{\gamma v}{v^2 - \mu^2} e_a - \frac{\gamma \mu}{v^2 - \mu^2} E_a +\frac{\gamma \mu}{v^2 - \mu^2} Z_a,\\
e_a' =& -\frac{\gamma\mu}{v^2 - \mu^2} e_a + \frac{\gamma v}{v^2 - \mu^2} E_a -\frac{\gamma v}{v^2 - \mu^2} Z_a.\\
\end{split}
\end{equation}
For ease of notation, let us pick a system of units in which the unit of time is equal to the mean duration of runs (which is $\gamma^{-1}$), and the unit of length is equal
 to the mean distance traveled by the RTP during a run (which is $v\gamma^{-1}$). This choice is equivalent to setting the mean duration of runs and the internal velocity to $1$:
 \begin{equation}
 \begin{split}
     \gamma :=& 1,\\
    v:=& 1.
  \end{split}   
 \end{equation}
 The free parameters of the model are therefore the drift $\mu$ and the parameter $\varphi$ of the distribution of tumble times.\\

 In these units, Eq. (\ref{combEvol}) becomes the linear ordinary differential equation 
 \begin{equation}\label{ODEExit}
\begin{split}
\frac{d}{dx} \vec{\mathcal{E}}(x) =& A \vec{\mathcal{E}}(x ),\\   
{\mathrm{with}}\;\;\; \vec{\mathcal{E}}(x) :=&
\begin{pmatrix}
 Z_a(x)\\
 E_a(x)\\
 e_a(x)
\end{pmatrix},\qquad
A := \begin{pmatrix}
\frac{\varphi}{\mu} &-\frac{\varphi}{\mu} & 0\\
 \frac{\mu}{1-\mu^2} & - \frac{\mu}{1 - \mu^2} & \frac{1}{1- \mu^2}\\
   -\frac{1}{1- \mu^2} & \frac{1}{1- \mu^2} &  -\frac{\mu}{1- \mu^2}
\end{pmatrix}.
\end{split}
\end{equation}
 The solution, given the initial position $x$ and the initial velocity state, yields the exit probabilities in our notations as:
 \begin{equation}\label{compProbE}
 \begin{split}
\mathcal{E}_a(x,0;[a,b],\varphi) =& 
 Z_a(x),\\
\mathcal{E}_a(x,+;[a,b],\varphi) =& E_a(x) + e_a(x),\\
\mathcal{E}_a(x,-;[a,b],\varphi) =& E_a(x) - e_a(x).
\end{split}
 \end{equation}
 The matrix $A$ identified in the evolution equation  (\ref{ODEExit}) happens to be diagonalizable. The eigenvalues and eigenvectors are derived in Appendix \ref{eigenvectors} (see Eqs (\ref{lambdapmExpr},\ref{uv+-}). This fact enables us to integrate Eq. \eqref{ODEExit} in Section \ref{sec:solutionproba}.

\subsection{Mean first-passage times}\label{subsec:mfp}

Let us collect the terms of order one in $s$ in the Taylor expansion around $s=0$ of Eq. (\ref{LapSDtau}). This yields the evolution equation of the mean first-passage times at $a$:
\begin{equation}
\begin{split}
 Z_a=& - \mu \frac{dU}{ d x} +\varphi U -\varphi  T ,\\
  E_a =& - \mu \frac{d T}{d x} - \frac{d t}{d x} + T -U,\\
  e_a =& - \mu \frac{d t}{d x} - \frac{d T}{d x} + t,
\end{split}
\end{equation}
 with the notations
\begin{equation}\label{UTt}
\begin{split}
U(x) :=& \int_0^\infty t \tau_a(x,t) dt = \frac{\partial \widetilde{\tau_a}(x,s)}{\partial s}|_{s=0},\\
 T(x) :=& \frac{1}{2}\int_0^\infty t \left(\rho_a(x,+,t) + \rho_a(x,-,t) \right) dt 
 = \frac{1}{2}\left( \frac{\partial \widetilde{\rho_a}(x,+,s)}{\partial s}|_{s=0}  + \frac{\partial \widetilde{\rho_a}(x,-,s)}{\partial s}|_{s=0} \right),\\ 
 t(x) :=& \frac{1}{2}\int_0^\infty t \left(\rho_a(x,+,t) - \rho_a(x,-,t) \right) dt
 = \frac{1}{2}\left(  \frac{\partial \widetilde{\rho_a}(x,+,s)}{\partial s}|_{s=0}  - \frac{\partial \widetilde{\rho_a}(x,-,s)}{\partial s}|_{s=0}  \right).
\end{split}
\end{equation}

 Assuming the functions $Z_a,E_a, e_a$ are known from the solution of Eq. (\ref{ODEExit}), the above system of equations is linear in the unknowns $U,T,t$:
\begin{equation}
\begin{split}
 Z_a =& - \mu \frac{dU}{ d x} +\varphi U -\varphi  T,\\
 \mu E_a - e_a=& (1 - \mu^2) \frac{d T}{d x} +\mu T -\mu U - t,\\
  \mu e_a - E_a =& (1- \mu^2) \frac{d t}{d x} +\mu t - T + U.
\end{split}
\end{equation}

Let us rewrite the system in vector form:
\begin{equation}\label{ODE}
\begin{split}
\frac{d}{dx} \vec{M}(x) =& A \vec{M}(x ) + \vec{C}(x),\\   
{\mathrm{with}}\;\;\; \vec{M}(x) :=&
\begin{pmatrix}
U(x)\\
 T(x)\\
 t(x)
\end{pmatrix},\qquad
\vec{C}(x) :=
\begin{pmatrix}
 -\frac{1}{\mu} Z_a(x)\\
 \frac{1}{\mu^2 - 1}(e_a(x) - \mu E_a(x) )\\
 \frac{1}{\mu^2 - 1}(E_a(x) - \mu e_a(x))
\end{pmatrix},
\end{split}
\end{equation}
 and $A$ is the matrix identified in the evolution equation of the  exit probability in Eq. (\ref{ODEExit}).\\

The components of the vector $\vec{M}(x)$, given the initial position $x$, yield the conditional mean first-passage times of interest (defined in Eq. (\ref{timesOfInterestDef})) as
\begin{equation}\label{timesOfInterest}
\begin{split}
\mathcal{T}_a(x,0;[a,b],\varphi) =& \frac{U(x)}{\mathcal{E}_a(x,0;[a,b],\varphi)} =\frac{U(x)}{Z_a(x)} = \frac{M_1(x)}{Z_a(x)}, \\
\mathcal{T}_a(x,+;[a,b],\varphi) =&   \frac{T(x) + t(x)}{\mathcal{E}_a(x,+;[a,b],\varphi)}   =\frac{T(x) +t(x)}{ E_a(x) + e_a(x)}=  \frac{M_2(x) + M_3(x)}{\mathcal{E}_2(x) + \mathcal{E}_3(x)},\\
\mathcal{T}_a(x,-;[a,b],\varphi) =&  \frac{T(x) - t(x)}{\mathcal{E}_a(x,-;[a,b],\varphi)}   = \frac{T(x) -t(x)}{ E_a(x) - e_a(x)} = \frac{ M_2(x) - M_3(x)}{\mathcal{E}_2(x) - \mathcal{E}_3(x)}.\\
\end{split}
\end{equation}
 Eq. \eqref{ODE} is solved explicitly in terms of the eigenvalues and eigenvectors of the matrix $A$ in Section \ref{sec:solutiontime}.\\

\section{Solution of the evolution equation for the exit probability}
\label{sec:solutionproba}

The evolution equation Eq. \eqref{ODEExit} is written in terms of the matrix $A$, whose first two columns are negative to each other. Hence, $0$ is an obvious eigenvalue of $A$.  Moreover, $A$ is diagonalizable. Its spectrum is derived in Appendix \ref{diagonalization} (see Eqs (\ref{lambdapmExpr},\ref{AVLambdaV})). 
 The three eigenvalues are expressed as follows:
\begin{equation}
\begin{split}
\lambda_1 :=& 0,\\
\lambda_2 :=& \lambda_+ =   \frac{\varphi(1- \mu ^2)\, -2\,\mu ^2 + \Delta}{2\,\mu\left(1 -\mu^2\right)},\\
\lambda_3 :=& \lambda_- =  \frac{\varphi(1- \mu ^2)\,  -2\,\mu ^2 - \Delta}{2\,\mu\left(1 -\mu^2\right)},\\
{\mathrm{with}}\;\;
\Delta =& \sqrt{\varphi ^2(1 - \mu^2 )^2+4\,\mu ^2}.
\end{split}
\end{equation}
Note that $\lambda_+$ (resp. $\lambda_-$) is positive if the drift is positive (resp. negative). We can write $A$ in terms of the eigenvalues and the presentation matrix $V$ of the  associated eigenvectors as
\begin{equation}\label{ADiag}
A = V \begin{pmatrix}
  0 & 0 & 0\\
    0 & \lambda_+ & 0\\
    0 & 0 & \lambda_-
\end{pmatrix}
V^{-1},
\end{equation}
with 
\begin{equation}
V:= 
\begin{pmatrix}
1 & u_+ & u_-\\
1 & v_+ & v_-\\
0 & 1 &  1  
\end{pmatrix},
%=  \begin{pmatrix}
%1 &   -\frac{\varphi \,\left(\varphi -\mu ^2\,\varphi +\Delta+2\right)}{2\,\mu \,\left(\varphi +1\right)}   &  -\frac{\varphi \,\left(\varphi -\mu ^2\,\varphi -\Delta+2\right)}{2\,\mu \,\left(\varphi +1\right)}\\
%1 &   -\frac{\varphi +\mu ^2\,\varphi -\Delta}{2\,\mu \,\left(\varphi +1\right)}   &  -\frac{\varphi +\mu ^2\,\varphi +\Delta}{2\,\mu \,\left(\varphi +1\right)}\\
%0 & 1 &  1  
%\end{pmatrix}
\end{equation}
where the coordinates $u_\pm$ and $v_\pm$ are expressed in terms of the drift $\mu$ and the rate $\varphi$ in Eq. \eqref{uv+-} of Appendix \ref{diagonalization}.\\

 The evolution equation is readily integrated as 
 \begin{equation}\label{ExExpr}
      \vec{\mathcal{E}}(x) = \exp( (x-a) A) \vec{\mathcal{E}}(a). 
 \end{equation}
The diagonalization as expressed in Eq. \eqref{ADiag} induces an expression 
 for the matrix exponential
\begin{equation}\label{qMatrixDef}
 Q(y):= \exp( yA) = \sum_{k=1}^3 q^{(k)} e^{\lambda_k y},
\end{equation}
 where $q^{(1)}$, $q^{(2)}$, $q^{(3)}$ are three matrices of size 3 expressed in terms of the coordinates of the eigenvectors of $A$: 
 \begin{equation}\label{qMatrixDefPrime}
q^{(l)}_{ik} := V_{il}  (V^{-1})_{lk},\;\;\;\;i,k,l\in 
\{1,2,3\}.
 \end{equation} 
  These entries are expressed in  Appendix \ref{diagonalization} (Eq. (\ref{qExpr})) in terms of the coordinates $u_\pm$ and $v_\pm$ of the eigenvectors of $A$. 
 The integration of the evolution equations is straightforward. It will prove easier to use the quantities $u_\pm$ and $v_\pm$ and to postpone the substitution in terms of the parameters $\mu$ and $\varphi$. One of the reasons for this is that $|u_+|$ becomes large in the limit of large $\varphi$, while $u_-,v_+,v_-$ go to finite limits (see Eq. (\ref{equivEntriesInf})). Expressing the solution in terms of $u_\pm$ and $v_\pm$ will make it easier to check consistency with the case of instantaneous tumble as described in  Eq. (\ref{TGE}) (this is be done in Sections \ref{checkENeg},\ref{checkEPos}). We conclude that
 \begin{equation}\label{ExExprq}
      \vec{\mathcal{E}}(x) =  \sum_{k=1}^3 e^{\lambda_k( x-a)} q^{(k)}\vec{\mathcal{E}}(a).
 \end{equation} 
 In particular, the values  $\vec{\mathcal{E}}(b)$ and  $\vec{\mathcal{E}}(a)$ are related by
\begin{equation}\label{EbEa}
\vec{\mathcal{E}}(b) = 
 \sum_{k=1}^3 e^{\lambda_k( b-a)} q^{(k)}\vec{\mathcal{E}}(a). 
\end{equation}
 The boundary conditions can be encoded by parametrizing $\vec{\mathcal{E}}(a)$ and  $\vec{\mathcal{E}}(b)$
 in a way that takes into account  the sign of the drift (this is done in Sections \ref{subsubsec:icndproba},\ref{subsubsec:icpdproba}). 

\subsection{Negative drift}\label{subsec:ndproba}

Consider the case of a negative drift. A particle living on the segment $[a,b]$ that starts its motion at $a$ in a tumbling state or in a negative velocity state  leaves the system immediately. On the other hand, a particle that starts at  $b$ with a positive velocity leaves the system immediately. Let us  express the vector $\vec{\mathcal{E}}(a)$ in terms of this unknown constant as 

%Let us introduce the shorthand notation for the exit probability of a particle starting at $a$ in a positive velocity state,
%\begin{equation}
% {\mathcal{E}}_a(a,+;[a,b],\varphi):= {\mathcal{E}}_a(a,+;[a,b],\varphi).
%\end{equation}

\begin{equation}\label{paramNeg}
\vec{\mathcal{E}}(a) = 
\begin{pmatrix}
 1\\
 \frac{{\mathcal{E}}_a(a,+;[a,b],\varphi) + 1}{2}\\
  \frac{{\mathcal{E}}_a(a,+;[a,b],\varphi) - 1}{2}
\end{pmatrix}.
%\qquad
%\vec{\mathcal{E}}(b) = 
%\begin{pmatrix}
% Z_a(b)\\
% E_a(b)\\
% -E_a(b)
%\end{pmatrix}
\end{equation}
Substituting the above expression of into the solution found in Eq. \eqref{ExExprq}, using linearity to express $q^{(k)}\vec{\mathcal{E}}(a)$ yields 
\begin{equation}\label{vecExExprPre}
\begin{split}
 {\vec{\mathcal{E}}}(x) = & \sum_{k=1}^3 e^{\lambda_k(x-a)} \left[ q^{(k)}
 \begin{pmatrix}
1\\
1\\
0 
\end{pmatrix} 
+\frac{{\mathcal{E}}_a(a,+;[a,b],\varphi) - 1}{2}  q^{(k)} \begin{pmatrix}
0\\
1\\
1 
\end{pmatrix} \right].
\end{split}
\end{equation}

The following matrix products are readily evaluated using the expression of the matrices $(q^{(i)})_{1\leq i \leq 3}$ in terms of the coordinates $u_\pm,v_\pm$ given in Eq. \eqref{uv+-}:
\begin{equation}\label{algIdq}
\begin{split}
q^{(1)}\begin{pmatrix}
1\\
1\\
0 
 \end{pmatrix} =& \begin{pmatrix}
1\\
1\\
0 
 \end{pmatrix},\qquad q^{(2)}\begin{pmatrix}
1\\
1\\
0 
 \end{pmatrix} = q^{(3)}\begin{pmatrix}
1\\
1\\
0 
 \end{pmatrix} = \vec{0},\\
q^{(1)}\begin{pmatrix}
0\\
1\\
1 
\end{pmatrix}=& \frac{u_- - u_+ - u_- v_+ + u_+ v_-}{u_- - u_+ - v_- + v_+}
\begin{pmatrix}
1\\
1\\
0 
\end{pmatrix},\\
q^{(2)}\begin{pmatrix}
0\\
1\\
1 
\end{pmatrix}=& \frac{ u_- - v_- + 1}{u_- - u_+ - v_- + v_+}
\begin{pmatrix}
u_+\\
v_+\\
 1
\end{pmatrix},\\
q^{(3)}\begin{pmatrix}
0\\
1\\
1 
\end{pmatrix}=& -\frac{u_+ - v_+ + 1 }{u_- - u_+ - v_- + v_+}
\begin{pmatrix}
 u_-\\
 v_-\\
 1 
\end{pmatrix}.\\
\end{split}
\end{equation}

Using linearity, the expression of $\vec{\mathcal{E}}(x)$ in terms of the integration constant ${\mathcal{E}}_a(a,+;[a,b],\varphi)$ becomes
\begin{equation}\label{vecExExpr}
\begin{split}
 {\vec{\mathcal{E}}}(x)   
=& 
\begin{pmatrix}
1\\
1\\
0 
\end{pmatrix} +\frac{{\mathcal{E}}_a(a,+;[a,b],\varphi) - 1}{2} \sum_{k=1}^3 e^{\lambda_k(x-a)} q^{(k)}
 \begin{pmatrix}
0\\
1\\
1 
\end{pmatrix} \\
=&  \begin{pmatrix}
1\\
1\\
0 
\end{pmatrix} +\frac{{\mathcal{E}}_a(a,+;[a,b],\varphi) - 1}{2( u_- - u_+ - v_- + v_+)} \left[
(u_- - u_+ - u_- v_+ + u_+ v_-)
\begin{pmatrix}
1\\
1\\
0 
\end{pmatrix}\right.\\
& \left.
+ e^{\lambda_+(x-a)} (u_- - v_- + 1)
\begin{pmatrix}
u_+\\
v_+\\
 1
\end{pmatrix}
- e^{\lambda_-(x-a)} (u_+ - v_+ + 1)
\begin{pmatrix}
 u_-\\
 v_-\\
 1 
\end{pmatrix}
\right].\\
%=&  \begin{pmatrix}
%1\\
%1\\
%0 
%\end{pmatrix} +\frac{{\mathcal{E}}_a(a,+;[a,b],\varphi) - 1}{2( u_- - u_+ - v_- + v_+)}
%\begin{pmatrix}
% u_- - u_+ - u_- v_+ + u_+ v_- +   e^{\lambda_+(x-a)} (u_- - v_- + 1)u_+  - e^{\lambda_-(x-a)} (u_+ - v_+ + 1) u_- \\ 
% u_- - u_+ - u_- v_+ + u_+ v_- +  e^{\lambda_+(x-a)} (u_- - v_- + 1)v_+   -  e^{\lambda_-(x-a)} (u_+ - v_+ + 1) v_-\\
%      e^{\lambda_+(x-a)} (u_- - v_- + 1) - e^{\lambda_-(x-a)} (u_+ - v_+ + 1)
%\end{pmatrix}.
\end{split}
\end{equation}

Using the notations introduced in Eqs (\ref{notationsE},\ref{ODEExit}), the exit probabilities are therefore expressed in terms of the integration constant ${\mathcal{E}}_a(a,+;[a,b],\varphi)$ as
{\small{
\begin{equation}\label{ExitProbaNegDriftuv}
\begin{split}
{\mathcal{E}}_a(x,0;[a,b],\varphi) =1 +& \frac{{\mathcal{E}}_a(a,+;[a,b],\varphi) - 1}{2( u_- - u_+ - v_- + v_+)}\left[    u_- - u_+ - u_- v_+ + u_+ v_-  \right.\\
&\left.+   e^{\lambda_+(x-a)} (u_- - v_- + 1)u_+  - e^{\lambda_-(x-a)} (u_+ - v_+ + 1) u_-\right],\\
{\mathcal{E}}_a(x,+;[a,b],\varphi) =1 +& \frac{{\mathcal{E}}_a(a,+;[a,b],\varphi) - 1}{2( u_- - u_+ - v_- + v_+)}\left[ u_- - u_+ - u_- v_+ + u_+ v_- \right.\\
&\left. +  e^{\lambda_+(x-a)} (u_- - v_- + 1)(v_+ + 1 )   -  e^{\lambda_-(x-a)} (u_+ - v_+ + 1) (v_- + 1 )\right],\\
{\mathcal{E}}_a(x,-;[a,b],\varphi) = 1  +& \frac{{\mathcal{E}}_a(a,+;[a,b],\varphi) - 1}{2( u_- - u_+ - v_- + v_+)} \left[  u_- - u_+ - u_- v_+ + u_+ v_- \right.\\
&\left. +  e^{\lambda_+(x-a)} (u_- - v_- + 1)(v_+ - 1 )   -  e^{\lambda_-(x-a)} (u_+ - v_+ + 1) (v_- - 1 )\right].
\end{split}
\end{equation}
}}

To obtain these exit probabilities in terms of the parameters $\mu$ and $\varphi$, we substitute the expressions of the coordinates $u_\pm,v_\pm$ (see Eq. \eqref{uv+-}).  We obtain
\begin{equation}\label{resExitNegativeDrift}
\begin{split}
\mathcal{E}_a(x,0;[a,b],\varphi)& = 1 + \frac{{\mathcal{E}}_a(a,+;[a,b],\varphi) - 1}{2}\left[ \frac{\varphi( \mu + 1)}{\mu(\varphi + 1 )} \right.\\
&+\frac{\varphi \,\left(\mu +1\right)\,\left(-2\,\mu +\varphi (1-\mu)^2  -\Delta\right)}{2\,\mu \,\left(\varphi +1\right)\, \Delta}e^{\lambda_+(x-a)} \\
&\left.+  \frac{\varphi \,\left(\mu +1\right)\,\left(2\,\mu - \varphi (1-\mu)^2  -\Delta\right)}{2\,\mu \,\left(\varphi +1\right)\, \Delta}
 e^{\lambda_-(x-a)} \right],\\
\mathcal{E}_a(x,+;[a,b],\varphi)&  = 1 + \frac{{\mathcal{E}}_a(a,+;[a,b],\varphi) - 1}{2}\left[ \frac{\varphi( \mu + 1)}{\mu(\varphi + 1 )}\right.\\
&+ \frac{4\mu^2 + 2\varphi \mu( \mu - 1)+\varphi^2 (1-\mu -\mu^2 + \mu^3) +  (2\mu -\varphi + \mu \varphi) \Delta}{2\,\mu \,\left(\varphi +1\right)\, \Delta}
e^{\lambda_+(x-a)}\\ 
&\left.
+\frac{-[4\mu^2 + 2\varphi \mu( \mu - 1)+\varphi^2 (1-\mu -\mu^2 + \mu^3)] +  (2\mu -\varphi + \mu \varphi) \Delta}{2\,\mu \,\left(\varphi +1\right)\, \Delta}
 e^{\lambda_-(x-a)} \right],\\
\mathcal{E}_a(x,-;[a,b],\varphi) & = 1 + \frac{{\mathcal{E}}_a(a,+;[a,b],\varphi) - 1}{2}\left[  \frac{\varphi( \mu + 1)}{\mu(\varphi + 1 )} \right.\\
& +\frac{ \varphi \,\left(\mu +1\right)\,\left(  -2\mu^2 +\varphi( 1-\mu ^2)  +  \Delta \right)}{2\,\mu \,\left(\varphi + 1\right)\, \Delta}
  e^{\lambda_+(x-a)} \\
&\left.+ \frac{\varphi \,\left(\mu +1\right)\,\left(  2\mu^2 -\varphi( 1-\mu ^2)  +  \Delta \right)}{2\,\mu \,\left(\varphi +1\right)\, \Delta}
 e^{\lambda_-(x-a)}\right],\\
 &\shoveright{\qquad\qquad\hspace{5cm}\text{with}\quad \Delta = \sqrt{\varphi ^2(1 - \mu^2 )^2+4\,\mu ^2}.}
\end{split}
\end{equation}

To obtain the full expression of the exit probabilities, we must calculate the integration constant ${\mathcal{E}}_a(a,+;[a,b],\varphi)$.\\

\subsubsection{Integration constant}\label{subsubsec:icndproba}
 To fix  the integration constant ${\mathcal{E}}_a(a,+;[a,b],\varphi)$, the parametrization of the vector $\vec{\mathcal{E}}(a)$ introduced in Eq. (\ref{paramNeg}) can be complemented by a parametrization of $\vec{\mathcal{E}}(b)$ 
using the boundary conditions. Indeed, a particle starting at coordinate $b$ leaves the system immediately (through $b$, not $a$). Hence, using the notations introduced in Eq. (\ref{notationsE}),
 \begin{equation}
 \mathcal{E}_a(b,+;[a,b],\varphi) = 0.
 \end{equation}
 The vector  $\vec{\mathcal{E}}(b)$ can therefore be written in terms of $Z_a(b)$ and $E_a(b)$, as
\begin{equation} 
\vec{\mathcal{E}}(b) = \begin{pmatrix}
 Z_a(b)\\
 E_a(b)\\
 -E_a(b)
\end{pmatrix}.
 \end{equation}
The solution of the evolution equation relates $\vec{\mathcal{E}}(b)$ and  $\vec{\mathcal{E}}(a)$ as stated in Eq. (\ref{EbEa}). It  therefore induces the linear system
\begin{equation}
\begin{pmatrix}
 Z_a(b)\\
 E_a(b)\\
 -E_a(b)
\end{pmatrix} = \exp\left( (b-a) A \right)
\begin{pmatrix}
 1\\
 \frac{{\mathcal{E}}_a(a,+;[a,b],\varphi) + 1}{2}\\
  \frac{{\mathcal{E}}_a(a,+;[a,b],\varphi) - 1}{2}
\end{pmatrix}.
\end{equation}
Let us use the shorthand notation $Q_{ij}$ for the entry of the matrix $Q(b-a) = \exp( (b-a)A)$ at row $i$ and column $j$:
\begin{equation}\label{shorthandQ}
Q_{ij}:= \sum_{k=1}^3 q^{(k)}_{ij} e^{\lambda_k(b-a)}. 
\end{equation}
 The above system is equivalently written as
\begin{equation}
S_-\begin{pmatrix}
 {\mathcal{E}}_a(a,+;[a,b],\varphi)\\
 Z_a(b)\\
 E_a(b)
\end{pmatrix}
   = 
\begin{pmatrix} 
Q_{11} + \frac{Q_{12} - Q_{13}}{2}\\
Q_{21} + \frac{Q_{22} - Q_{23}}{2}\\
Q_{31} + \frac{Q_{32} - Q_{33}}{2}
\end{pmatrix}, \qquad {\mathrm{with}}\quad S_- := 
\begin{pmatrix} 
-\frac{1}{2}(Q_{12} + Q_{13} ) & 1 & 0\\
-\frac{1}{2}(Q_{22} + Q_{23} )& 0 & 1\\
 -\frac{1}{2}(Q_{32} + Q_{33} ) & 0 & -1
\end{pmatrix}.
\end{equation}
 The integration constant ${\mathcal{E}}_a(a,+;[a,b],\varphi)$ is therefore evaluated as the first component of the vector
\begin{equation}
\begin{pmatrix}
 {\mathcal{E}}_a(a,+;[a,b],\varphi)\\
 Z_a(b)\\
 E_a(b)
\end{pmatrix} = (S_-)^{-1}
    \begin{pmatrix} 
Q_{11} + \frac{Q_{12} - Q_{13}}{2}\\
Q_{21} + \frac{Q_{22} - Q_{23}}{2}\\
Q_{31} + \frac{Q_{32} - Q_{33}}{2}
\end{pmatrix}.
\end{equation}

The matrix $S_-$ is readily inverted as
\begin{equation}
{\small{
\begin{split}
(S_-)^{-1} = \frac{1}{Q_{22} + Q_{23}+ Q_{32} + Q_{33}}
\begin{pmatrix}
  0 & -2 & -2 \\
    Q_{22} + Q_{23} + Q_{32} + Q_{33} & -(Q_{12} + Q_{13}) & -(Q_{12} + Q_{13}) \\
 0 & (Q_{32} + Q_{33}) & -(Q_{22} + Q_{23})
\end{pmatrix}.
\end{split}
}}
\end{equation}
The constant ${\mathcal{E}}_a(a,+;[a,b],\varphi)$ is therefore obtained as 
{\tiny{
\begin{equation}\label{EaPlusNegDrift}
\begin{split}
{\mathcal{E}}_a(a,+;[a,b],\varphi)=&  \frac{-2Q_{21} - 2Q_{31} - Q_{22}  + Q_{23}  -Q_{32} + Q_{33} }{Q_{22} + Q_{23} + Q_{32} + Q_{33}}\\
=&  \frac{\sum_{k=1}^3 [-2q^{(k)}_{21} - 2q^{(k)}_{31} - q^{(k)}_{22}  + q^{(k)}_{23}  - q^{(k)}_{32} + q^{(k)}_{33}]e^{\lambda_k( b-a )} }{ \sum_{k=1}^3  [ q^{(k)}_{22} + q^{(k)}_{23} + q^{(k)}_{32}  + q^{(k)}_{33} ] e^{\lambda_k( b-a )}}\\
=& \frac{ -u_- + u_+ + 2 v_- - 2v_+ - u_- v_+ + u_+ v_-) + e^{\lambda_+(b-a)} (v_+ + 1)( u_- - v_- + 1 )  - e^{\lambda_-(b-a)} (v_- + 1)( u_+ - v_+ + 1 ) }{   u_- - u_+ - u_- v_+ + u_+ v_-   +
e^{\lambda_+(b-a)} (v_+ + 1)( u_- - v_- + 1 ) 
- e^{\lambda_-(b-a)} (v_- + 1)( u_+ - v_+ + 1 )   }.
\end{split}
\end{equation}
}}
 In the last step we used Eq. (\ref{qExpr})  for the expression of the matrices $(q^{(k)})_{1\leq k\leq 3}$ in terms of the components of the eigenvectors of the matrix $A$.\\

%%%%%%%%%%%%%%%%%%%%%%%%%%%%%%%%%%%%%%%

 Substituting the expression of the components of the eigenvectors in terms of the drift and rate $\varphi$ into Eq. (\ref{EaPlusNegDrift}) yields
  \begin{equation}\label{EaPlusNegDriftExpl}
\begin{split}
\mathcal{E}_a(a,+;[a,b], \varphi)  =&\frac{\frac{(\varphi - 2\mu -\mu \varphi)\Delta }{\mu^2 \left(\varphi +1\right)}+X_+ e^{\lambda_+(b-a)}
+X_- e^{\lambda_-(b-a)}   }{\frac{\varphi \left(\mu +1\right)\Delta}{\mu^2 \left(\varphi +1\right)} + X_+ e^{\lambda_+(b-a)} + X_- e^{\lambda_-(b-a)} },\\
 \text{where} \qquad
X_+ =& \frac{ 4\mu^2 + 2\varphi \mu( \mu - 1)+\varphi^2 (1-\mu -\mu^2 + \mu^3) +  (2\mu -\varphi + \mu \varphi) \Delta}{2\,\mu ^2\,\left(\varphi +1\right)},\\
X_- =& \frac{-[4\mu^2 + 2\varphi \mu( \mu - 1)+\varphi^2 (1-\mu -\mu^2 + \mu^3)] +  (2\mu -\varphi + \mu \varphi) \Delta}{2\,\mu ^2\,\left(\varphi +1\right)}.
\end{split}
\end{equation}
This probability of first-return to $a$, together with Eq. (\ref{resExitNegativeDrift}), is the solution of the evolution equation on $[a,b]$ with a negative drift. Consistency with Eq. (\ref{TGE}) is checked in Section \ref{checkENeg}.\\

\subsubsection{Limit of a half-line} \label{subsubsec:limitndproba}
 For a negative drift, the largest eigenvalue is the positive $\lambda_-$ (see Eq. \eqref{lambdaPMSigns}). In the limit of large $b$, 
 \begin{equation}
e^{\lambda_-(b-a)}\gg 1 \gg e^{\lambda_+(b-a)},
\end{equation}
 and 
 the dominant terms in the numerator and denominator in the expression of $E_a(a, +)$ in Eq. (\ref{EaPlusNegDrift}) are therefore of order $O( e^{\lambda_-(b-a)})$. We read off the large-$b$ limit of ${\mathcal{E}}_a(a,+;[a,b],\varphi)$ as
\begin{equation}
\begin{split}
\mathcal{E}_a(x,+;[a,+\infty[, \varphi) \underset{b\to +\infty}{\sim}& 1, \qquad (\mu < 0).
\end{split}
\end{equation}
 Substituting into the solution found in Eq. (\ref{ExExpr}) yields
\begin{equation}\label{almostSurely}
 \mathcal{E}_a(x,0;[a,+\infty[, \varphi) = \mathcal{E}_a(x,+;[a,+\infty[, \varphi) =  \mathcal{E}_a(x,-;[a,+\infty[, \varphi)  = 1,\qquad (x\geq a ).
\end{equation}
 The particle therefore exits the system almost surely for any initial value of the internal velocity state, and for any value of the parameter $\varphi$. This is intuitive since the particle moves towards $a$ during tumble events, and the RTP with instantaneous tumbles is known  to exit the half-line almost surely if the drift is negative (see Eq. (\ref{exitOrdinary}) in the case $\mu<0$).\\

\subsubsection{Consistency checks}\label{checkENeg}
The quantities $u_+,u_-, v_+,v_-$ are components of the eigenvectors of the matrix $A$, expressed in terms of the parameters $\mu$ and $\varphi$ in Eq. \eqref{uv+-}. Their behavior in the limit of instantaneous tumbles is studied in Appendices \ref{appShortTumble}). Let us check that the expression we obtained for the probability of return to $a$, or ${\mathcal{E}}_a(a,+;[a,b],\varphi)$, is  consistent with known results in the limit of instantaneous tumbles.\\

In the limit of large rate $\varphi$, the component $u_+$ becomes large, while $u_-$, $v_-$, $v_+$ go to finite limits (given in Eq. \eqref{equivEntriesInf}). Moreover $\lambda_+$ becomes large and negative (see Eq. (\ref{lambdaLim}). Hence,  the expression of the probability ${\mathcal{E}}_a(a,+;[a,b],\varphi)$ obtained in Eq. (\ref{EaPlusNegDrift}) in the case of a negative drift has the following limit when $\varphi$ goes to infinity: 
 \begin{equation}
\begin{split}
 {\mathcal{E}}_a(a,+;[a,b],\varphi) \underset{\varphi \to +\infty}{\sim}& \frac{v_- +1 - (v_-+1) e^{-\frac{\mu}{1-\mu^2}(b-a)}}{ v_- - 1 - (v_- + 1) e^{-\frac{\mu}{1-\mu^2}(b-a)}} \sim \frac{1 -\mu+ (\mu - 1) e^{-\frac{\mu}{1-\mu^2}(b-a)}}{ \mu + 1 + (\mu - 1) e^{-\frac{\mu}{1-\mu^2}(b-a)}}.\\ 
  \end{split}   
 \end{equation}
In this limit,  the exit probability of a particle starting its motion at $a$ in a positive velocity state becomes
\begin{equation}\label{consistencyEaaLargephi}
    \mathcal{E}_a(a, +; [a,b], \infty):=\underset{\varphi \to \infty}{\lim} {\mathcal{E}}_a(a,+;[a,b],\varphi)= \frac{1 -\mu+ (\mu - 1) e^{-\frac{\mu}{1-\mu^2}(b-a)}}{ \mu + 1 + (\mu - 1) e^{-\frac{\mu}{1-\mu^2}(b-a)}}.
\end{equation}
 This is the result of \cite{gueneau2024relating} stated in Eq. (\ref{TGE}) at coordinate $x=a$.\\

Moreover, using again Eqs (\ref{equivEntriesInf},\ref{lambdaLim}), the expression of the exit probabilities 
 ${\mathcal{E}}_a(x,+;[a,b],\varphi)$ and  ${\mathcal{E}}_a(x,+;[a,b],\varphi)$ given in Eq. (\ref{ExitProbaNegDriftuv}) have the following large-$\varphi$ limits, obtained by suppressing the term $e^{\lambda_+(x-a)}$ and collecting the coefficient of $u_+$ in the numerators and denominators: 
\begin{equation}
\begin{split}
\mathcal{E}_a(x,\pm;[a,b],\varphi ) \underset{\varphi\to \infty}{\sim}&
1 - \frac{1}{2}\left(\mathcal{E}_a(a, +; [a,b], \infty) - 1\right)[ v_- 1 -( v_- \pm 1 ) e^{\lambda_-(x-a)} ]
\\ 
\underset{\varphi\to \infty}{\sim} & 
\frac{(\mu - 1 ) e^{-\frac{\mu(b-a)}{1-\mu^2}}  -(\pm \mu -1)  e^{ -\frac{\mu(x-a)}{1-\mu^2}}}{ \mu + 1 + (\mu - 1) e^{-\frac{\mu(b-a)}{1-\mu^2}}}.
\end{split}
\end{equation}
This reproduces Eq. (\ref{TGE}) for any coordinate $x\geq a$.\\

 The quantity ${\mathcal{E}}_a(x,0;[a,b],\varphi)$ is undefined in the ordinary model of RTP with instantaneous tumbles, which has only two internal velocity states, but the large-$\varphi$ limit of Eq. (\ref{ExitProbaNegDriftuv}) yields
\begin{equation}
\begin{split}
\mathcal{E}_a(x,0;[a,b],\varphi ) \underset{\varphi\to \infty}{\sim}&
1 - \frac{1}{2}\left(\mathcal{E}_a(a, +; [a,b], \infty) - 1\right)[ v_- - 1 -u_- e^{\lambda_-(x-a)} ]\\
\underset{\varphi\to \infty}{\sim}&
\frac{(\mu - 1 ) e^{-\frac{\mu(b-a)}{1-\mu^2}}  +  e^{ -\frac{\mu(x-a)}{1-\mu^2}}}{ \mu + 1 + (\mu - 1) e^{-\frac{\mu(b-a)}{1-\mu^2}}}.\\
\end{split}
\end{equation}
We recognize the average of the exit probabilities for each of the non-zero initial velocity states, in the limit of instantaneous tumble:
\begin{equation}
    \mathcal{E}_a(x,0;[a,b],\infty ) = \frac{1}{2}\left[   \mathcal{E}_a(x,+;[a,b],\infty )  +  \mathcal{E}_a(x,-;[a,b],\infty ) \right]. 
\end{equation}
 This is intuitively consistent: in the limit of instantaneous tumble, a particle starting its motion in a tumbling state instantaneously picks a random non-zero internal velocity state.\\

\subsection{Positive drift}\label{sec:pdproba}

In the presence of a positive drift, a particle that starts at $b$ in a tumbling state leaves the system immediately (through $b$, not $a$). On the other hand, if it starts at $a$ with a negative internal velocity (or at $b$ with a positive internal velocity), it leaves the system immediately, as in the case of a negative drift. The vectors $\vec{\mathcal{E}}(a)$ and $\vec{\mathcal{E}}(b)$ may therefore be parametrized as follows:  
\begin{equation}\label{defEaPosDrift}
\vec{\mathcal{E}}(a) = 
\begin{pmatrix}
 Z_a(a)\\
 \frac{{\mathcal{E}}_a(a,+;[a,b],\varphi) + 1}{2}\\
 \frac{{\mathcal{E}}_a(a,+;[a,b],\varphi) - 1 }{2}
\end{pmatrix},\qquad
\vec{\mathcal{E}}(b) = 
\begin{pmatrix}
 0\\
 E_a(b)\\
 -E_a(b)
\end{pmatrix}.
\end{equation}
With this parametrization, the solution displayed in Eq. (\ref{ExExprq}) becomes (using linearity of $q^{(k)}$ and  the algebraic identities of Eq.\eqref{algIdq})
{\footnotesize{
\begin{equation}\label{vecExExprPosDrift}
\begin{split}
 {\vec{\mathcal{E}}}(x) =&  \sum_{k=1}^3 e^{\lambda_k(x-a)} q^{(k)} \vec{\mathcal{E}}(a) \\
 %=&
 %\sum_{k=1}^3 e^{\lambda_k(x-a)} q^{(k)} 
% \left( 
 %(Z_a(a) - 1)   
 %\begin{pmatrix}
%1\\
%0\\
%0 
% +
% \begin{pmatrix}
%1\\
%1\\
%0 
%\end{pmatrix} 
%+ \frac{{\mathcal{E}}_a(a,+;[a,b],\varphi) - 1}{2} 
%\begin{pmatrix}
%0\\
%1\\
%1 
%\end{pmatrix} \right)\\    
=&  \sum_{k=1}^3 e^{\lambda_k(x-a)}\begin{pmatrix}
1\\
1\\
0 
\end{pmatrix} +
 (Z_a(a) - 1 ) \sum_{k=1}^3 e^{\lambda_k(x-a)} q^{(k)}
 \begin{pmatrix}
1\\
0\\
0 
\end{pmatrix} 
+
\frac{{\mathcal{E}}_a(a,+;[a,b],\varphi) - 1}{2} \sum_{k=1}^3 e^{\lambda_k(x-a)} q^{(k)}
 \begin{pmatrix}
0\\
1\\
1 
\end{pmatrix} \\
=&  \begin{pmatrix}
1\\
1\\
0 
\end{pmatrix} +\frac{{\mathcal{E}}_a(a,+;[a,b],\varphi) - 1}{2( u_- - u_+ - v_- + v_+)} \left[
(u_- - u_+ - u_- v_+ + u_+ v_-)
\begin{pmatrix}
1\\
1\\
0 
\end{pmatrix}\right.\\
& \left.
+ e^{\lambda_+(x-a)} (u_- - v_- + 1)
\begin{pmatrix}
u_+\\
v_+\\
 1
\end{pmatrix}
- e^{\lambda_-(x-a)} (u_+ - v_+ + 1)
\begin{pmatrix}
 u_-\\
 v_-\\
 1 
\end{pmatrix}
\right]\\
&+ \frac{Z_a(a) - 1}{u_- - u_+ -v_- + v_+} \left[
(v_+ - v_-)
\begin{pmatrix}
1\\
1\\
0 
\end{pmatrix}
- e^{\lambda_+(x-a)}
\begin{pmatrix}
u_+\\
v_+\\
0 
\end{pmatrix}
+e^{\lambda_-(x-a)}
\begin{pmatrix}
u_-\\
v_-\\
0 
\end{pmatrix}
\right].
%=&  {\tiny{\begin{pmatrix}
%1\\
%1\\
%0 
%\end{pmatrix} +\frac{{\mathcal{E}}_a(a,+;[a,b],\varphi) - 1}{2( u_- - u_+ - v_- + %v_+)}
%\begin{pmatrix}
% u_- - u_+ - u_- v_+ + u_+ v_- +   e^{\lambda_+(x-a)} (u_- - v_- + 1)u_+  - e^{\lambda_-(x-a)} (u_+ - v_+ + 1) u_- \\ 
% u_- - u_+ - u_- v_+ + u_+ v_- +  e^{\lambda_+(x-a)} (u_- - v_- + 1)v_+   -  e^{\lambda_-(x-a)} (u_+ - v_+ + 1) v_-\\
 %     e^{\lambda_+(x-a)} (u_- - v_- + 1) - e^{\lambda_-(x-a)} (u_+ - v_+ + 1)
%\end{pmatrix}}}.
\end{split}
\end{equation}
}} 
From the definition of $\vec{\mathcal{E}}$ in Eqs (\ref{ODEExit},\ref{compProbE}), the exit probabilities are expressed as follows:
\begin{equation}\label{ZeePosDrift}
\begin{split}
{\mathcal{E}}_a(x,0;[a,b],\varphi) = 1 + &\frac{{\mathcal{E}}_a(a,+;[a,b],\varphi) - 1}{2( u_- - u_+ - v_- + v_+)}\left[    u_- - u_+ - u_- v_+ + u_+ v_-\right.\\
&\left.+   e^{\lambda_+(x-a)} (u_- - v_- + 1)u_+  - e^{\lambda_-(x-a)} (u_+ - v_+ + 1) u_-\right]\\
&+ \frac{Z_a(a) - 1}{u_- - u_+ - v_- + v_+}\left[ v_+ - v_- - u_+ e^{\lambda_+(x-a)} + u_- e^{\lambda_-(x-a)} \right] ,\\
{\mathcal{E}}_a(x,+;[a,b],\varphi) = 1 + &\frac{{\mathcal{E}}_a(a,+;[a,b],\varphi) - 1}{2( u_- - u_+ - v_- + v_+)}\left[ u_- - u_+ - u_- v_+ + u_+ v_- \right.\\
&\left.+  e^{\lambda_+(x-a)} (u_- - v_- + 1)(v_+ + 1 )   -  e^{\lambda_-(x-a)} (u_+ - v_+ + 1) (v_- + 1 )\right],\\
&+ \frac{Z_a(a) - 1}{u_- - u_+ - v_- + v_+}\left[ v_+ - v_- - v_+ e^{\lambda_+(x-a)} + v_- e^{\lambda_-(x-a)} \right],\\
{\mathcal{E}}_a(x,-;[a,b],\varphi) = 1  +& \frac{{\mathcal{E}}_a(a,+;[a,b],\varphi) - 1}{2( u_- - u_+ - v_- + v_+)}\left[  u_- - u_+ - u_- v_+ + u_+ v_-\right.\\
&\left.+  e^{\lambda_+(x-a)} (u_- - v_- + 1)(v_+ - 1 )   -  e^{\lambda_-(x-a)} (u_+ - v_+ + 1) (v_- - 1 )\right]\\
&+ \frac{Z_a(a) - 1}{u_- - u_+ - v_- + v_+}\left[  -  e^{\lambda_+(x-a)} + e^{\lambda_-(x-a)} \right].
\end{split}
\end{equation}

To obtain the full expression of the exit probabilities, we must fix the integration constants  ${\mathcal{E}}_a(a,+;[a,b],\varphi)$ and $Z_a(a)={\mathcal{E}}_a(a,0;[a,b],\varphi)$ (the exit probabilities of a particle starting its motion at the left end of the segment $[a,b]$ with positive and zero internal velocity respectively).

 \subsubsection{Integration constants}\label{subsubsec:icpdproba}

To fix the integration constants ${\mathcal{E}}_a(a,+;[a,b],\varphi)$ and $Z_a(a)$, let us use the parametrization of the vectors $\vec{\mathcal{E}}(a)$ and $\vec{\mathcal{E}}(b)$ introduced in Eq. \eqref{defEaPosDrift}. It induces the linear system
\begin{equation}
\begin{pmatrix}
  0\\
 \frac{1}{2}{\mathcal{E}}_a(b,-;[a,b],\varphi) \\
 -\frac{1}{2}{\mathcal{E}}_a(b,-;[a,b],\varphi) 
\end{pmatrix} = \exp( (b-a) A ) 
\begin{pmatrix}
 Z_a(a)\\
 \frac{{\mathcal{E}}_a(a,+;[a,b],\varphi) + 1}{2}\\
  \frac{{\mathcal{E}}_a(a,+;[a,b],\varphi) - 1}{2}
\end{pmatrix}.
\end{equation}
 using again the shorthand notation $Q_{ij}$ introduced in Eq. \eqref{shorthandQ} for the entries of the matrix $Q(b-a) = \exp( (b-a) A)$, the above system is equivalently written as
\begin{equation}\label{defSPlus}
S_+\begin{pmatrix}
 Z_a( a )\\
 {\mathcal{E}}_a(a,+;[a,b],\varphi)\\
 {\mathcal{E}}_a(b,-;[a,b],\varphi) \\
\end{pmatrix}
   = 
\begin{pmatrix} 
 \frac{Q_{12} - Q_{13}}{2}\\
 \frac{Q_{22} - Q_{23}}{2}\\
 \frac{Q_{32} - Q_{33}}{2}
\end{pmatrix}, \qquad {\mathrm{with}}\quad S_+ := 
\begin{pmatrix} 
-Q_{11} &-\frac{1}{2}(Q_{12} + Q_{13} )  & 0\\
-Q_{21} &-\frac{1}{2}(Q_{22} + Q_{23} )  & 1\\
-Q_{31} & -\frac{1}{2}(Q_{32} + Q_{33}) & -1
\end{pmatrix}.
\end{equation}
Inverting the matrix $S_+$ yields
{\tiny{
\begin{equation}
\begin{split}
&(S_+)^{-1} = \frac{1}{
Q_{12} Q_{21} - Q_{11} Q_{22} - Q_{11} Q_{23} + Q_{13} Q_{21} - Q_{11}Q_{32} + Q_{12}Q_{31} - Q_{11} Q_{33} + Q_{13} Q_{31}}\\
&\times\begin{pmatrix}
Q_{22}  + Q_{23}  + Q_{32} + Q_{33} &  - Q_{12} - Q_{13}&     - Q_{12} - Q_{13}\\
- 2( Q_{21} + Q_{31}) &    2   Q_{11} &    2    Q_{11}\\
 Q_{21} Q_{32} - Q_{22} Q_{31} + Q_{21} Q_{33}  -  Q_{23} Q_{31} & Q_{12} Q_{31}  - Q_{11} Q_{32} - Q_{11} Q_{33} + Q_{13} Q_{31}  & Q_{11} Q_{22} - Q_{12}Q_{21}  + Q_{11}Q_{23} - Q_{13} Q_{21}\\
 \end{pmatrix}.
\end{split}
\end{equation}
}}
Hence the expression of the probability of exit given a nonnegative initial internal velocity state:
{\small{
\begin{equation}\label{ZEEabxpPositiveDrift}
\begin{split}
Z_a(a) =& \frac{( Q_{22} + Q_{23} + Q_{32} + Q_{33})( Q_{12} - Q_{13} ) -( Q_{12} + Q_{13}) (Q_{22} - Q_{23}+Q_{32} -Q_{33})}{2[ Q_{12} Q_{21}  - Q_{11}Q_{22} - Q_{11} Q_{23} + Q_{13}Q_{21} - Q_{11} Q_{32} +  Q_{12} Q_{31} - Q_{11} Q_{33}+ Q_{13}Q_{31}]},\\
\mathcal{E}_a( a, +; [a,b],\varphi ) =& \frac{(-Q_{21} -Q_{31})( Q_{12} - Q_{13} ) + Q_{11} (Q_{22} - Q_{23}+Q_{32} -Q_{33})}{Q_{12}Q_{21} - Q_{11}Q_{22} - Q_{11} Q_{23} + Q_{13}Q_{21} - Q_{11}Q_{32} + Q_{12} Q_{31} - Q_{11}Q_{33}+ Q_{13}Q_{31}}.
\end{split}
\end{equation}
}}
 As $Q_{ij} = \sum_{m=1}^3 e^{\lambda_m(b-a)} q_{ij}^{(m)}$, substituting into Eq. (\ref{ZeePosDrift}) yields the exit probabilities through $a$ given initial position and velocity states. This expression is not particularly illuminating, but it allows to check consistency with the statement of Eq. (\ref{TGE}) by taking the limit 
  of instantaneous tumble as a consistency check (see Eq. (\ref{consistencyEPosInterval})).\\

\subsubsection{Limit of a half-line}
\label{subsubsec:limitpdproba}
As the drift is positive, the largest eigenvalue of the matrix $A$ is the positive $\lambda_+$ (see Eq. \eqref{lambdaPMSigns}). Therefore, we have for large segments
\begin{equation}
e^{\lambda_+(b-a)}\gg 1 \gg e^{\lambda_-(b-a)}.
\end{equation}
Asymptotic expansions relevant to the large-$b$ limit are shown in the Appendix.
 Both the numerator and the denominator in the expressions of $Z_a(a)$ and ${\mathcal{E}}_a(a,+;[a,b],\varphi)$ are $O( e^{\lambda_+(b-a)})$. 
Using Eqs (\ref{detSPlusEq},\ref{NumEaEqPos}) yields 

\begin{equation}\label{ZaaPlusPosDrift}
Z_a(a)  \underset{b\to +\infty}{\sim} \frac{u_-}{v_- - 1},
\end{equation}

\begin{equation}\label{EaaPlusPosDrift}
   {\mathcal{E}}_a(a,+;[a,b],\varphi) \underset{b\to +\infty}{\sim}  \frac{v_- + 1}{v_- - 1}.
\end{equation}
Substituting into Eq. (\ref{defEaPosDrift}) yields the expression of the vector 
 $\vec{\mathcal{E}}(a)$ in the large-$b$ limit:
\begin{equation}
 \vec{\mathcal{E}}(a) \underset{b\to +\infty}{\sim} \frac{1}{v_- - 1}
    \begin{pmatrix}
      u_-\\
      v_-\\
      1 \\
    \end{pmatrix}.
\end{equation}

 The expression of the vector of exit probabilities follows:
\begin{equation}\label{vecExitPositiveDrift}
   \vec{\mathcal{E}}(x) \underset{b\to +\infty}{\sim} \frac{1}{v_- - 1}\sum_{k=1}^3 e^{\lambda_k(x-a)}q^{(k)} \begin{pmatrix}
     u_-\\
     v_-\\
     1\\
    \end{pmatrix}.
 \end{equation}
It is easy to check (see Eq. (\ref{idAlgoP}) in the Appendix) that the vector $(u_-,v_-,1)^T$  is annihilated by $q^{(1)}$ and $q^{(2)}$ and mapped to itself by $q^{(3)}$. Hence (with the notation $\lambda_- = \lambda_3$), 
\begin{equation}\label{ExSymbPositiveDrift}
\begin{split}
 Z_a(x ) \underset{b\to +\infty}{\sim}& \frac{u_-}{v_- - 1}e^{\lambda_-(x-a)},\\
 \mathcal{E}_a(x,+; [a,b],\varphi) \underset{b\to +\infty}{\sim}& \frac{v_- + 1 }{v_- - 1}    e^{\lambda_-(x-a)},\\
  \mathcal{E}_a(x,-; [a,b],\varphi)  \underset{b\to +\infty}{\sim}& e^{\lambda_-(x-a)}.\\
\end{split} 
\end{equation}

\begin{figure}
      \centering
	   \begin{subfigure}{0.57\linewidth}
		\includegraphics[width=\linewidth]{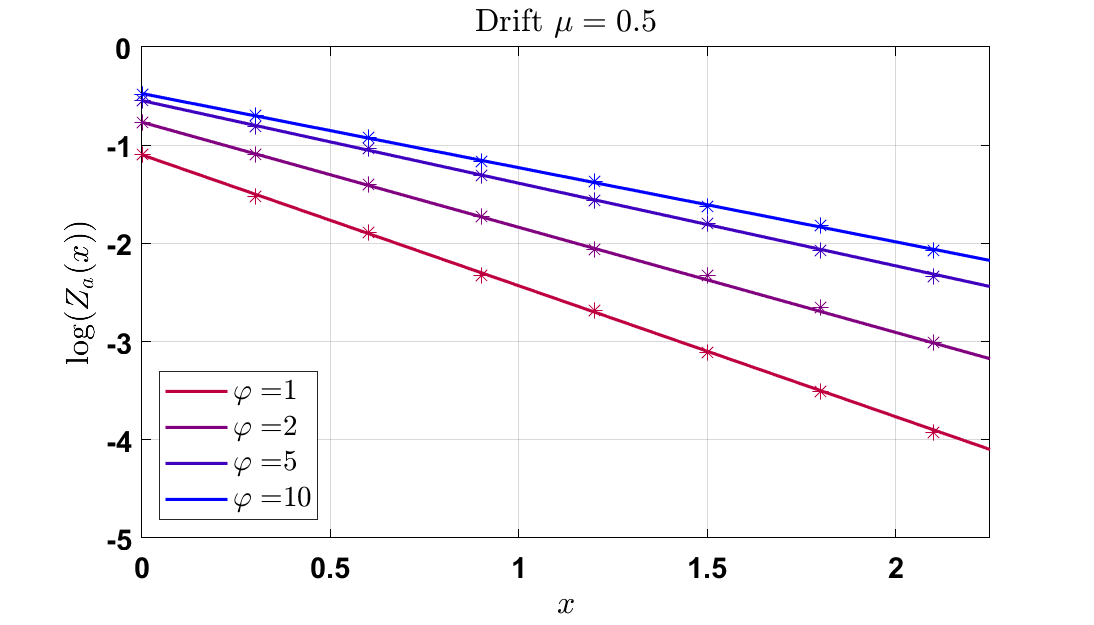}
		\caption{}
		\label{figExitZPositiveDrift}
	   \end{subfigure}
    \vfill   
	   \begin{subfigure}{0.57\linewidth}
		\includegraphics[width=\linewidth]{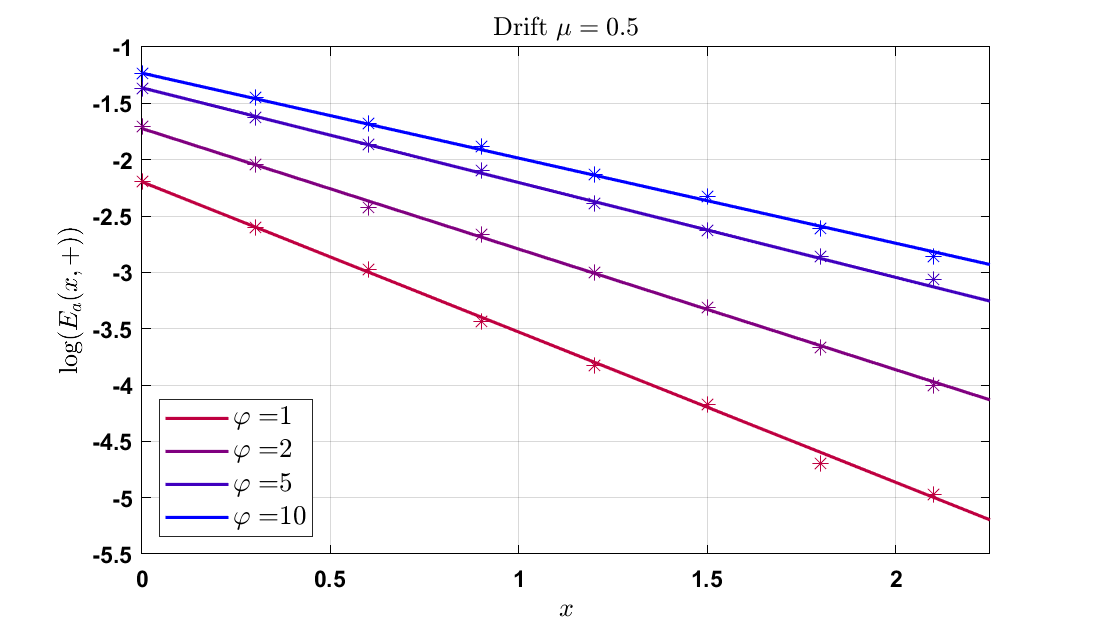}
		\caption{}
		\label{figExitPlusPositiveDrift}
	    \end{subfigure}
	\vfill
	     \begin{subfigure}{0.57\linewidth}
		 \includegraphics[width=\linewidth]{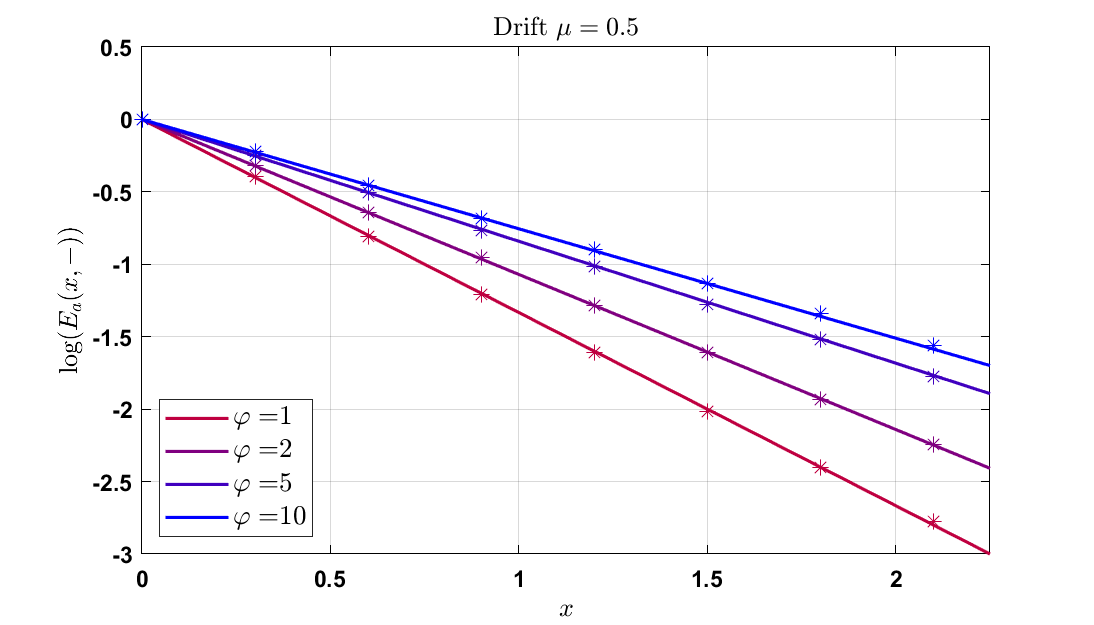}
		 \caption{}
		 \label{figExitMinusPositiveDrift}
	      \end{subfigure}
	\caption{The probability of exit from the half-line $[a,+\infty[$ for a positive subcritical drift, given an initial state (a) with zero internal velocity (tumbling), (b) with positive internal velocity, (c) with negative internal velocity. The stars correspond to the results of direct numerical simulations with 100,000 particles (the simulation was run until the time reached ten times the value of the mean conditional first-passage time calculated in Section 5).}
	\label{fig:subfigures4}
\end{figure}

Substituting the expressions of the eigenvalues and eigenvectors of $A$ in terms of the positive drift $\mu$ and the rate $\varphi$ (given in Eqs (\ref{lambdapmExpr},\ref{uv+-})) yields the result announced in Eq. (\ref{exitFinite}) in the case of a positive drift:
\begin{equation}\label{ProbExitPosFinal}
\begin{split}
 \mathcal{E}_a(x,0;[a,\infty[,\varphi) =&   \frac{2+ \varphi( 1 - \mu^2)  \left( 1 - \sqrt{1 + \frac{4\mu^2}{(1-\mu^2)^2 \varphi^2} }\right)}{  1+ \mu^2 + 2\mu\left( 1+\frac{1}{\varphi}\right) +(1-\mu^2 ) \sqrt{1 + \frac{4\mu^2}{(1-\mu^2)^2 \varphi^2} }} \\
 &\times \exp\left(  
\left[-\frac{\mu}{1-\mu^2} +\frac{\varphi}{2\mu} \left( 1 - \sqrt{1 + \frac{4\mu^2}{(1-\mu^2)^2 \varphi^2} }\right)  \right](x-a) 
\right),\\
 \mathcal{E}_a(x,+;[a,\infty[,\varphi) =&\frac{ 1+ \mu^2 - 2\mu\left( 1+\frac{1}{\varphi}\right) +(1-\mu^2 ) \sqrt{1 + \frac{4\mu^2}{(1-\mu^2)^2 \varphi^2} }}{ 1+ \mu^2 + 2\mu\left( 1+\frac{1}{\varphi}\right) +(1-\mu^2 ) \sqrt{1 + \frac{4\mu^2}{(1-\mu^2)^2 \varphi^2} }}\\
&\times 
\exp\left(  
\left[-\frac{\mu}{1-\mu^2} +\frac{\varphi}{2\mu} \left( 1 - \sqrt{1 + \frac{4\mu^2}{(1-\mu^2)^2 \varphi^2} }\right)  \right](x-a) 
\right),
\\
\mathcal{E}_a(x,-;[a,\infty[,\varphi) =& \exp\left(  
\left[-\frac{\mu}{1-\mu^2} +\frac{\varphi}{2\mu} \left( 1 - \sqrt{1 + \frac{4\mu^2}{(1-\mu^2)^2 \varphi^2} }\right)  \right](x-a) 
\right),   \qquad(\mu>0).
\end{split}
\end{equation}
The values are plotted as a function of $x$ for $\mu = 0.5$ and various values of the rate $\varphi$ in Figs \ref{figExitZPositiveDrift}, \ref{figExitPlusPositiveDrift}, \ref{figExitMinusPositiveDrift}, together with the results of direct simulations.\\

\subsubsection{Consistency checks}\label{checkEPos}

$\bullet$ {\bf{On the segment  $[a,b]$.}}
Let us take the limit of the integration constants $\mathcal{E}_a(a,0;[a,b],\varphi)$ and  $\mathcal{E}_a(a,+;[a,b],\varphi)$  found in Eq. (\ref{ZEEabxpPositiveDrift}) when the  tumble becomes instantaneous.  
 In this limit, the eigenvalue $\lambda_+$ goes to $+\infty$, and  
 the eigenvalue $\lambda_-$ goes to the finite limit $-\mu/(1-\mu^2)$, while the matrix $q^{(i)}$ goes to the limit $\qinst^{(i)}$ (displayed in Eq. (\ref{qLimExpr})). 
  The following  asymptotic expansions are worked out in the Appendix (see Eqs (\ref{denomCom},\ref{numer1},\ref{numer2})):
\begin{equation}
\begin{split}
Q_{12} Q_{21}  - Q_{11}Q_{22} - Q_{11} Q_{23} +& Q_{13}Q_{21} - Q_{11} Q_{32} +  Q_{12} Q_{31} - Q_{11} Q_{33} + Q_{13}Q_{31}\\
=& e^{\lambda_+(b-a)} \left[ -1 - \frac{1}{\mu} + \left(   -1 + \frac{1}{\mu} \right) e^{-\frac{\mu(b-a)}{1-\mu^2}} \right] + o( e^{\lambda_+(b-a)}).
\end{split}
\end{equation}
\begin{equation}
\begin{split}
( Q_{22} + Q_{23} + Q_{32} + Q_{33})&( Q_{12} - Q_{13} ) -( Q_{12} + Q_{13}) (Q_{22} - Q_{23}+Q_{32} -Q_{33})\\
&= e^{\lambda_+(b-a)}[ -2\mu^{-1} + 2( \mu^{-1} - 1)   e^{-\frac{\mu(b-a)}{1-\mu^2}} ] + o( e^{\lambda_+(b-a)} ).
\end{split}
\end{equation}
   \begin{equation}
\begin{split}
(-Q_{21} -Q_{31})&( Q_{12} - Q_{13} ) + Q_{11} (Q_{22} - Q_{23}+Q_{32} -Q_{33})\\
=& e^{\lambda_+(b-a)}(1-\mu^{-1})( 1 - e^{-\frac{\mu(b-a)}{1-\mu^2}}) + o( e^{\lambda_+(b-a)}) .
\end{split}
\end{equation}
 In particular, there are no terms of order $e^{2\lambda_+(b-a)}$ in the above quantities. The leading terms in the numerator and  denominator of the expressions of $Z_a(a)$ and $E_a(a,+)$ are therefore of order $e^{\lambda_+(b-a)}$ in the limit of large $\varphi$. Hence the limit of $Z_a(a)$ and $E_a(a,+)$ are expressed by taking the quotient of the terms of order $e^{\lambda_+(b-a)}$ in the numerator and denominator.
 Taking quotients yields the limits:
\begin{equation}\label{consistencyEPosInterval}
\begin{split}
\mathcal{E}_a( a, 0; [a,b],\infty) =& \frac{1 + (\mu - 1 ) e^{-\frac{\mu(b-a)}{1-\mu^2}}}{ 1 + \mu + (\mu - 1) e^{-\frac{\mu(b-a)}{1-\mu^2}}} = \frac{1}{2}\left[1 + \mathcal{E}_a( a, +; [a,b],\infty)   \right], \\
\mathcal{E}_a( a, +; [a,b],\mu,1,\infty) =& \frac{( 1-\mu)( 1 - e^{-\frac{\mu(b-a)}{1-\mu^2}})}{ 1 + \mu + (\mu - 1) e^{-\frac{\mu(b-a)}{1-\mu^2}}}, 
\end{split}
\end{equation}
 which is consistent with Eq. (\ref{TGE}) in our system of units.\\

$\bullet$ {\bf{On the half-line $[a,\infty[$.}} The large-$\varphi$ limit of Eq. (\ref{ProbExitPosFinal}) is readily evaluated as 
\begin{equation}\label{chekEPosHalfLineInst}
\begin{split}
\underset{\varphi\to \infty}{\lim}  \mathcal{E}_a(x,0;[a,\infty[,\varphi) =& \frac{1}{1+\mu}\exp\left( -\frac{\mu}{1-\mu^2}(x-a) \right),\\
\underset{\varphi\to \infty}{\lim}  \mathcal{E}_a(x,+;[a,\infty[,\varphi)=&  \frac{1 - \mu}{1 + \mu}\exp\left( -\frac{\mu}{1-\mu^2}(x-a) \right) ,\\
\underset{\varphi\to \infty}{\lim}  \mathcal{E}_a(x,-;[a,\infty[,\varphi) =& \exp\left( -\frac{\mu}{1-\mu^2}(x-a) \right).
\end{split}
\end{equation}
 The limits of $\mathcal{E}_a(x,+;[a,\infty[,\varphi)$ and $\mathcal{E}_a(x,-;[a,\infty[,\varphi)$ are consistent with Eq. (\ref{exitOrdinary}) in the case of a positive drift. On the other hand, the exit
 probability of a particle starting in a tumbling state at coordinate $x$ is the simple average of the exit probabilities of particles starting at $x$ with positive and negative internal velocities. Indeed, in the limit of instantaneous tumbles, the tumbling particle  draws its internal velocity state at time $0$ in $\{ -1,+1\}$.\\

\section{Solution of the evolution equation for the conditional mean first-passage time}\label{sec:solutiontime}

 Using the expression of the vector $\vec{\mathcal{E}}(x)$ obtained in the previous section (Eq. \eqref{ExExpr}), the evolution equation for the mean first-passage times (Eq. \eqref{ODE}) reads
 \begin{equation}
 \begin{split}
\frac{d\vec{M}}{dx}(x) =& A\vec{M}(x) + R\vec{\mathcal{E}}(x)\\
=& A\vec{M}(x) + R Q(x-a)\vec{\mathcal{E}}(a),\\
 \end{split}
 \end{equation}
with the notation
\begin{equation}
R := \begin{pmatrix}
-\frac{1}{\mu} & 0 & 0 \\
0 & \frac{\mu}{1-\mu^2} &  -\frac{1}{1-\mu^2}   \\
0 &  -\frac{1}{1-\mu^2}    &    \frac{\mu}{1-\mu^2}
\end{pmatrix},
\end{equation}
and as in the previous section $Q(y) = \exp(yA) = \sum_{k=1}^3 q^{(k)} e^{\lambda_k y}$. 
Integrating between $a$ and $x$ (for $x$ in $]a,b[$) and substituting the expression of the matrices $Q$ in terms of the eigenvalues of the matrix $A$ given in Eq. (\ref{qMatrixDef}), we obtain
\begin{equation}\label{exprMGen}
\begin{split}
\vec{M}(x) =& Q(x-a)  \vec{M}(a) + \int_a^x dy Q(x-y) R Q(y-a) \vec{\mathcal{E}}(a)\\
=& \sum_{k=1}^3  e^{\lambda_k(x-a)}  q^{(k)}  \vec{M}(a) + \int_a^x dy  \left( \sum_{i=1}^3 q^{(i)} e^{\lambda_i(x-y)} \right) R    \left( \sum_{j=1}^3 q^{(j)} e^{\lambda_j(y-a)} \right) \vec{\mathcal{E}}(a)\\
=& \sum_{k=1}^3 e^{\lambda_k(x-a)}  q^{(k)}  \vec{M}(a) + \sum_{i=1}^3  \sum_{j=1}^3 \int_a^x dy e^{\lambda_i(x-y)} e^{\lambda_j(y-a)}  q^{(i)} R q^{(j)}\vec{\mathcal{E}}(a)\\
=& \sum_{k=1}^3 e^{\lambda_k(x-a)}  q^{(k)}  \vec{M}(a) + \sum_{i=1}^3  \sum_{j=1}^3 G_{ji}(x) W^{(i,j)}\vec{\mathcal{E}}(a),\\
\end{split}
\end{equation}
where $G_{ji}(x)$ is a scalar function, and $W^{(i,j)}$ is a three-by-three matrix (independent of the space coordinate $x$):
\begin{equation}\label{defW}
  G_{ji}(x) :=   \int_a^x dy e^{\lambda_i(x-y)} e^{\lambda_j(y-a)},\qquad
  W^{(i,j)} := q^{(i)} R q^{(j)},\qquad (1\leq i,j\leq 3).
\end{equation}
 The matrices $W^{(i,j)}$ are expressed in terms of the components of the eigenvectors of $A$ in  Eq. (\ref{WExpl}) of Appendix \ref{AppW}.\\

Working out the integrals in the above expression of $\left( G_{ij}(x)\right)_{1\leq i,j \leq 3}$ yields
\begin{equation}
\begin{split}
G_{11}(x) &= x - a,\qquad G_{22}(x) = (x-a) e^{\lambda_+(x-a)},\qquad  G_{33}(x) = (x-a) e^{\lambda_-(x-a)},\\
 G_{12}(x) &= G_{21}(x) = \frac{1}{\lambda_+}( e^{\lambda_+(x-a)} - 1 ),\qquad
 G_{13}(x) = G_{31}(x) = \frac{1}{\lambda_-}( e^{\lambda_-(x-a)} - 1 ),\\
 G_{23}(x) &= G_{32}(x) = \frac{1}{\lambda_+ - \lambda_-}( e^{\lambda_+(x-a)} - e^{\lambda_-(x-a)} ).\\  
\end{split}
\end{equation}

Grouping the terms proportional to each of the functions $e^{\lambda_i(x-a)}$ for $i$ in $[1..3]$, the solution becomes

%\begin{equation}\label{MxGen}
%\begin{split}
%\vec{M}(x) =& q^{(1)}\vec{M}(a)  + \left[(x-a) W^{(1,1)}
%-\frac{1}{\lambda_-}( W^{(1,3)} + W^{(3,1)})
%-\frac{1}{\lambda_+}( W^{(1,2)} + W^{(2,1)})\right]\vec{\mathcal{E}}(a)
%\\
%&+ e^{\lambda_+(x-a)} \left\{ q^{(2)}\vec{M}(a) %+ \left[ (x-a) W^{(2,2)} + \frac{1}{\lambda_+}( W^{(1,2)} + W^{(2,1)})
% + \frac{1}{\lambda_+ - \lambda_-}( W^{(2,3)} + W^{(3,2)}) \right] \vec{\mathcal{E}}(a)
% \right\}\\
%&+ e^{\lambda_-(x-a)} \left\{ q^{(3)}\vec{M}(a) + \left[
% (x-a)W^{(3,3)} + \frac{1}{\lambda_-}( W^{(1,3)} + W^{(3,1)}) 
% - \frac{1}{\lambda_+ - \lambda_-}( W^{(2,3)} + %W^{(3,2)})\right] \vec{\mathcal{E}}(a)
% \right\}.    
%\end{split}
%\end{equation}
%The solution therefore takes the form
\begin{equation}\label{JDef}
\begin{split}
\vec{M}(x) =&  Q(x-a) \vec{M}(a) + \vec{J}(x)\\  
=&\vec{M}(a )+ e^{\lambda_-(x-a)} q^{(2)}\vec{M}(a) + e^{\lambda_-(x-a)} q^{(3)}\vec{M}(a)  + \vec{J}(x),
\end{split}
\end{equation}
 where the vector ${\vec{J}}(x)$ is a linear function of the vector $\vec{\mathcal{E}}(a)$, expressed as follows:
\begin{equation}\label{defJ}
\vec{J}(x) := \Gamma(x) \vec{\mathcal{E}}(a),\\
\end{equation}
with
\begin{equation}
\begin{split}
\Gamma(x):=& (x-a) W^{(1,1)}
-\frac{1}{\lambda_-}( W^{(1,3)} + W^{(3,1)})
-\frac{1}{\lambda_+}( W^{(1,2)} + W^{(2,1)})
\\
&+ e^{\lambda_+(x-a)} 
  \left[ (x-a) W^{(2,2)} + \frac{1}{\lambda_+}( W^{(1,2)} + W^{(2,1)})
 + \frac{1}{\lambda_+ - \lambda_-}( W^{(2,3)} + W^{(3,2)}) \right]
\\
&+ e^{\lambda_-(x-a)} \left[
 (x-a)W^{(3,3)} + \frac{1}{\lambda_-}( W^{(1,3)} + W^{(3,1)}) 
 - \frac{1}{\lambda_+ - \lambda_-}( W^{(2,3)} + W^{(3,2)})\right].   
\end{split}    
\end{equation}
 The vector $\vec{J}(x)$ depends on the sign of the drift through the vector $ \vec{\mathcal{E}}(a)$. Given the sign of the drift, the boundary conditions can be encoded by parametrizing the vectors $\vec{M}(a)$ and $\vec{M}(b)$.\\ 

\subsection{Negative drift}\label{subsec:ndtime}
\subsubsection{Integration constant}\label{subsubsec:icndtime}
 A particle starting its motion at $a$ in a zero or negative internal velocity state leaves the system immediately through $a$, hence the mean first-passage time through $a$ is zero. Hence $Z_a(a) = 0$ and $M_2(a) -M_3(a) = 0$. Moreover, a particle starting its motion at $b$ in a positive velocity state has a positive total velocity (because drift is subcritical). It leaves the system immediately (through $b$). Hence, the probability of leaving the system through $a$ is zero, and $M_2(b)+M_3(b) =0$. 
The boundary conditions therefore imply the following form for the vectors  $\vec{M}(a)$ and $\vec{M}(b)$: 
\begin{equation}
\vec{M}(a) = 
\begin{pmatrix}
  0\\
 M_2( a )\\
M_2( a )
\end{pmatrix},\qquad
\vec{M}(b) = 
\begin{pmatrix}
 M_1(b)\\
 M_2( b )\\
 -M_2( b )
\end{pmatrix}.
\end{equation}

As $\vec{M}(b)$ is expressed in terms of  $\vec{M}(a)$ by Eq. \eqref{JDef},  the following linear system is satisfied by the unknown parameters $M_2(a)$, $M_1(b)$ and $M_3(b)$:
\begin{equation}
\begin{split}
T_-\begin{pmatrix}
 M_2(a)\\
  M_1(b)\\
 M_2(b)
\end{pmatrix}
   =& 
\begin{pmatrix} 
J_1(b)\\
J_2(b)\\
J_3(b)
\end{pmatrix},\\ 
{\mathrm{with}}\quad T_- :=& 
\begin{pmatrix} 
  -(Q_{12}+ Q_{13})(b-a)  & 1 & 0 \\
    -(Q_{22}+ Q_{23})(b-a)& 0  & 1\\
  -(Q_{32}+ Q_{33})(b-a)   &  0  &  -1 
\end{pmatrix},
\end{split}
\end{equation}
 with the shorthand notation $Q_{ij}=\exp\left((b-a)A\right)$ introduced in Eq. (\ref{shorthandQ}).\\

Hence, the integration constant $M_2(a)$ is obtained as the first component of the vector
\begin{equation}
\begin{split}
\begin{pmatrix}
 M_2(a)\\
 M_1(b)\\
 M_2(b)
\end{pmatrix}
   =& \left(T_-\right)^{-1} 
\begin{pmatrix} 
J_1(b)\\
J_2(b)\\
J_3(b)
\end{pmatrix}.
\end{split}    
\end{equation}

Inverting the matrix $T_-$ yields
{\tiny{
\begin{equation}
\begin{split}
(T_-)^{-1} =& \frac{1}{Q_{22}(b-a) + Q_{23}( b-a) + Q_{32}(b-a) + Q_{33}( b-a)}\\
&\times\begin{pmatrix}
  0 & -1 & -1 \\
    Q_{22}(b-a) + Q_{23}( b-a) + Q_{32}(b-a) + Q_{33}( b-a) & -(Q_{12}(b-a) + Q_{13}( b-a)) & -(Q_{12}(b-a) + Q_{13}( b-a)) \\
 0 & (Q_{32}(b-a) + Q_{33}( b-a)) & -(Q_{22}(b-a) + Q_{23}( b-a))
\end{pmatrix}.
\end{split}
\end{equation}
}}
Hence the expression of the integration constant
\begin{equation}\label{M2aNegDrift}
\begin{split}
M_2( a ) =& \frac{1}{2}\mathcal{T}_a(a, +; [a,b],\varphi) = -\frac{J_2(b) + J_3(b)}{Q_{22}(b-a) + Q_{23}( b-a) + Q_{32}(b-a) + Q_{33}( b-a)}\\
=&- \frac{\sum_{j=1}^3 \left( \Gamma_{2j}(b) +\Gamma_{3j}(b) \right) \mathcal{E}_j(a)}{ \sum_{k=1}^3 \left[  q^{(k)}_{22} + q^{(k)}_{23} + q^{(k)}_{32} + q^{(k)}_{33} \right] e^{\lambda_k(b-a)}},
 \end{split}   
\end{equation}
with the notations introduced in Eq. (\ref{defJ}). This expression is amenable to numerical evaluation, but studying the limit of a half-line ($b\to \infty$) is more informative as allows to generalize the known result displayed in Eq. (\ref{MFPTOrdinary}).\\

\subsubsection{Limit of a half-line}\label{subsubsec:limitndtime}
% With the expression of the vector of exit probabilities in the case of negative drift on a half-line (almost-sure exit),
%\begin{equation}
%\vec{J}( b ) = \Gamma(b) \vec{\mathcal{E}}(a) = 
%\Gamma 
% \begin{pmatrix}
%1\\
%1\\
% 0\\
%\end{pmatrix}
%=
%\begin{pmatrix}
%\Gamma_{11}(b) + \Gamma_{12}(b)\\
%\Gamma_{21}(b) + \Gamma_{22}(b)\\
%\Gamma_{31}(b) + \Gamma_{32}(b)\\
%\end{pmatrix}
%\end{equation}

In the limit where $b$ goes to infinity, the leading term in the denominator in the expression of $M_2(a)$ in Eq. (\ref{M2aNegDrift}) is $O( e^{\lambda_-(b-a)})$. Indeed, $\lambda_-$ is positive, and the coefficient of $e^{\lambda_-(b-a)}$ in the denominator reads (using the expression of $q^{(3)}$ in Eq. (\ref{qExpr}))
\begin{equation}
q^{(3)}_{22} + q^{(3)}_{23} + q^{(3)}_{32} + q^{(3)}_{33} = -\frac{(v_- + 1)( u_+ - v_+ + 1)}{u_- - u_+ - v_- + v_+}.
\end{equation}

On the other hand, the particle exits the system $[a,+\infty[$ almost surely if the drift is negative   (the vector $\vec{\mathcal{E}}(a)$ goes to $(1,1,0)^T$).  
Evaluating the terms proportional to $e^{\lambda_-(b-a)}$ in the numerator yields:
\begin{equation}
\begin{split}
&M_2( a )\underset{b\to +\infty}{\sim} -\frac{ \Gamma_{21}(b) + \Gamma_{22}(b) + \Gamma_{31}(b) + \Gamma_{32}(b) }{(q^{(3)}_{22} + q^{(3)}_{23} + q^{(3)}_{32} + q^{(3)}_{33}) e^{\lambda_-(b-a)}}\\
&=-\frac{ 1}{q^{(3)}_{22} + q^{(3)}_{23} + q^{(3)}_{32} + q^{(3)}_{33}}\times \left[ (b-a)  (W^{(3,3)}_{21} +W^{(3,3)}_{22}+
W^{(3,3)}_{31} +W^{(3,3)}_{32})  \right.\\
&  + \left.  \frac{1}{\lambda_-}( (W^{(1,3)} +W^{(3,1)})_{21} + (W^{(1,3)} +W^{(3,1)})_{22} + (W^{(1,3)} +W^{(3,1)})_{31} + (W^{(1,3)} +W^{(3,1)})_{32}) \right.\\
& +  \left.  -\frac{1}{\lambda_+ - \lambda_-}(  (W^{(2,3)} +W^{(3,2)})_{21} + (W^{(2,3)} +W^{(3,2)})_{22} + (W^{(2,3)} +W^{(3,2)})_{31} + (W^{(2,3)} +W^{(3,2)})_{32})  \right].\\
 \end{split}   
\end{equation}
 Based on the expression of the matrices of $(W^{(i,j)})_{1\leq i,j\leq 3}$ in Eq. (\ref{WExpl}), it is easy to check that the following combinations of entries are zero:
\begin{equation}\label{checkneg}
\begin{split}
W^{(3,3)}_{21} +W^{(3,3)}_{22} =& W^{(3,3)}_{31} +W^{(3,3)}_{32} = 0,\\
W^{(1,3)}_{21} +W^{(1,3)}_{22} =& W^{(1,3)}_{31} = W^{(1,3)}_{32} = 0,\\
W^{(2,3)}_{21} +W^{(2,3)}_{22} =& W^{(2,3)}_{31} +W^{(2,3)}_{32} = 0,\\
W^{(3,2)}_{21} +W^{(3,2)}_{22} =& W^{(3,2)}_{31} +W^{(3,2)}_{32} = 0,\\
\end{split}
\end{equation}
 This leads to simplifications (in particular, there is no term of order $(b-a) e^{\lambda_-(b-a)}$ in the numerator), 
  so that 
 \begin{equation}\label{M2aNegAlg}
\begin{split}
M_2( a ) \underset{b\to +\infty}{\sim}& - \frac{ W^{(3,1)}_{21}  + W^{(3,1)}_{22} + W^{(3,1)}_{31}  +W^{(3,1)}_{32}}{\lambda_-( q^{(3)}_{22} + q^{(3)}_{23} + q^{(3)}_{32} + q^{(3)}_{33})}= \frac{\mu v_+ - \mu u_+ + 1}{\lambda_- \mu( \mu^2 - 1) (u_+ - v_+ + 1)}.\\
\end{split}
\end{equation}
 Substituting the entries of $q^{(3)}$ in Eq. (\ref{qExpr}) and $W^{(3,1)}$ in  Eq. (\ref{WExpl}) yields

\begin{equation}
\begin{split}
 M_2( a ) \underset{b\to +\infty}{\sim} \frac{2\mu[ \varphi( 1-\mu^2) +2 + \sqrt{\varphi^2(1-\mu^2)^2 +4\mu^2}  ]}{  
  ( \varphi( 1-\mu^2)  - 2\mu^2 - \sqrt{\varphi ^2(1 - \mu^2 )^2+4\,\mu ^2}  )(  \varphi( 1-\mu^2) - 2\mu+\sqrt{\varphi ^2(1 - \mu^2 )^2+4\,\mu ^2}   ) }.
 \end{split}
\end{equation}

Having worked out the integration constant $\vec{M}(a)$ (in the case of a negative drift), we obtain the
  expression of $\vec{M}(x)$ in the limit of a half-line from Eq. (\ref{JDef}) as follows:
{\footnotesize{
\begin{equation}
\begin{split}
\vec{M}(x) &\underset{b\to \infty} \sim  q^{(1)}
\begin{pmatrix}
0\\
M_2(a)\\
M_2(a)
\end{pmatrix}  
+ \left[(x-a) W^{(1,1)}
-\frac{1}{\lambda_-}( W^{(1,3)} + W^{(3,1)})
-\frac{1}{\lambda_+}( W^{(1,2)} + W^{(2,1)})\right]
\begin{pmatrix}
1\\
1\\
0
\end{pmatrix}
\\
&+ e^{\lambda_+(x-a)} \left\{ q^{(2)}
\begin{pmatrix}
0\\
M_2(a)\\
M_2(a)
\end{pmatrix}  
 + \left[ (x-a) W^{(2,2)} + \frac{1}{\lambda_+}( W^{(1,2)} + W^{(2,1)})
 + \frac{1}{\lambda_+ - \lambda_-}( W^{(2,3)} + W^{(3,2)}) \right] 
 \begin{pmatrix}
1\\
1\\
0
\end{pmatrix}
 \right\}\\
&+ e^{\lambda_-(x-a)} \left\{ q^{(3)}
\begin{pmatrix}
0\\
M_2(a)\\
M_2(a)
\end{pmatrix}   + \left[
 (x-a)W^{(3,3)} + \frac{1}{\lambda_-}( W^{(1,3)} + W^{(3,1)}) 
 - \frac{1}{\lambda_+ - \lambda_-}( W^{(2,3)} + W^{(3,2)})\right] 
 \begin{pmatrix}
1\\
1\\
0
\end{pmatrix}
 \right\}.    
\end{split}
\end{equation}
}}
With the integration constant obtained in Eq. (\ref{M2aNegAlg}) and the matrices of Eq. (\ref{WExpl}), one can check that the following algebraic identities hold: 
\begin{equation}
\begin{split}
W&^{(3,3)}  \begin{pmatrix}
 1\\
 1\\
 0
 \end{pmatrix} = 
 W^{(2,2)}  \begin{pmatrix}
 1\\
 1\\
 0
 \end{pmatrix} = \vec{0},\\
& ( W^{(2,3)} + W^{(3,2)} ) 
\begin{pmatrix}
 1\\
 1\\
 0
 \end{pmatrix} = \vec{0},\\
& q^{(3)}
 \begin{pmatrix}
 0\\
 M_2(a)\\
 M_2(a)
 \end{pmatrix}
  + \frac{1}{\lambda_-} ( W^{(1,3)} +   W^{(3,1)}  ) 
  \begin{pmatrix}
 1\\
 1\\
 0
 \end{pmatrix} = \vec{0}.
 \end{split}
\end{equation}

The expression of $\vec{M}(x)$ therefore simplifies. In particular, the coefficient of $e^{\lambda_-(x-a)}$ is equal to zero.  Calculating the remaining matrix products (using Eq. (\ref{algIdq}) yields 
\begin{equation}
\begin{split}
\vec{M}(x) =& M_2(a) q^{(1)}
\begin{pmatrix}
 0\\
 1\\
 1\\
 \end{pmatrix} + 
\left[(x-a) W^{(1,1)}
-\frac{1}{\lambda_-}( W^{(1,3)} + W^{(3,1)})
-\frac{1}{\lambda_+}( W^{(1,2)} + W^{(2,1)})\right]\begin{pmatrix}
 1\\
 1\\
 0\\
 \end{pmatrix}
\\
&+ e^{\lambda_+(x-a)} \left[ M_2(a) 
 q^{(2)} 
  \begin{pmatrix}
 0\\
 1\\
 1\\
 \end{pmatrix}
 +
 \frac{1}{\lambda_+}( W^{(1,2)} + W^{(2,1)})
 \begin{pmatrix}
 1\\
 1\\
 0\\
 \end{pmatrix} \right]
\\
=& -  (x-a)   \frac{v_- - v_+ + \mu^2 (u_- -  u_+ - v_- + v_+) + \mu (u_- v_+ - u_+ v_-)}{\mu( \mu^2 - 1) (u_- - u_+ - v_- + v_+ )}
   \begin{pmatrix}
 1\\
 1\\
 0\\
 \end{pmatrix}\\
& \frac{\mu v_+ - \mu u_+ + 1}{\lambda_- \mu( \mu^2 - 1) (u_+ - v_+ + 1)} \times\frac{u_- - u_+ - u_- v_+ + u_+ v_-}{u_- - u_+ - v_- + v_+} 
\begin{pmatrix}
 1\\
 1\\
 0\\
 \end{pmatrix}\\
 &
- \frac{\mu v_+ - \mu u_+ + 1}{\lambda_- \mu( \mu^2 - 1) (u_- - u_+ - v_- + v_+ )}
   \begin{pmatrix}
 u_-\\
 v_-\\
 1\\
 \end{pmatrix} \\
&+\frac{\mu v_- - \mu u_- + 1}{\lambda_+ \mu( \mu^2 - 1) (u_- - u_+ - v_- + v_+ )}
\begin{pmatrix}
 u_+\\
 v_+\\
 1\\
 \end{pmatrix}
\\
&+ e^{\lambda_+(x-a)} \left[
 \frac{\mu v_+ - \mu u_+ + 1}{\lambda_- \mu( \mu^2 - 1) (u_+ - v_+ + 1)}  \times \frac{u_- - v_- + 1}{u_- - u_+ - v_- + v_+} \begin{pmatrix}
 u_+\\
 v_+\\
 1\\
 \end{pmatrix}\right.\\
 &\left.
    -\frac{\mu v_- - \mu u_- + 1}{\lambda_+ \mu( \mu^2 - 1) (u_- - u_+ - v_- + v_+ )} \begin{pmatrix}
 u_+\\
 v_+\\
 1\\
 \end{pmatrix} \right],
\end{split}
\end{equation}
 where we have used the algebraic identities of Eq. (\ref{algIdq}) to evaluate $q^{(2)}(0,1,1)^T$. 
 There are therefore corrections to the affine form of the conditional mean first-passage time displayed in Eq. (\ref{MFPTOrdinary}), but these corrections vanish exponentially with the distance to $a$, because the eigenvalue $\lambda_+$ is negative if the drift is negative. The expression of some of the coefficient in terms of $\varphi$ are quite involved, but the coefficient of $(x-a)$ is $-\mu^{-1}$, for any value of the mean tumble time. Indeed, thanks to the identity of Eq. (\ref{coeffxUniv}),
 \begin{equation}
   \frac{v_- - v_+ + \mu^2 (u_- -  u_+ - v_- + v_+) + \mu (u_- v_+ - u_+ v_-)}{\mu( \mu^2 - 1) (u_- - u_+ - v_- + v_+ )} = \frac{1}{\mu}.
\end{equation}

%
%bidule = \frac{1}{\mu( \mu^2 - 1) (u_+ - v_+ + 1) (u_- - u_+ - v_- + v_+)} 
%\left(  \frac{1}{\lambda_-}(\mu v_+ - \mu u_+ + 1)( u_- - v_- + 1)  - 
%   \frac{1}{\lambda_+}(\mu v_- - \mu u_- + 1)( u_+ - v_+ + 1)\right)

From the definition of the vector $\vec{M}(x)$ (in Eqs (\ref{UTt},\ref{ODE},\ref{timesOfInterest}), and the fact that the particle exits the half-line almost surely in the presence of a negative drift (see Eq. \eqref{almostSurely}):
\begin{equation}
\begin{split}
\mathcal{T}_a(x,0;[a,\infty[,\varphi) =& \frac{M_1(x)}{\mathcal{E}_a(x,0;[a,\infty[,\varphi) } = M_1(x),\\
\mathcal{T}_a(x,+;[a,\infty[,\varphi) =& \frac{M_2(x) + M_3(x)}{\mathcal{E}_a(x,+;[a,\infty[,\varphi) } = M_2(x) + M_3(x),\\
\mathcal{T}_a(x,-;[a,\infty[,\varphi) =& \frac{M_2(x) - M_3(x)}{\mathcal{E}_a(x,+;[a,\infty[,\varphi) } = M_2(x) - M_3(x).
\end{split}
\end{equation}
 Rewriting the above using the expression of $\vec{M}(x)$ we have just found yields
\begin{equation}\label{MFPTNegDriftAlg}
\begin{split}
\mathcal{T}_a(x,0;[a,\infty[,\varphi) =  & - \frac{1}{\mu}(x-a)\\
&\frac{\mu v_+ - \mu u_+ + 1}{\lambda_- \mu( \mu^2 - 1) (u_- - u_+ - v_- + v_+)} \times\frac{u_- - u_+ - u_- v_+ + u_+ v_-}{u_+ - v_+ + 1}  \\
&- \frac{\mu v_+ - \mu u_+ + 1}{\lambda_- \mu( \mu^2 - 1) (u_- - u_+ - v_- + v_+ )} u_- + \frac{\mu v_- - \mu u_- + 1}{\lambda_+ \mu( \mu^2 - 1) (u_- - u_+ - v_- + v_+ )} u_+\\
 & + e^{\lambda_+(x-a)} \frac{1}{\mu( \mu^2 - 1) (u_+ - v_+ + 1) (u_- - u_+ - v_- + v_+)} \\
 &\times
\left(  \frac{1}{\lambda_-}(\mu v_+ - \mu u_+ + 1)( u_- - v_- + 1)  - 
   \frac{1}{\lambda_+}(\mu v_- - \mu u_- + 1)( u_+ - v_+ + 1)\right) u_+,\\
\mathcal{T}_a(x,+;[a,\infty[,\varphi) =  & - \frac{1}{\mu}(x-a)\\
& \frac{\mu v_+ - \mu u_+ + 1}{\lambda_- \mu( \mu^2 - 1) (u_- - u_+ - v_- + v_+)} \times\frac{u_- - u_+ - u_- v_+ + u_+ v_-}{u_+ - v_+ + 1} \\
&- \frac{\mu v_+ - \mu u_+ + 1}{\lambda_- \mu( \mu^2 - 1) (u_- - u_+ - v_- + v_+ )} (v_- +1) + \frac{\mu v_- - \mu u_- + 1}{\lambda_+ \mu( \mu^2 - 1) (u_- - u_+ - v_- + v_+ )} (v_+ + 1)\\
&+ e^{\lambda_+(x-a)} \frac{1}{\mu( \mu^2 - 1) (u_+ - v_+ + 1) (u_- - u_+ - v_- + v_+)} \\
&\times
\left(  \frac{1}{\lambda_-}(\mu v_+ - \mu u_+ + 1)( u_- - v_- + 1)  - 
   \frac{1}{\lambda_+}(\mu v_- - \mu u_- + 1)( u_+ - v_+ + 1)\right) (v_+ + 1 ),\\
\mathcal{T}_a(x,-;[a,\infty[,\varphi) = & - \frac{1}{\mu}(x-a)\\
& \frac{\mu v_+ - \mu u_+ + 1}{\lambda_- \mu( \mu^2 - 1) (u_- - u_+ - v_- + v_+)} \times\frac{u_- - u_+ - u_- v_+ + u_+ v_-}{u_+ - v_+ + 1}  \\
&- \frac{\mu v_+ - \mu u_+ + 1}{\lambda_- \mu( \mu^2 - 1) (u_- - u_+ - v_- + v_+ )} (v_- -1) + \frac{\mu v_- - \mu u_- + 1}{\lambda_+ \mu( \mu^2 - 1) (u_- - u_+ - v_- + v_+ )} (v_+ - 1)\\
& + e^{\lambda_+(x-a)} \frac{1}{\mu( \mu^2 - 1) (u_+ - v_+ + 1) (u_- - u_+ - v_- + v_+)} \\
&\times
\left(  \frac{1}{\lambda_-}(\mu v_+ - \mu u_+ + 1)( u_- - v_- + 1)  - 
   \frac{1}{\lambda_+}(\mu v_- - \mu u_- + 1)( u_+ - v_+ + 1)\right) (v_+ - 1),
\end{split}
\end{equation}

\begin{figure}
      \centering
	   \begin{subfigure}{0.57\linewidth}
		\includegraphics[width=\linewidth]{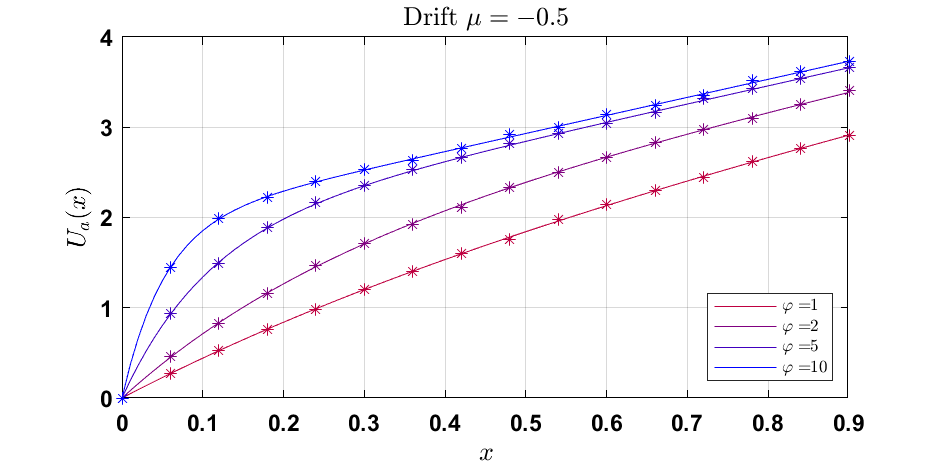}
		\caption{}
		\label{figTimesZeroNegativeDrift}
	   \end{subfigure}
    \vfill   
	   \begin{subfigure}{0.57\linewidth}
		\includegraphics[width=\linewidth]{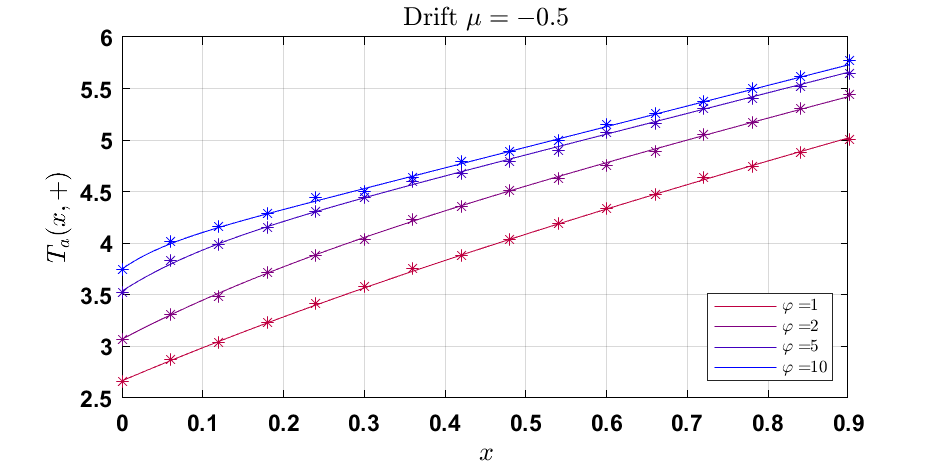}
		\caption{}
		\label{figTimesPlusNegativeDrift}
	    \end{subfigure}
	\vfill
	     \begin{subfigure}{0.57\linewidth}
		 \includegraphics[width=\linewidth]{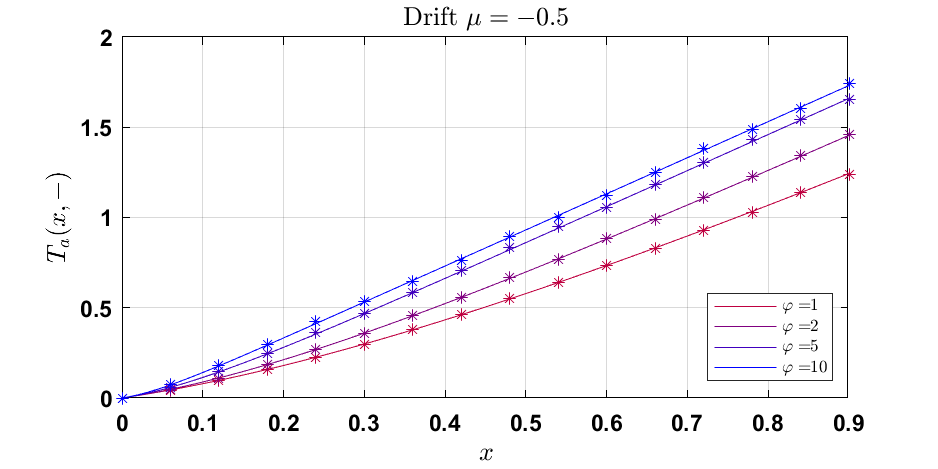}
		 \caption{}
		 \label{figTimesMinusNegativeDrift}
	      \end{subfigure}
	\caption{The mean first passage-time at $a$ of a particle starting in the  half-line $[a,+\infty[$ (in the presence of a negative drift), given an initial state (a) with zero internal velocity (tumbling), (b) with positive internal velocity, (c) with negative internal velocity. The stars correspond to the results of direct numerical simulations with 100,000 particles (the simulation was run until the last particle left the system through $a$).}
\end{figure}

\begin{figure}
      \centering
	   \begin{subfigure}{0.54\linewidth}
		\includegraphics[width=\linewidth]{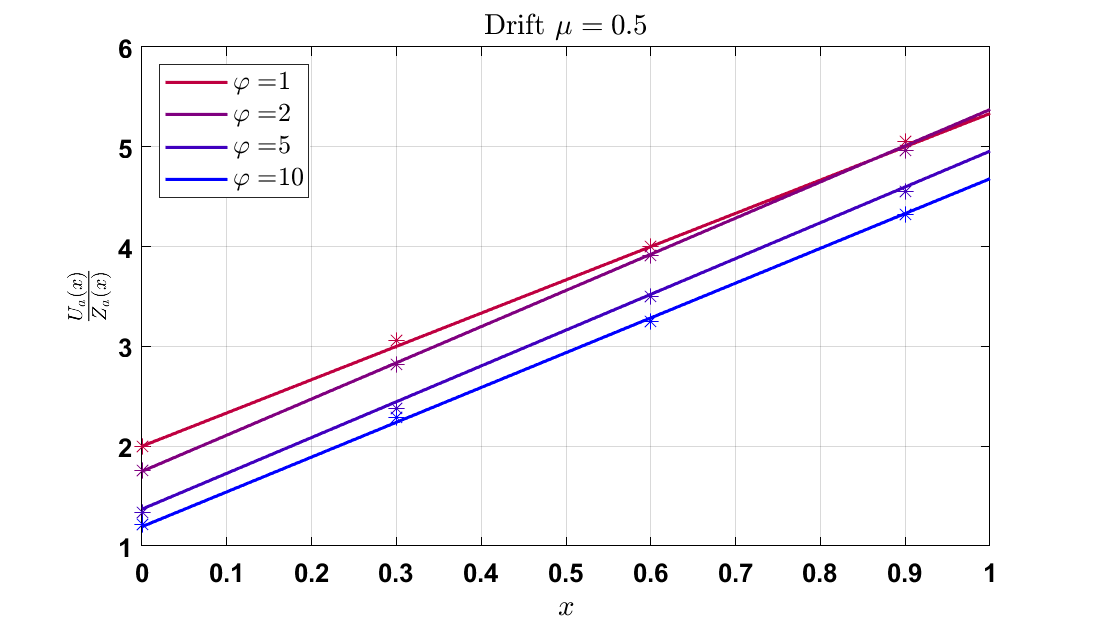}
		\caption{}
		%\label{figTimesTumblePositiveDrift}
	   \end{subfigure}
    \vfill   
	   \begin{subfigure}{0.54\linewidth}
		\includegraphics[width=\linewidth]{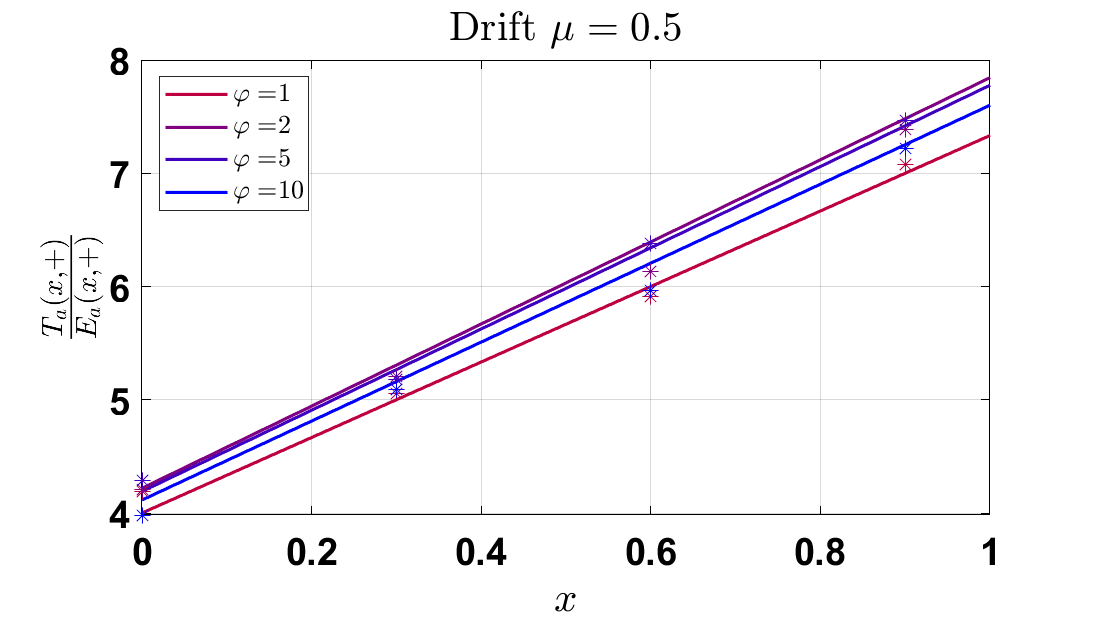}
		\caption{}
		%\label{figTimesPlusPositiveDrift}
	    \end{subfigure}
	\vfill
	     \begin{subfigure}{0.54\linewidth}
		 \includegraphics[width=\linewidth]{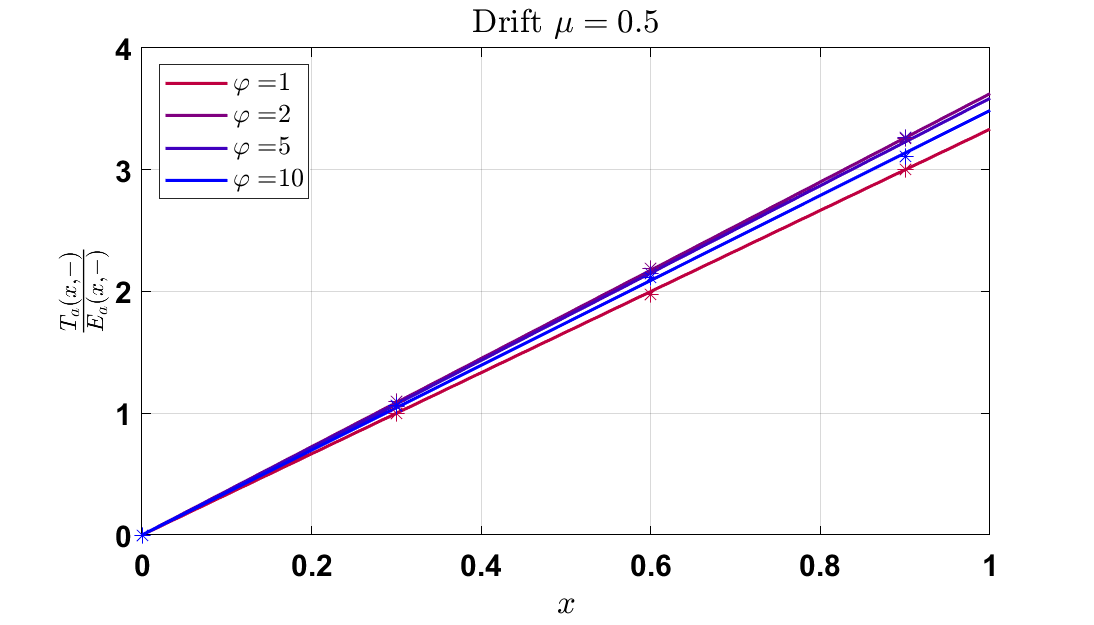}
		 \caption{}
		 %\label{figTimesMinusPositiveDrift}
	      \end{subfigure}
	\caption{The conditional mean first passage-time at $a$ of a particle starting in the  half-line $[a,+\infty[$ (in the presence of a positive drift), given an initial state (a) with zero internal velocity (tumbling), (b) with positive internal velocity, (c) with negative internal velocity. The stars correspond to the results of direct numerical simulations with 100,000 particles (the simulation was run until the time reached ten times the predicted value of the mean conditional first-passage time).}
	\label{figTimesMinusPositiveDrift}
\end{figure}

The above symbolic expression can readily be used to check consistency with the model of an RTP on a half-line with instantaneous tumble in the presence of a negative drift (see Eq. (\ref{consistencyTHalfLineNeg})).\\

%\begin{equation}
%\frac{\mu v_+ - \mu u_+ + 1}{\lambda_- \mu( \mu^2 - 1 )( u_- - u_+ - v_- + v_+)} = \frac{\varphi( 1  -%\mu ^2) +\Delta}{2\,\mu \,\Delta},
%\end{equation}
 
%\begin{equation}
%\frac{\mu v_- - \mu u_- + 1}{\lambda_+ \mu( \mu^2 - 1 )( u_- - u_+ - v_- + v_+)} = 
%\frac{\varphi( 1  -\mu ^2) -\Delta}{2\,\mu \,\Delta},
%\end{equation}

%\begin{equation}
%\frac{u_- - u_+ - u_- v_+ + u_+ v_-}{u_+ - v_+ + 1} =
%\frac{\Delta[2\,\mu - \varphi(1-\mu^2)+\Delta] }{2\,\mu ^2\,\left(-1+\mu \right)\,\left(1+\varphi \right)}
%\end{equation}

Substituting the expressions of the eigenvalues and of the components of the eigenvectors (using Eqs (\ref{coeffMinus},\ref{coeffPlus},\ref{nextInCommon})) yields 
%\begin{equation}
%\begin{split}
%\frac{\mu v_+ - \mu u_+ + 1}{\lambda_- \mu( \mu^2 - 1 )( u_- - u_+ - v_- + v_+)} \times &\frac{u_- - u_+ - u_- v_+ + u_+ v_-}{u_+ - v_+ + 1}\\
%&= \frac{ ( \varphi ( 1 - \mu^2) + \Delta )( -2 \mu + \varphi  ( 1 - \mu^2) - \Delta ) }{4 \mu^3 (1-\mu)  (1+\varphi) },\\
%\frac{u_- - v_- + 1}{ u_+ - v_+ + 1} =&\frac{2\,\mu -\varphi(1 -\mu ^2)+\Delta}{2\,\mu -\varphi(1 -\mu ^2)-\Delta },
%\end{split}
%\end{equation}
%and
{\footnotesize{
\begin{equation}\label{closedFormMinus}
\begin{split}
\mathcal{T}_a(x,0;[a,\infty[,\varphi) 
&=  -\frac{1}{\mu}(x-a) + \frac{1}{\mu^2(1+\varphi)}\left[\frac{ ( \varphi ( 1 - \mu^2) + \Delta )( -2 \mu + \varphi  ( 1 - \mu^2) - \Delta ) }{4 \mu (1-\mu)  } + \varphi \right]\\
&+e^{\lambda_+(x-a)}\left[ \frac{2\mu + \Delta }{\varphi( 1 - \mu^2)} +  \frac{\varphi( 1-\mu^2) - \Delta }{2\mu}
\right] \frac{\varphi \,\left(2 + \varphi( 1 -\mu ^2) +\Delta\right)}{2\,\mu\Delta \,\left(1+\varphi \right)},\\
%-e^{\lambda_+(x-a)}  \Lambda \frac{\varphi \,\left(\varphi( 1  -\mu ^2) +\Delta+2\right)}{2\,\mu \,\left(\varphi +1\right)},\\
\mathcal{T}_a(x,+;[a,\infty[,\varphi) &= -\frac{1}{\mu}(x-a) 
 + \frac{1}{\mu^2(1+\varphi)}\left[\frac{ ( \varphi ( 1 - \mu^2) + \Delta )( -2 \mu + \varphi  ( 1 - \mu^2) - \Delta ) }{4 \mu (1-\mu)  } +\varphi(1-\mu) - \mu\right]\\
&+ e^{\lambda_+(x-a)}\left[ 
\frac{2\mu + \Delta }{\varphi( 1 - \mu^2)} +  \frac{\varphi( 1-\mu^2) - \Delta }{2\mu}
\right] \frac{-2\,\mu +\varphi( 1 - \mu)^2  -\Delta}{2\,\mu \Delta \,\left(1+\varphi \right)},\\
%- e^{\lambda_+(x-a)} \Lambda \left(\frac{\varphi +\mu ^2\,\varphi -\Delta}{2\,\mu \,\left(\varphi +1\right)} - 1 \right),\\
\mathcal{T}_a(x,-;[a,\infty[,\varphi) &= -\frac{1}{\mu}(x-a)+
 \frac{1}{\mu^2(1+\varphi)}\left[\frac{ ( \varphi ( 1 - \mu^2) + \Delta )( -2 \mu + \varphi  ( 1 - \mu^2) - \Delta ) }{4 \mu (1-\mu) } +\varphi(1+\mu) + \mu \right]\\
&+ e^{\lambda_+(x-a)}\left[ \frac{2\mu + \Delta }{\varphi( 1-\mu^2)} +  \frac{\varphi( 1-\mu^2) - \Delta }{2\mu}
\right] \frac{2\,\mu +\varphi(1+\mu)^2  -\Delta}{2\,\mu \Delta \,\left(1+\varphi \right)}
,\\
%- e^{\lambda_+(x-a)} \Lambda\left(\frac{\varphi( 1  +\mu ^2) -\Delta}{2\,\mu \,\left(\varphi +1\right)} + 1 \right),\\
 \text{with}\quad \Delta =& \sqrt{\varphi^2(1-\mu^2)^2 +4\mu^2}.
\end{split}
\end{equation}
}} 
 This is the explicit form of the result announced in Eq. (\ref{resMeanAnnounced}) in the case of a negative drift, with a constant term and exponential corrections in $e^{\lambda_+(x-a)}$ weighted by coefficients depending on the drift and the on the rate $\varphi$.  
These mean first-passage times are plotted for a few values of  the rate $\varphi$ on Figs \ref{figTimesZeroNegativeDrift},\ref{figTimesPlusNegativeDrift},\ref{figTimesMinusNegativeDrift}, together with the results of direct numerical simulations.

%$$\frac{2\,\mu +\Delta}{\varphi \,\left(-1+\mu ^2\right)\,\Delta}$$

%$$\frac{\mu ^2\,\varphi -\varphi +\Delta}{2\,\mu \,\Delta}$$

\subsubsection{Consistency checks}

  It is easy to take the large-$\varphi$ limit of the mean first-passage times, using their expression in Eq. (\ref{MFPTNegDriftAlg}). In this limit, $\lambda_+$ goes to minus infinity, and $u_+$ becomes large, while $\lambda_-$, $u_-$, $v_+$ and $v_-$ go to a finite limit (see Eqs (\ref{limsLambda},\ref{equivEntriesInf})). The terms proportional to $e^{\lambda_+(x-a)}$ therefore vanish for any $x>a$ and we obtain
\begin{equation}\label{consistencyTHalfLineNeg}
\begin{split}
\mathcal{T}_a(x,0;[a,\infty[,\varphi) \underset{\varphi\to \infty}{\sim}& 
\frac{1}{\lambda_-(\mu^2 - 1)}(v_- - u_- - 1) -  \frac{1}{\mu}(x-a)\\
\underset{\varphi\to \infty}{\sim}& -\frac{1}{\mu} - \frac{1}{\mu}(x-a),
\\
\mathcal{T}_a(x,+;[a,\infty[,\varphi) \underset{\varphi\to \infty}{\sim}& \frac{1}{\lambda_-(\mu^2 - 1)}(v_- - 1) - \frac{1}{\lambda_-(\mu^2 - 1)}(v_- + 1)-  \frac{1}{\mu}(x-a)\\
\underset{\varphi\to \infty}{\sim}& -\frac{2}{\mu} - \frac{1}{\mu}(x-a),\\
\mathcal{T}_a(x,-;[a,\infty[,\varphi)  \underset{\varphi\to \infty}{\sim}& 
- \frac{1}{\mu}(x-a).
\end{split}    
\end{equation}
The last two equivalents are the known expressions of  $\mathcal{T}_a(x,+;[a,\infty[,\infty)$ and $\mathcal{T}_a(x,-;[a,\infty[,\infty)$ reported in Eq. (\ref{MFPTOrdinary}) in the case of a negative drift. The first is their average: 
\begin{equation}
\mathcal{T}_a(x,0;[a,\infty[,\infty) = \frac{1}{2}\left[  \mathcal{T}_a(x,+;[a,\infty[,\infty) + \mathcal{T}_a(x,-;[a,\infty[,\infty)\right],
\end{equation}
 because in the limit of instantaneous tumble ($\varphi\to\infty$), a particle starting at $x$ in a tumbling state instantaneously picks a random non-zero velocity state and starts running.\\

\subsection{Positive drift}\label{subsec:pdtime}
\subsubsection{Integration constants}\label{subsubsec:icpdtime}
 In the presence of a positive drift, a particle starting at $b$ in a tumbling state or in a positive internal velocity state immediately leaves the system.  There are no events contributing to the exit of the particle through $a$. Therefore,
\begin{equation}
M_1(b) = 0,\qquad\text{and}\quad M_2( b) + M_3(b) = 0.
\end{equation}
 On the other hand, a particle  starting at $a$ in a negative velocity state or immediately leaves the system. Hence,
 \begin{equation}
 M_2( a) - M_3(a) = 0.
\end{equation}

 The quantities $\vec{M}(a)$ and $\vec{M}(b)$ can therefore be expressed as follows: 
\begin{equation}
\vec{M}(a) = 
\begin{pmatrix}
 M_1( a )\\
 M_2( a )\\
M_2( a )
\end{pmatrix},\qquad
\vec{M}(b) = 
\begin{pmatrix}
 0 \\
 M_2(b)\\
 -M_2(b)\\
\end{pmatrix}.
\end{equation}
On the other hand, the vectors $\vec{M}(b)$ and $\vec{M}(a)$ are related by the solution of the evolution equation, according to Eq. (\ref{JDef}): 
\begin{equation}
\begin{pmatrix}
 0 \\
 M_2(b)\\
 -M_2(b)\\
\end{pmatrix}
 = \exp( (b-a) A)\begin{pmatrix}
  Z_a( a )\\
 M_2( a )\\
M_2( a )
\end{pmatrix} + \vec{J}(b).
\end{equation}

Hence, with the shorthand notation $Q_{ij}=\exp\left((b-a)A\right)$ introduced in Eq. (\ref{shorthandQ}), we obtain the linear system satisfied by the unknown parameters $M_1(a),M_2(a),M_2(b)$:
\begin{equation}\label{defTPlus}
\begin{split}
T_+\begin{pmatrix}
 M_1(a)\\
 M_2(a)\\
 M_2(b)
\end{pmatrix}
   =& 
\begin{pmatrix} 
J_1(b)\\
J_2(b)\\
J_3(b)
\end{pmatrix},\\ 
{\mathrm{with}}\quad T_+ :=& 
\begin{pmatrix} 
 -Q_{11} & -(Q_{12}+ Q_{13})  & 0 \\
 -Q_{21} &  -(Q_{22}+ Q_{23})  & 1\\
 -Q_{31} & -(Q_{32}+ Q_{33})   &  -1 
\end{pmatrix}.
\end{split}
\end{equation}
Hence, the two unknown constants are obtained as the first two components of the vector
\begin{equation}
\begin{split}
\begin{pmatrix}
  M_1(a)\\
 M_2(a)\\
 M_2(b)
\end{pmatrix}
   =& T_+^{-1} 
\begin{pmatrix} 
J_1(b)\\
J_2(b)\\
J_3(b)
\end{pmatrix}.
\end{split}
\end{equation}
Inverting the matrix $T_+$,
{\tiny{
\begin{equation}
\begin{split}
&(T_+)^{-1} = \frac{1}{Q_{12} Q_{21} - Q_{11} Q_{22} - Q_{11} Q_{23} + Q_{13}Q_{21} - Q_{11}Q_{32} + Q_{12}Q_{31} - Q_{11}Q_{33} + Q_{13}Q_{31}}\\
&\times\begin{pmatrix}
Q_{22} + Q_{23} + Q_{32} + Q_{33} &  - Q_{12} - Q_{13} &  - Q_{12} - Q_{13}\\
 - Q_{21} - Q_{31} &    Q_{11} &    Q_{11} \\
Q_{21} Q_{32} - Q_{22}Q_{31} + Q_{21}Q_{33} - Q_{23}Q_{31}& Q_{12}Q_{31} - Q_{11}Q_{32} - Q_{11}Q_{33} + Q_{13}Q_{31}& Q_{11}Q_{22} - Q_{12}Q_{21} + Q_{11}Q_{23} - Q_{13}Q_{21}
\end{pmatrix}.
\end{split}
\end{equation}
}}

Hence the two constants
\begin{equation}\label{M12aSegmentPosDrift}
\begin{split}
M_1(a) =& \frac{(Q_{22} + Q_{23} + Q_{32} + Q_{33})J_1(b)   - (Q_{12} + Q_{13})( J_2(b) + J_3(b) )}{\det( T_+)},\\
M_2(a) =& \frac{ -( Q_{21} + Q_{31}  )J_1(b) + Q_{11}( J_2(b) + J_3(b) )}{\det( T_+)}.
\end{split}
\end{equation}
 These two constants are  enough to express $\vec{M}(x)$ for any $x$ in the segment $[a,b]$, according to the solution of the evolution equation in Eq. (\ref{JDef}). To parallel the results obtained in the case of a negative drift, we are instructed to evaluate the components $J_2(b)$ and $J_3(b)$ and to study the limit of the above two integration constants when the coordinate $b$ goes to infinity.

\subsubsection{Limit of a half-line}\label{subsubsec:limitpdtime}
 For a positive drift, the largest eigenvalue is the positive $\lambda_+$, and in the limit of  large $b$,
\begin{equation}
    e^{\lambda_+(b-a)} \gg 1 \gg e^{\lambda_-(b-a)}.
\end{equation}

  We have to extract the dominant term of the numerator and denominator in the expressions of the integration constants $M_1(a)$ and $M_2(a)$ given in Eq. \eqref{M12aSegmentPosDrift}. The determinant of the matrix $T_+$ satisfies
\begin{equation}
\det( T_+ ) = \frac{1}{2} \det( S_+),
\end{equation}
   where $S_+$ is the matrix we encountered when calculating the integration constants for the exit probability. Indeed, $T_+$ is obtained from $S_+$ by multiplying its second column by a factor of $1/2$ (see Eqs (\ref{defSPlus},\ref{defTPlus})). An asymptotic expansion of $\det( T_+ )$
    is therefore obtained from Eq. (\ref{detSPlusAsymp}) as
\begin{equation}
\det( T_+) = \frac{(v_- - 1)(v_+ - u_+ + 1)}{2(u_- - u_+ - v_- + v_+)}e^{\lambda_+(b-a)} + e^{\lambda_+(b-a)}.
\end{equation}

 To study the numerators, we need to express the vector $\vec{J}(b) = \Gamma(b) \vec{\mathcal{E}}(a)$, where $\Gamma(b)$ is the matrix defined in Eq. (\ref{defJ}). The algebra is shown in Appendix \ref{appVecMa}, and yields

\begin{equation}\label{M12aExprPosDrift}
\begin{split}
M_1( a ) \underset{b\to\infty}{\sim}& \frac{1}{\mu(1-\mu^2)( v_- -1)^2(u_- - u_+ - v_- + v_+)}\\
&\times\left[  (u_- - v_- + 1)\left(\frac{1}{\lambda_-}   (\mu u_- - \mu u_+ - u_- v_- + u_- v_+ + \mu u_+ v_-^2 - \mu u_- v_- v_+ )\right)\right.\\
&\left.+ ( u_- - u_+ - u_- v_+ + u_+ v_-) \left( \frac{1}{\lambda_+ - \lambda_-}    (- \mu v_-^2 + \mu u_- v_- + \mu - u_-) \right) \right],\\
M_2( a )
\underset{b\to\infty}{\sim}& \frac{1}{\mu(1-\mu^2)( v_- -1)^2(u_- - u_+ - v_- + v_+)}\\
& \times\left[ \frac{1}{\lambda_-}   (\mu u_- - \mu u_+ - u_- v_- + u_- v_+ + \mu u_+ v_-^2 - \mu u_- v_- v_+)  \right.\\
&\left.+ (v_- - v_+)\left( \frac{1}{\lambda_+ - \lambda_-}    (- \mu v_-^2 + \mu u_- v_- + \mu - u_-) \right) \right].\\
\end{split}
\end{equation}
We have therefore expressed all the components of the vector $\vec{M}(a)$.\\

Let us express the vector $\vec{M}(x)$ in the limit of a half-line. Replacing the vector $\vec{\mathcal{E}}(a)$ with its large-$b$ limit, using the identities
\begin{equation}\label{algId}
\begin{split}
   W^{(1,1)}  
   \begin{pmatrix}
  u_-\\
  v_-\\
  1
   \end{pmatrix} =& 0,\qquad  
   W^{(2,2)}  
   \begin{pmatrix}
  u_-\\
  v_-\\
  1
  \end{pmatrix} = 0,\\ \qquad
  W^{(3,3)} 
   \begin{pmatrix}
  u_-\\
  v_-\\
  1
  \end{pmatrix} 
  = &-
\frac{(\mu - u_- + \mu^2 u_- - \mu^2 u_+ - \mu^2 v_- + \mu^2 v_+ + \mu u_+ v_- - \mu v_- v_+)}{\mu (\mu^2 - 1) (v_- - 1)(u_- - u_+ - v_- + v_+)}
  \begin{pmatrix}
  u_-\\
  v_-\\
  1
  \end{pmatrix}\\
  =& \Xi \vec{\mathcal{E}}(a).
  \end{split}
\end{equation}
with the notation
\begin{equation}
\begin{split}
\Xi =& 
\frac{\mu - u_- + \mu^2( u_- - u_+ - v_- +  v_+) + \mu v_-( u_+  -  v_+)}{\mu (1-\mu^2 )(u_- - u_+ - v_- + v_+)}\\
%& \frac{\varphi(1-\mu^2) + 2\mu^2 + \mu^2\sqrt{\varphi^2(1-\mu^2)^2 + 4\mu^2}}{\mu(1-\mu^2)\sqrt{\varphi^2(1-\mu^2)^2 + 4\mu^2}}.
\end{split}
\end{equation}

Hence the simplification of the expression of $\vec{M}(x)$:
{\footnotesize{
\begin{equation}
\begin{split}
\vec{M}(x) \underset{b\to +\infty}{\sim}& q^{(1)} \vec{M}(a) -\frac{1}{\lambda_-}( W^{(1,3)} + W^{(3,1)})\vec{\mathcal{E}}(a)
\\
&+ e^{\lambda_+(x-a)} 
  \left\{ q^{(2)} \vec{M}(a) 
 + \frac{1}{\lambda_+ - \lambda_-}( W^{(2,3)} + W^{(3,2)}) \vec{\mathcal{E}}(a)\right\}
\\
&+ e^{\lambda_-(x-a)} \left\{ q^{(3)} \vec{M}(a) + 
 (x-a)\Xi \vec{\mathcal{E}}(a) + \left[\frac{1}{\lambda_-}( W^{(1,3)} + W^{(3,1)}) 
 - \frac{1}{\lambda_+ - \lambda_-}( W^{(2,3)} + W^{(3,2)})\right] \vec{\mathcal{E}}(a) \right\}.\\
\end{split}
\end{equation}
}}
 Moreover, the expressions of $M_1(a)$ and $M_2(a)$ imply that the coefficients of  $e^{0(x-a)}$ and $e^{\lambda_+(x-a)}$ both vanish:
\begin{equation}
\begin{split}
& q^{(1)} \vec{M}(a)  \underset{b\to +\infty}{\sim} \frac{1}{\lambda_-}( W^{(1,3)} + W^{(3,1)})\vec{\mathcal{E}}(a),
\\
&q^{(2)} \vec{M}(a) 
  \underset{b\to +\infty}{\sim} - \frac{1}{\lambda_+ - \lambda_-}( W^{(2,3)} + W^{(3,2)}) \vec{\mathcal{E}}(a).
\end{split}    
\end{equation}
 On the other hand, the operator $Q(0) = \exp( 0 A)$ is the identity, hence
\begin{equation}
  q^{(3)} \vec{M}(a) + \left[\frac{1}{\lambda_-}( W^{(1,3)} + W^{(3,1)}) 
 - \frac{1}{\lambda_+ - \lambda_-}( W^{(2,3)} + W^{(3,2)})\right] \vec{\mathcal{E}}(a) = ( q^{(1)} + q^{(2)} + q^{(3)})  \vec{M}(a) =  \vec{M}(a).
\end{equation}

 We therefore obtain the expression of $\vec{M}(x)$ in the large-$b$ limit as 
 \begin{equation}
\begin{split}
\vec{M}(x)  \underset{b\to +\infty}{\sim}
& e^{\lambda_-(x-a)} \left[ \vec{M}(a) + 
 (x-a)\Xi \vec{\mathcal{E}}(a)  \right].\\
\end{split}
\end{equation}

 The following combinations follow:
\begin{equation}
\begin{split}
M_1( x ) \underset{b\to\infty}{\sim}&   e^{\lambda_-(x-a)}\left(U( a ) +  \frac{\Xi(x-a)}{ v_- - 1} u_-\right), \\
M_2(x) + M_3(x) \underset{b\to\infty}{\sim}& e^{\lambda_-(x-a)}\left(2M_2( a ) +  \frac{\Xi(v_- + 1)}{v_- - 1} (x-a) \right), \\
M_2(x) - M_3(x)  \underset{b\to\infty}{\sim}&  e^{\lambda_-(x-a)} \Xi(x-a).\\ 
\end{split}    
\end{equation}
Moreover, using Eq. (\ref{ExSymbPositiveDrift}) for the expression of the exit probabilities, we obtain the large-$b$ limits of the mean first-passage times through $a$, conditioned on the exit of the system: 
 \begin{equation}\label{XiFormal}
\begin{split}
\frac{M_1( x )}{Z_a(x)}  \underset{b\to +\infty}{\sim}&  \frac{v_- - 1}{u_-}M_1(a) +\Xi(x-a),\\
\frac{T_a( x,+ )}{E_a(x,+)}  \underset{b\to +\infty}{\sim}& 2\frac{v_- - 1}{v_- + 1}M_2(a) +\Xi (x-a),\\ 
 \frac{T_a( x,- )}{E_a(x,-)}  \underset{b\to +\infty}{\sim}&  \Xi (x-a),\\
 \text{with}\qquad \Xi =&  -\frac{\mu - u_- + \mu^2( u_- - u_+ - v_- +  v_+) + \mu v_-( u_+  -  v_+)}{\mu (\mu^2 - 1)(u_- - u_+ - v_- + v_+)}
\end{split}    
\end{equation}

In terms of the parameters $\mu$ and $\varphi$, the constant terms in the above affine expressions become
\begin{equation}
\begin{split}
{\mathcal{T}}_a( a, 0; [a,\infty[, \varphi ) =&  \underset{b\to\infty}{\lim} \frac{M_1(a)}{Z_a(a)}\\
=& \frac{1}{\mu(1-\mu^2)u_-( v_- -1)(u_- - u_+ - v_- + v_+)}\\
&\times\left[  (u_- - v_- + 1)\left(\frac{1}{\lambda_-}   (\mu u_- - \mu u_+ - u_- v_- + u_- v_+ + \mu u_+ v_-^2 - \mu u_- v_- v_+ )\right)\right.\\
&\left.+ ( u_- - u_+ - u_- v_+ + u_+ v_-) \left( \frac{1}{\lambda_+ - \lambda_-}    (- \mu v_-^2 + \mu u_- v_- + \mu - u_-) \right) \right],\\
{\mathcal{T}}_a( a, +; [a,\infty[, \varphi ) =&  \underset{b\to\infty}{\lim} \left( 2\frac{M_2(a)}{{\mathcal{E}}_a(a,+;[a,b],\varphi)}\right)\\
=& \frac{2}{\mu(1-\mu^2)( v_- +1)( v_- -1)(u_- - u_+ - v_- + v_+)}\\
& \times\left[ \frac{1}{\lambda_-}   (\mu u_- - \mu u_+ - u_- v_- + u_- v_+ + \mu u_+ v_-^2 - \mu u_- v_- v_+)  \right.\\
&\left.+ (v_- - v_+)\left( \frac{1}{\lambda_+ - \lambda_-}    (- \mu v_-^2 + \mu u_- v_- + \mu - u_-) \right) \right].\\
\end{split}    
\end{equation}
The quantities $\lambda_-,\lambda_+, u-_, u_+, v_-, v_+$ are related to the variables $\mu$ and $\varphi$ by the following identities:
{\tiny{
\begin{equation}
\begin{split}
\Xi = \frac{\mu - u_- + \mu^2( u_- - u_+ - v_- +  v_+) + \mu v_-( u_+  -  v_+)}{\mu (1 -  \mu^2 )(u_- - u_+ - v_- + v_+)}\\
=& \frac{\varphi(1-\mu^2) + 2\mu^2 + \mu^2\Delta}{\mu(1-\mu^2)\Delta},\\
\mu u_- - \mu u_+ - u_- v_- + u_- v_+ + \mu u_+ v_-^2 - \mu u_- v_- v_+ =&
  \frac{\varphi \,\left(\mu ^2-1\right)\,\Delta}{\mu ^2\,\left(\varphi +1\right)},\\
- \mu v_-^2 + \mu u_- v_- + \mu - u_- =& \frac{\varphi \,\left(\mu ^2+1\right)\,\left(\varphi -\mu ^2\,\varphi -\Delta+2\right)}{2\,\mu \,\left(\varphi +1\right)},\\
u_- - v_- + 1 =& \frac{2\,\mu -\varphi(1 -\mu ^2) +\Delta}{2\,\mu },\\
 u_- - u_+ - u_- v_+ + u_+ v_- =&   \frac{\varphi \,\left(\mu +1\right)\,\Delta}{\mu ^2\,\left(\varphi +1\right)} ,\\
v_- - v_+ =&   -\frac{\Delta}{\mu \,\left(\varphi +1\right)} ,\\
u_-( v_- -1)(u_- - u_+ - v_- + v_+) =& \frac{\varphi \,\Delta\,\left(2 + \varphi( 1  -\mu ^2) -\Delta\right)\,\left(2\,\mu +\varphi( 1+\mu)^2  +\Delta\right)}{4\,\mu ^3\,{\left(\varphi +1\right)}^2},\\
( v_- +1)( v_- -1)(u_- - u_+ - v_- + v_+) =& \frac{\Delta \,\left(-2\,\mu +\varphi( 1-\mu)^2 +\Delta\right)\,\left(2\,\mu + \varphi( 1+\mu)^2  +\Delta\right)}{4\,\mu ^3\,{\left(\varphi +1\right)}^2}\\
\text{with}\qquad\Delta =& \sqrt{4\mu^2 + (\varphi(1-\mu^2))^2}.\\
\end{split}
\end{equation}
}}
Substituting, we obtain the expressions of the constants ${\mathcal{T}}_a( a, 0; [a,\infty[, \varphi )$ and  ${\mathcal{T}}_a( a, +; [a,\infty[, \varphi )$ in terms of the drift $\mu$ and the rate $\varphi$. 
To sum up, the mean first-passage time at $a$ of the RTP moving on $[a,\infty[$ with finite mean tumble duration $\varphi^{-1}$ (conditioned on the exit of the particle) has an affine dependence on the initial coordinate $x$, for each possible value of the initial velocity state:

\begin{equation}\label{affineFormPlus}
\begin{split}
&{\mathcal{T}}_a( x, 0; [a,\infty[, \varphi ) =  {\mathcal{T}}_a( a, 0; [a,\infty[, \varphi )  +
 \Xi (x-a),\\
&{\mathcal{T}}_a( x, +; [a,\infty[, \varphi )= {\mathcal{T}}_a( a, +; [a,\infty[, \varphi ) +
 \Xi (x-a),\\   
&{\mathcal{T}}_a( x, -; [a,\infty[, \varphi ) =   \Xi(x-a),\\
\end{split}
\end{equation}
with
\begin{equation}
\Xi =  \frac{\varphi(1-\mu^2) + 2\mu^2 + \mu^2\Delta}{\mu(1-\mu^2)\Delta}, \quad \text{where} \quad \Delta = \sqrt{4\mu^2 + (\varphi(1-\mu^2))^2},
\end{equation}
\begin{equation}\label{closedFormPlus}
\begin{split}
&{\mathcal{T}}_a( a, 0; [a,\infty[, \varphi )
= \frac{4\,{\left(\varphi +1\right)}}{  \Delta\,\left(2 + \varphi( 1  -\mu ^2) -\Delta\right)\,\left(2\,\mu +\varphi( 1+\mu)^2  +\Delta\right)}\\
&\times\left[  (2\,\mu -\varphi(1 -\mu ^2) +\Delta)\left(\frac{ \left(\mu ^2-1\right)\,\Delta}{\varphi( 1- \mu^2) -2\mu^2 - \Delta}   \right) + \frac{ \mu +1 }{\varphi +1}  \left(   \frac{\varphi \,\left(\mu ^2+1\right)\,\left(\varphi( 1 -\mu ^2)+2 -\Delta\right)}{2} 
\right) \right],\\
&{\mathcal{T}}_a( a, +; [a,\infty[, \varphi )  
= \frac{8 \,\mu\, \left(\varphi +1\right)}{ \Delta \,\left(-2\,\mu +\varphi( 1-\mu)^2 +\Delta\right)\,\left(2\,\mu + \varphi( 1+\mu)^2  +\Delta\right)}\\
 &\times\left[ \frac{2 \varphi \,\left(\mu ^2-1\right)\,\Delta }{\varphi( 1- \mu^2) -2\mu^2 - \Delta}  -  \frac{\varphi \,\left(\mu ^2+1\right)\,\left(\varphi( 1 -\mu ^2)+2 -\Delta\right)}{2\left(\varphi +1\right) }   \right].\\
\end{split}    
\end{equation}
 This is the result reported in Eq. (\ref{resMeanAnnounced}) of the introduction, in the case of a positive drift. The three quantities are plotted in Fig. (\ref{figTimesMinusPositiveDrift}) as a function of the coordinate $x$.\\

\subsubsection{Consistency checks}\label{checkTPos}

In the limit of instantaneous tumble events $(\varphi\to\infty)$, the quantities $u_+$ and $\lambda_+$ go to infinity, hence the equivalents:
\begin{equation}
  U( a) \underset{\varphi \to +\infty, b \to +\infty}{\sim} M_2(a) \sim \frac{1}{\mu(1-\mu^2)\lambda_-( v_- - 1)^2}\mu( 1 - v_-^2)\sim \frac{1-\mu}{\mu( 1 + \mu)}.
\end{equation}
 Moreover, we know from the consistency checks for the exit probability on the half-line  (Eq. (\ref{chekEPosHalfLineInst})) that $\lim_{\varphi\to\textbf{}\infty}\mathcal{E}_a(a,+;[a,\infty[,\varphi) = (1-\mu)(1+\mu)^{-1}$. Hence,
 \begin{equation}
  \mathcal{T}_a(a,+,[a,b],\varphi) \underset{\varphi \to +\infty, b \to +\infty}{\sim} \frac{2}{\mu},
\end{equation}
This is the value of  $\mathcal{T}_a(a,+,[a,\infty[,\infty)$ reported in Eq. (\ref{MFPTOrdinary}).
 In the limit of instantaneous tumble events, a particle starting tumbling at $a$ starts running with an internal velocity state drawn uniformly, which intuitively makes $M_1(a)$ very close to $M_2(a)$.\\

As the dependence of the conditional mean first-passage time at $a$ is affine, we just have to check  the limit of the slope $\Xi$. Indeed, as $u_+$ becomes large in this limit, Eq. (\ref{XiFormal}) implies
\begin{equation}
\Xi \underset{\varphi\to\infty}{\sim}  \frac{-\mu^2 + \mu v_-}{\mu(\mu^2-1)} \sim \frac{1+\mu^2}{\mu(1-\mu^2)}.
\end{equation}
 This limit also follows from a direct calculation using the explicit formula for $\Xi$ in Eq. (\ref{affineFormPlus}) in terms of $\varphi$ and $\mu$. 
 It reproduces the slope in the affine expression of $\mathcal{T}_a(x,+;[a,\infty[,\infty)$ displayed in  Eq. (\ref{MFPTOrdinary}).\\

\section{Milne interpolation length}\label{sec:Milne}

 If the particle living on the half-line $[a,\infty[$ starts its motion at $a$ with a nonnegative total velocity $\sigma(0) \times 1 + \mu$, its mean lifetime (conditioned on exiting the system) is positive. This induces a coordinate, call it $x_{\sigma(0)}(\varphi,\mu)$ such that the conditional mean-first passage time evaluated at this coordinate equals zero. This coordinate is defined implicitly by the condition
\begin{equation}
 \mathcal{T}_a( x_{\sigma(0)}, \sigma(0);[a,\infty[,\varphi) = 0,
\end{equation} 
 with the conditions
 \begin{equation}
     \sigma(0) \in \{0,+1 \} \quad \text{if} \quad \mu>0, \quad \sigma(0) = -1 \quad {\text{if}}  \quad \mu<0.
 \end{equation}
 By analogy with neutron scattering, the length 
  \begin{equation}
 \xi_{\sigma(0)}(\varphi,\mu) := | x_{\sigma(0)}(\varphi,\mu) -a|
  \end{equation}
 has been named the Milne interpolation length \cite{le2019noncrossing,de2021survival}. It depends on the drift $\mu$ and on the rate $\varphi$ as well as on the initial velocity state, but for brevity we omit this dependence from the notation. The formulas obtained for $\mathcal{T}_a$ allow to work out a closed-form expression of the Milne interpolation lengths. 

 \subsection{Negative drift}
 There is only one Milne interpolation length, $\xi_+$. According to Eq. (\ref{closedFormMinus}), it satisfies
 \begin{equation}\label{QPlusFPlus}
 \begin{split}
   0 =& +\frac{1}{\mu}  \xi_+(\varphi,\mu)  + Q_+ + F_+ e^{-\lambda_+ \xi_+(\varphi,\mu)},\\
{\text{with}}\qquad Q_+ =&  \frac{1}{\mu^2(1+\varphi)}\left[\frac{ ( \varphi ( 1 - \mu^2) + \Delta )( -2 \mu + \varphi  ( 1 - \mu^2) - \Delta ) }{4 \mu (1-\mu)  } +\varphi(1-\mu) - \mu\right],\\
F_+ =&\left[ 
\frac{2\mu + \Delta }{\varphi( 1 - \mu^2)} +  \frac{\varphi( 1-\mu^2) - \Delta }{2\mu}
\right] \frac{-2\,\mu +\varphi( 1 - \mu)^2  -\Delta}{2\,\mu \Delta \,\left(1+\varphi \right)}.
\end{split}
 \end{equation}
This can be equivalently rewritten as 
\begin{equation}
(\lambda_+\xi_+(\varphi,\mu) + \mu \lambda_+ Q_+) e^{\lambda_+\xi_+(\varphi,\mu) + \mu \lambda_+ Q_+} = - \lambda_+ F_+ e^{\mu \lambda_+ Q_+},
\end{equation}
 which is solved explicitly in terms of the Lambert $W$ function (defined  functionally for nonnegative $y$ by $W(y) = x$ if $y = x e^x$) as 
\begin{equation}\label{MilneNegDrift}
\begin{split}
\lambda_+( \xi_+(\varphi,\mu) + \mu Q_+) =& W\left( -\mu\lambda_+ F_+ e^{\mu\lambda_+ Q_+}\right),\\ 
{\text{hence}},\quad \xi_+(\varphi,\mu) =& \frac{1}{\lambda_+}  W\left( -\mu\lambda_+ F_+ e^{\mu\lambda_+ Q_+}\right) -\mu Q_+.
\end{split}
\end{equation}

After some algebra, one can express $F_+$ in terms of the two positive quantities $-2\mu$ and $\varphi(1 - \mu^2)$, in product form:  
\begin{equation}\label{manif}
\begin{split}
F_+ =&\frac{(\Delta + 2\mu - \varphi( 1 - \mu^2 ) )( -2\mu + \varphi( 1 - \mu)^2 -\Delta) }{ \varphi( 1 - \mu^2 )( -2\mu)^2(1+\varphi)}\\
=& \frac{(\Delta +2\mu - \varphi( 1 - \mu^2 ))( -2\mu + \varphi( 1 - \mu^2)\frac{2 - 2\mu}{2+2\mu} -\Delta) }{ 4\mu^2\varphi( 1 - \mu^2 )(1+\varphi)}\\
=& \frac{(\Delta  +2\mu  -\varphi( 1 - \mu^2 )) (-\Delta   -2\mu + \varphi( 1 - \mu^2 ) - \varphi( 1 - \mu^2 )\frac{2\mu}{1+\mu})}{4\mu^2\varphi(1-\mu^2)(1+\varphi)}.
\end{split}
\end{equation}
On the other hand, the quantity $\Delta - ( -2\mu + \varphi( 1 - \mu^2 ))$ is negative. Indeed, in terms of the two positive quantities $m:= -2\mu$ and $f:=\varphi(1-\mu^2)$, 
\begin{equation}
\begin{split}
\Delta - ( -2\mu + \varphi( 1 - \mu^2 )) =& \sqrt{m^2 + f^2} - m - f\\
=& m\left(  \sqrt{1 + \left(\frac{f}{m}\right)^2}  - 1 - \frac{f}{m} \right).
\end{split}    
\end{equation}
 On the other hand, the function $g(x):=\sqrt{1+x^2} - 1 - x$ satisfies $g'(x) = \frac{x-\sqrt{1+x^2}}{\sqrt{1+x^2}}$, which is negative for $x>0$. As $g(0) = 0$, the quantity $g(f/m)$ is negative. The quantity $F_+$ as expressed in Eq. (\ref{manif})
  is therefore manifestly negative. The argument of the Lambert function in Eq. (\ref{MilneNegDrift}) is therefore positive (let us recall that $\lambda_+$ is negative in the case of a negative drift $\mu$). The value of the Lambert function is therefore uniquely determined.\\

Moreover, the quantity $Q_++F_+$ is positive as it equals the mean first-return time to the origin 
  $\mathcal{T}(a,+;[a,\infty[,\varphi)$, hence 
  \begin{equation}\label{QF}
      Q_+ = \mathcal{T}(a,+;[a,\infty[,\varphi) -F_+> 0.
  \end{equation}
The quantity $\mu\lambda_+ Q_+$ is therefore positive, and the argument of the Lambert function goes to infinity in the limit of instantaneous tumble:
\begin{equation}
    -\mu \lambda_+ F_+ e^{\mu \lambda_+ Q_+}\underset{\varphi \to \infty}{ \longrightarrow} +\infty.
\end{equation}
%As the Lambert function goes logarithmically to infinity, 
%\begin{equation}
% W( x )  \underset{x\to\infty}{=}  \ln( x ) - \ln\left ( \ln(x) \right)+ o( 1 ). 
%\end{equation}

From the expression of $F_+$ in Eq. (\ref{manif})  and the equivalent of $\lambda_+$ given in Eq. (\ref{limsLambda}), the argument of the Lambert function has the following equivalent in the limit of instantaneous tumble:
\begin{equation}
 - \mu\lambda_+ F_+ e^{\mu \lambda_+ Q_+} \underset{\varphi\to \infty}{\sim} \frac{1}{1+\mu} e^{\mu\lambda_+ Q_+}.
\end{equation}
On the other hand, the Lambert at large argument function goes logarithmically to infinity, 
\begin{equation}
 W( x )  \underset{x\to\infty}{=}  \ln( x ) + o( 1 ). 
\end{equation}
Hence, 
\begin{equation}
 W( - \mu\lambda_+ F_+ e^{\mu \lambda_+ Q_+})  \underset{\varphi\to \infty}{\sim} \mu\lambda_+ Q_+.
\end{equation}
Hence, the Milne extrapolation length goes to zero in the limit of instantaneous tumble:
\begin{equation}
  \underset{\varphi \to \infty}{\lim}\xi_+(\varphi,\mu)  = 0.
\end{equation}

%{\bf{Remark (limit of instantaneous tumble).}} As the eigenvalue $\lambda_+$ is large and negative, the exponential correction proportional to $e^{\lambda_+(x-a)}$ becomes dominant when $\varphi$ goes  to infinity. The Milne extrapolation length therefore goes to zero in the limit of large $\varphi$.

This limit is different from  the Milne length obtained in the mean first-passage time in the RTP model with instantaneous tumble (Eq. (\ref{MFPTOrdinary}) ):
\begin{equation}
 \underset{\varphi \to \infty}{\lim}\xi_+(\varphi,\mu) \neq -\mu \mathcal{T}_a(a,  +; [a,\infty[,\infty).
\end{equation}
 Intuitively, this is due to the fact that exponential corrections proportional to $e^{\lambda_+(x-a)}$ become large for any $x<a$ when the rate $\varphi$ goes to infinity.\\

 On the other hand, the quantities $\lambda_+$, $F_+$ and $Q_+$ go to finite limits when $\varphi$ goes to zero. Indeed, from Eqs (\ref{QPlusFPlus},\ref{manif})
\begin{equation}
\begin{split}
 \Delta \underset{\varphi \to 0}{=}& -2\mu + O(\varphi^2),\\
 F_+ \underset{\varphi \to 0}{\sim}& \frac{ -\varphi( 1 - \mu^2) \varphi( 1-\mu^2)\left( 1 - \frac{2\mu}{1+\mu}  \right)}{\varphi( 1-\mu^2)( -2\mu)^2}\\
 \sim& \frac{ - \varphi( 1-\mu^2)\left( 1 - \frac{2\mu}{1+\mu}  \right)}{( -2\mu)^2},\\
 Q_+ \underset{\varphi \to 0}{\sim}& - \frac{1}{\mu}.
\end{split}
\end{equation}
 Hence, using $W(0) = 0$ and continuity of the Lambert function at $0$, 
 \begin{equation}
 \begin{split}
 \underset{\varphi \to 0}{\lim} F_+ =& 0,\\
 \underset{\varphi \to 0}{\lim} -\mu Q_+ =& 1,\\
 \underset{\varphi \to 0}{\lim} \xi_+(\varphi,\mu) =& 1.\\
 \end{split}
 \end{equation}
 The long-tumble limit of the Milne extrapolation length is therefore independent of the value of the negative drift. This result can be recovered intuitively as follows. In the limit of long tumble, $F_+$ vanishes and according to Eq. (\ref{QF}), the quantity $Q_+$ coincides with  the mean first-return time $\mathcal{T}_a(a,+;[a,\infty[,\varphi)$. 
  In the limit of a long average tumble time, this mean first-return time  is the expectation of the sum of a run time of duration $\tau$ (with $\tau$ exponentially distributed), and a tumble time during which the particle travels from $( 1+\mu)\tau$ to $0$ at velocity $\mu$, for a time 
 $( 1+\mu)\tau/|\mu|$. Hence,
 \begin{equation}
    Q_+ \underset{\varphi \to 0}{\sim} \int_0^\infty \left( \tau +\frac{(1+\mu)\tau}{|\mu|}\tau  \right) e^{-\tau}d\tau = -\frac{1}{\mu}\int_0^\infty \tau e^{-\tau}d\tau = -\frac{1}{\mu}.
 \end{equation}
This is consistent with the above limit because exponential corrections vanish with $F_+$  and the slope $-1/\mu$ does not depend on the rate $\varphi$.\\

%CODE: Milne_length_all_signs
%The Milne extrapolation length is plotted for a couple of negative values of the drift on Fig. (\ref{figMilneNegativeDrift}).\\

%CODE: Milne_length
\begin{figure}
\begin{center}
    \includegraphics[width=0.8\textwidth]{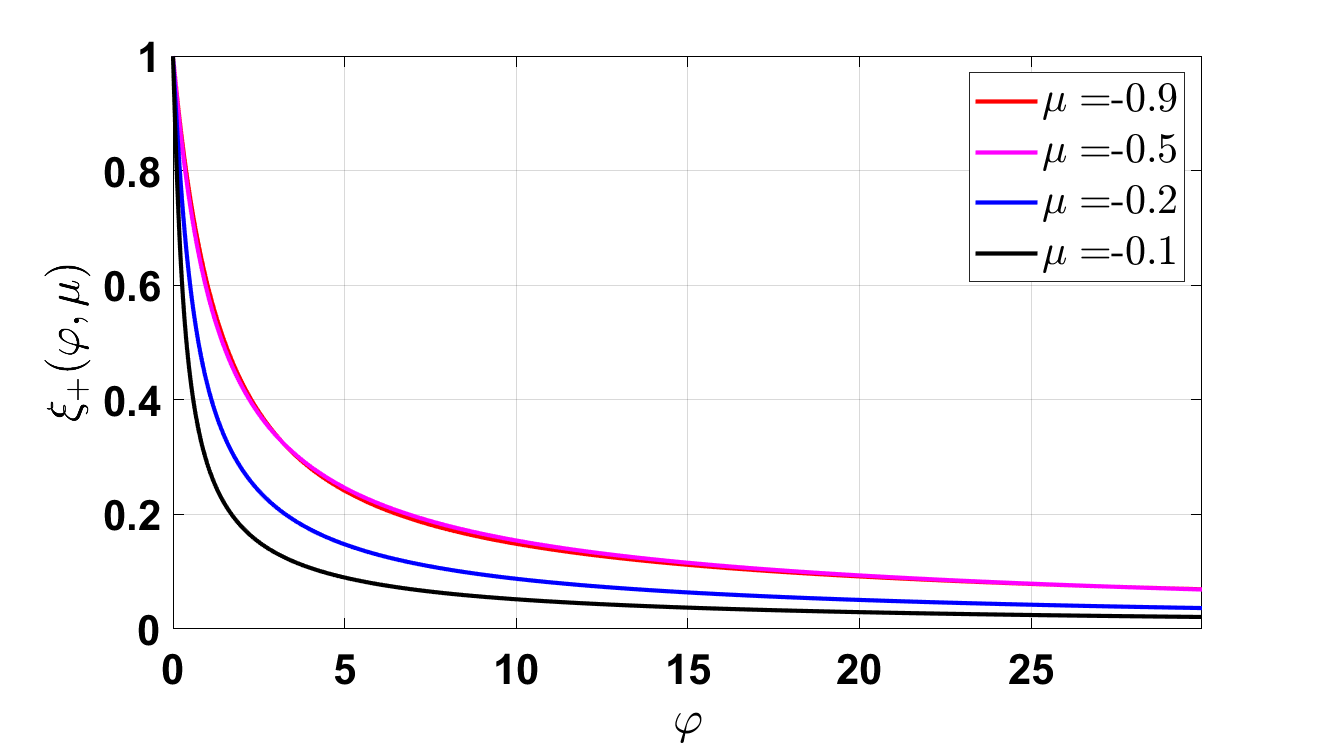}
\end{center}
\caption{The Milne length for a particle starting its motion in a positive internal velocity state, with a negative drift.}
\label{figMilneNegativeDrift}
\end{figure}

 \subsection{Positive drift}
  There are two Milne interpolation lengths $\xi_+(\varphi,\mu)$ and $\xi_0(\varphi,\mu)$. Due to the affine form of the conditional mean first-passage times, their explicit expressions follow immediately:

%\begin{equation}
%\begin{split}
%\xi_0(\varphi,\mu) =& \frac{1}{\Xi}\mathcal{T}_a(a,0;[a,\infty[,\varphi)\\
%=& \frac{4 \mu( 1 - \mu^2)\,{\left(\varphi +1\right)}}{  \left(     \varphi(1-\mu^2 )  + 2\mu^2  + \mu^2\Delta \right)\,\left(2 + \varphi( 1  -\mu ^2) -\Delta\right)\,\left(2\,\mu +\varphi( 1+\mu)^2  +\Delta\right)      }\\
%&\times\left[  (2\,\mu -\varphi(1 -\mu ^2) +\Delta)\left(\frac{ \left(\mu ^2-1\right)\,\Delta}{\varphi( 1- \mu^2) -2\mu^2 - \Delta}   \right) + \frac{ \mu +1 }{\varphi +1}  \left(   \frac{\varphi \,\left(\mu ^2+1\right)\,\left(\varphi( 1 -\mu ^2)+2 -\Delta\right)}{2} 
%\right) \right],\\
%    \xi_+(\varphi,\mu) =& \frac{1}{\Xi}\mathcal{T}_a(a,+;[a,\infty[,\varphi)\\
%=&     \frac{8 \mu^2( 1 - \mu^2)\, \left(\varphi +1\right)}{ \left(     \varphi(1-\mu^2 )  + 2\mu^2  + \mu^2\Delta \right)\,\left(-2\,\mu +\varphi( 1-\mu)^2 +\Delta\right)\,\left(2\,\mu + \varphi( 1+\mu)^2  +\Delta\right)}\\
%& \times\left[ \frac{2 \varphi \,\left(\mu ^2-1\right)\,\Delta }{\varphi( 1- \mu^2) -2\mu^2 - \Delta}  -  \frac{\varphi \,\left(\mu ^2+1\right)\,\left(\varphi( 1 -\mu ^2)+2 -\Delta\right)}{2\left(\varphi +1\right) }   \right].\\
%\end{split}
%  \end{equation}

  \begin{equation}
\begin{split}
\xi_+(\varphi,\mu) =& \frac{1}{\Xi}\mathcal{T}_a(a,+;[a,\infty[,\varphi)\\
=&    \frac{\mu(1-\mu^2)\Delta}{\varphi(1-\mu^2) + 2\mu^2 + \mu^2\Delta} \mathcal{T}_a(a,+;[a,\infty[,\varphi), \qquad{\text{with}}\quad \Delta = \sqrt{\varphi^2( 1 - \varphi^2)^2 + 4\mu^2},\\ 
\xi_0(\varphi,\mu) =& \frac{1}{\Xi}\mathcal{T}_a(a,0;[a,\infty[,\varphi)\\
=& \frac{\mu(1-\mu^2)\Delta}{\varphi(1-\mu^2) + 2\mu^2 + \mu^2\Delta}\mathcal{T}_a(a,0;[a,\infty[,\varphi).\\
\end{split}
  \end{equation}

As the dependence of $\mathcal{T}_a(a,+;[a,\infty[,\varphi)$
  on the distance to $a$ is affine in the case of a positive drift, the consistency checks of Section \ref{checkTPos} ensure that the limit of the Milne interpolation length is the value derived from the model of an RTP on a half-line with instantaneous tumble displayed  in Eq. (\ref{MFPTOrdinary}):
\begin{equation}
\begin{split}
\underset{\varphi \to \infty}{\lim} \xi_0( \varphi,\mu) =& \frac{1}{   \underset{\varphi\to\infty}{\lim}\Xi}\times  \underset{\varphi\to\infty}{\lim} \mathcal{T}_a(a,0;[a,\infty[, \varphi) = \frac{(1-\mu)(1-\mu^2)}{1+\mu^2},\\
 \underset{\varphi \to \infty}{\lim} \xi_+( \varphi,\mu) =&\frac{1}{ \underset{\varphi\to\infty}{\lim} \Xi}\times \underset{\varphi\to\infty}{\lim} \mathcal{T}_a(a,+;[a,\infty[, \varphi) = \frac{2(1-\mu^2)}{1+\mu^2}.\\
\end{split}
\end{equation}
On the other hand, both the conditional mean first-return times to $a$ 
 and the slope $\Xi$ enjoy a finite limit when the mean tumble time becomes long. Hence,
\begin{equation}
\begin{split}
 \underset{\varphi \to 0}{\lim} \xi_+( \varphi,\mu) =& \frac{1}{ \underset{\varphi\to 0}{\lim} \Xi}\times \underset{\varphi\to 0}{\lim} \mathcal{T}_a(a,+;[a,\infty[, \varphi) = \frac{4(1-\mu^2)}{1+\mu^2},\\
 \underset{\varphi \to 0}{\lim} \xi_0( \varphi,\mu) =& \frac{1}{ \underset{\varphi\to 0}{\lim}  \Xi}\times \lim_{\varphi \to 0} \mathcal{T}_a(a,0;[a,\infty[, \varphi) = \frac{1-\mu}{\mu}
\end{split}
\end{equation}

\section{Summary and outlook}
 We have worked out the Fokker--Planck equations satisfied by the densities of the first-passage times at the left end of a segment $[a,b]$ with two absorbing boundaries, for a run-and-tumble particle with exponentially-distributed tumble durations in the presence of a constant drift. The Laplace transform of these equations yields a system of ordinary differential equations, whose Taylor expansion induces evolution equations for the exit probabilities and mean first-passage times through $a$ (given the initial value of the internal velocity state). These systems of equations are linear. Diagonalizing the matrix of the system allows to solve the equations and to work out the large-$b$ limit. This limit yields generalizations of existing results for the RTP on a half-line with instantaneous tumble.\\

  If the drift is negative, the particle exits the system almost surely. Moreover, the mean first-passage time through $a$ is the sum of an affine 
   function (whose slope $\mu^{-1}$ for any value of the mean tumble duration $\varphi^{-1}$), and corrections that decrease exponentially with the distance to left-end $a$ of the system. 
    If the drift is positive, the exit probabilities decrease exponentially 
     with the distance to the left-end $a$ of the system, at a rate given by the negative eigenvalue of the matrix of the linear system. On the other hand, the mean first-passage time through $a$ (conditioned on the exit of the particle) is an affine function of the initial distance to $a$, but the slope depends on the rate $\varphi$.\\

 The most straightforward extension would be to evaluate the variance of the first-passage time by taking higher-order   
   terms in the Taylor expansion of the Laplace transform of the Fokker--Planck equations.
 Positive run-and-tumble times are  features of active particles modeled by run-and-tumble particles, which suggests natural extensions of existing works on the run-and-tumble particles. In particular, more general boundary conditions, such as hard or attractive boundaries \cite{angelani2015run,angelani2017confined,angelani2023one,bressloff2022encounter,bressloff2023encounter}, have been investigated analytically.
 A duality symmetry between the exit probability of the  RTP on a segment with absorbing walls and the stationary distribution of positions with hard walls has been identified in \cite{gueneau2024relating} in the presence of a force field. This is a special case of the Siegmund duality \cite{siegmund1976equivalence} (for reviews with a physical perspective, see \cite{gueneau2024siegmund,touzo2025exact,gueneau2025non}). It would be interesting to formulate this duality for an RTP with positive tumble duration.\\

 Moreover, non-equilibrium steady states have been computed exactly for RTPs subjected to stochastic resetting, both in dimension one  \cite{evans2018run,bressloff2025encounter} and in dimension two \cite{Santra2020}. Positive tumble times should also preserve stationary bound states of two RTPs \cite{le2021stationary}. Inclusion of spatially-varying potentials \cite{dhar2019run,nath2024survival,gueneau2024optimal,gueneau2024run,dutta2024harmonically,dutta2024inertial}, even in one dimension, would be technically more challenging. It is natural to expect that confining potential should lead to stationary states.\\

 %other boundary conditions

 %interactions, several particles (fertile site)

 %Higher dimension

 %Higher 

\begin{appendices}

\section{Eigenvalues and eigenvectors of the matrix in the evolution equation}\label{eigenvectors}
\subsection{Diagonalization of the matrix $A$}\label{diagonalization}
 Let $(\vec{u}_1, \vec{u}_2, \vec{u}_3)$ denote the canonical basis of $\mathbbm{R}^3$. By inspection of the first two columns of the matrix $A$ defined on Eq. (\ref{ODEExit}), the vector $\vec{u}_1 + \vec{u}_2$ is an eigenvector of $A$ associated with the eigenvalue $0$.\\
Let us parametrize the entries of $A$ as follows: 
\begin{equation}
\begin{split}
\alpha:=& \frac{1}{2(1 + \mu)},\qquad \beta:= \frac{1}{2(1- \mu )},\qquad \sigma := \frac{\varphi}{\mu},\\
\frac{\mu}{1 - \mu^2} =& -(\alpha -\beta),\qquad \frac{1}{1 - \mu^2} = \alpha +\beta,\\
A =& \begin{pmatrix}
 \sigma & -\sigma & 0\\
  -(\alpha - \beta)&    \alpha - \beta & \alpha +\beta\\
   - (\alpha + \beta) &  \alpha + \beta & \alpha - \beta
\end{pmatrix}.
\end{split}
\end{equation} 

The eigenvalues of $A$ are the roots of the polynomial $P_A(\lambda)$, which is obtained expanding the determinant w.r.t. the first row as 
\begin{equation}
\begin{split}
 P_A(\lambda) =& \det( A- \lambda I_3)\\
 %=& -(\alpha - \beta)[(\sigma - \lambda)(\alpha - \beta) - \sigma(\alpha - \beta) ] + (\alpha + \beta - \lambda)[(\sigma - \lambda)(\alpha + \beta - \lambda) -\sigma(\alpha - \beta)   ]\\
 %=& -\lambda[ \lambda^2  - (\sigma + 2(\alpha + \beta))\lambda - (\alpha - \beta)^2 + (\alpha + \beta)^2 + \sigma(\alpha + \beta)]\\
 =& -\lambda^3+ [2(\alpha - \beta) + \sigma]\lambda^2 + [4 \alpha\beta - \sigma(\alpha - \beta )]\lambda\\
 =& -\lambda( \lambda - \lambda_+)(\lambda - \lambda_-).   
\end{split}
\end{equation}
The following discriminant $\chi$ is needed:
\begin{equation}
\chi := [ 2(\alpha - \beta) + \sigma]^2 - 4( \sigma(\alpha -\beta) -4\alpha\beta) = 4(\alpha - \beta)^2 + \sigma^2 + 16\alpha\beta = 4(\alpha + \beta)^2 + \sigma^2. 
\end{equation}
This discriminant is positive, hence the roots $\lambda_+$ and $\lambda_-$ are real, and expressed as
\begin{equation}
\lambda_\pm := \frac{\sigma}{2} + (\alpha - \beta) \pm \frac{1}{2}\sqrt{ \sigma^2 + 4 (\alpha + \beta)^2 }. 
\end{equation}
Re-expressing the eigenvalues in terms of the original parameters $\varphi$ and $\mu$ yields
\begin{equation}\label{lambdapmExpr}
\begin{split}
\lambda_\pm =& \frac{\varphi(1- \mu ^2)\, \pm\sqrt{\varphi ^2(1 - \mu^2 )^2+4\,\mu ^2}-2\,\mu ^2}{2\,\mu\left(1 -\mu^2\right)}.\\
%=& \frac{\varphi(1- \mu ^2)\left[ 1 \pm \sqrt{ 1 + \frac{4\mu^2}{(1-\mu^2)^2 \varphi^2}} \right]\,-2\,\mu ^2}{2\,\mu \left(1 -\mu^2\right)}.
\end{split}
\end{equation}

%\begin{equation}
%    \lambda_+ = \frac{\varphi -\mu ^2\,\varphi +A-2\,\mu ^2}{2\,\left(\mu -\mu ^3\right)}.
%\end{equation}

{\bf{Remark (signs of the eigenvalues).}} The product of the eigenvalues $\lambda_+$ and $\lambda_-$ reads
\begin{equation}
\lambda_- \lambda_+ = \sigma(\alpha - \beta) - 4\alpha \beta = -\frac{\varphi}{\mu}\frac{\mu}{1-\mu^2} - 4\frac{1}{4(1-\mu^2)} = - \frac{\varphi + 1}{1-\mu^2}.
\end{equation}
 The eigenvalues therefore have opposite signs. Moreover, the denominator in their expressions has the sign of $\mu$, and $\lambda_+$ has a larger numerator than $\lambda_-$. Hence,
\begin{equation}\label{lambdaPMSigns}
\begin{split}
{\mathrm{if}}\quad \mu<0, \quad  \lambda_+<0<\lambda_-,\\
{\mathrm{if}}\quad \mu>0, \quad\lambda_-<0<\lambda_+.
\end{split}
\end{equation}
 
%2. In the limit of short tumble events, $\varphi \gg 1$, $\lambda_+$ becomes large and its sign is the opposite of the sign of $\mu$, and $\lambda_-$ goes to a finite limit whose sign is the opposite of the sign of $\mu$ (see Eq. \eqref{limsLambda}).\\

 The vector $(1,1,0)^T$ is associated with the eigenvalue 0. Let us look for an eigenvector associated with the eigenvalue $\lambda_\pm$ in the form $(u_\pm, v_\pm, 1)^T$.\\
This implies the linear system
 \begin{align}
 \sigma u_\pm - \sigma v_\pm = \lambda_\pm u_\pm,\\
 - 2( \alpha + \beta ) u_\pm + 2 ( \alpha +\beta)v_\pm + 2( \alpha - \beta) = \lambda_\pm,
 \end{align}
 whose solution reads
 \begin{equation}
\begin{split}
u_\pm =&  \frac{\sigma}{\alpha + \beta}\left( \frac{\alpha - \beta}{\lambda_\pm} - \frac{1}{2}\right)= -\frac{\mu \sigma}{\lambda_\pm} - \sigma( 1 - \mu^2)\\
v_\pm =& \frac{\sigma - \lambda_\pm}{\alpha + \beta} \left(  \frac{\alpha - \beta}{\lambda_\pm} - \frac{1}{2}  \right) = ( \sigma - \lambda_\pm )
\left( -\frac{\mu}{\lambda_\pm} + \mu^2 - 1\right).
\end{split}
 \end{equation}
Substituting the expressions of the eigenvalues and the expression of the parameters $\alpha$ and $\beta$ yields the following expression of the components of the eigenvector in terms of $\mu$ and $\varphi$:
\begin{equation}\label{uv+-}
\begin{split}
u_+ =& -\frac{\varphi \,\left(2 + \varphi( 1 -\mu ^2) +\Delta\right)}{2\,\mu \,\left(\varphi +1\right)}\\
%=& -\frac{\varphi( 1 - \mu^2) \left[ 1 +\sqrt{1 + \frac{4\mu^2}{\varphi^2(1-\mu^2)^2}}\right]+2}{2\,\mu \,\left( 1+\frac{1}{\varphi}\right)},\\
u_- =& -\frac{\varphi \,\left(2 + \varphi( 1 - -\mu ^2)-\Delta \right)}{2\,\mu \,\left(\varphi +1\right)}\\
%=& -\frac{\varphi( 1 - \mu^2) \left[ 1 -\sqrt{1 + \frac{4\mu^2}{\varphi^2(1-\mu^2)^2}}\right]+2}{2\,\mu \,\left(1 +\frac{1}{\varphi}\right)},\\
v_+ =& -\frac{\varphi +\mu ^2\,\varphi -\Delta}{2\,\mu \,\left(\varphi +1\right)}\\
%    =& -\frac{ 1 + \mu ^2  - (1-\mu^2)\sqrt{1 + \frac{4\mu^2}{\varphi^2( 1-\mu^2)^2}}}{2\,\mu \,\left(1 + \frac{1}{\varphi }\right)},\\
v_- =&-\frac{\varphi +\mu ^2\,\varphi +\Delta}{2\,\mu \,\left(\varphi +1\right)},\\
%=&-  \frac{ 1 + \mu ^2  + (1-\mu^2)\sqrt{1 + \frac{4\mu^2}{\varphi^2( 1-\mu^2)^2}}}{2\,\mu \,\left(1 + \frac{1}{\varphi }\right)}.\\
\text{where}\quad 
\Delta=&\sqrt{\varphi^2 (1-\mu^2)^2 + 4\mu^2}.
\end{split}
\end{equation}

Let us denote by $V$ the matrix whose column vectors are the eigenvectors $(1,1,0)^T$, $(u_+,v_+, 1)^T$ and $(u_-, v_-, 1)^T$:\\
\begin{equation}
V:= \begin{pmatrix}
1 & u_+ & u_-\\
1 & v_+ & v_-\\
0 & 1 &  1  
\end{pmatrix}.
\end{equation}
 To sum up, we have established that
 \begin{equation}\label{AVLambdaV}
AV = 
\begin{pmatrix}
0 & 0 & 0\\
0 & \lambda_+ & 0\\
0 & 0 & \lambda_-
\end{pmatrix}
V.
 \end{equation}

The inverse of $V$ is obtained from the comatrix of $V$ as
\begin{equation}
\begin{split}
V^{-1} =& \frac{1}{\det(V)}
\begin{pmatrix}
v_+ - v_- & -1 & 1 \\
-u_+ + u_- & 1 & -1 \\
 u_+v_- - v_+ u_- & - v_- + u- & v_+ - u_+
\end{pmatrix}^T\\
=& \frac{1}{v_+ - v_- - u_+ + u_-}
\begin{pmatrix}
v_+ - v_- & -u_+ + u_-  & u_+v_- - v_+ u_- \\
-1 & 1 & -v_- + u_- \\
 1 & -1 & v_+ - u_+
\end{pmatrix}.
\end{split}
\end{equation}

%\iffalse{\textcolor{red}{How are the following 4 terms found? The entries of the presentation matrix $V$ satisfy}}

%%%%%%%%%%%%%%%%%

%%%%%%%%%%%%%%%%%%%%%%

 The following combinations repeatedly occur in calculations:
 \begin{equation}
u_+ - v_+ = -\frac{1}{2\mu}[\varphi( 1 - \mu^2)   + \Delta],
\end{equation}
 \begin{equation}
u_- - v_- = -\frac{1}{2\mu}[\varphi( 1 - \mu^2)   - \Delta],
\end{equation}
\begin{equation}
u_- - v_- - v_- + v_+ =  \frac{1}{\mu}   \Delta,
\end{equation}
\begin{equation}
u_-v_+ - u_+ v_- = - \frac{\varphi }{\mu^2( \varphi + 1)}  \Delta,
\end{equation}
\begin{equation}
\frac{u_- - u_+ - u_- v_+ + u_+ v_-}{u_- - u_+ - v_- + v_+} = \frac{\varphi(\mu + 1)}{\mu(\varphi + 1)},
\end{equation}\
\begin{equation}\label{coeffxUniv}
\frac{v_- - v_+ + \mu^2 (u_- -  u_+ - v_- + v_+) + \mu (u_- v_+ - u_+ v_-)}{( \mu^2 - 1) (u_- - u_+ - v_- + v_+ )} = 1,
\end{equation}
\begin{equation}\label{coeffMinus}
\frac{\mu v_+ - \mu u_+ + 1}{\lambda_- \mu ( \mu^2 - 1) (u_- - u_+ - v_- + v_+ )} =
\frac{\Delta + \varphi( 1-\mu^2)}{2\mu \Delta},  
\end{equation}
\begin{equation}\label{coeffPlus}
\frac{\mu v_- - \mu u_- + 1}{\lambda_+ \mu ( \mu^2 - 1) (u_- - u_+ - v_- + v_+ )} =  \frac{-\Delta + \varphi( 1-\mu^2)}{2\mu \Delta},
\end{equation}
\begin{equation}\label{nextInCommon}
\frac{u_- - u_+ - u_- v_+ + u_+ v_-}{u_+ - v_+ + 1} =
\frac{\Delta[(2\,\mu - \varphi(1-\mu^2))+\Delta] }{2\,\mu ^2\,\left(-1+\mu \right)\,\left(1+\varphi \right)},
\end{equation}
\begin{equation}
\frac{ u_- - v_- + 1}{u_- - u_+ - v_- + v_+ } = \frac{ \Delta^2 +[ 2\mu - \varphi( 1-\mu^2 ) ]\Delta }{2\Delta^2},
\end{equation}
where 
\[\Delta=\sqrt{\varphi^2 (1-\mu^2)^2 + 4\mu^2}.\]

\iffalse \begin{equation}
\begin{split}
u_+ - v_+ =& -\frac{1}{2\mu}[\varphi( 1 - \mu^2)   + \sqrt{\varphi^2 (1-\mu^2)^2 + 4\mu^2}],\\
u_- - v_- =& -\frac{1}{2\mu}[\varphi( 1 - \mu^2)   - \sqrt{\varphi^2 (1-\mu^2)^2 + 4\mu^2}],\\
u_- - v_- - v_- + v_+ =&  \frac{1}{\mu}   \sqrt{\varphi^2 (1-\mu^2)^2 + 4\mu^2},\\
u_-v_+ - u_+ v_- =& - \frac{\varphi }{\mu^2( \varphi + 1)}  \sqrt{\varphi^2 (1-\mu^2)^2 + 4\mu^2},\\
\frac{u_- - u_+ - u_- v_+ + u_+ v_-}{u_- - u_+ - v_- + v_+} =& \frac{\varphi(\mu + 1)}{\mu(\varphi + 1)},\\
\frac{v_- - v_+ + \mu^2 (u_- -  u_+ - v_- + v_+) + \mu (u_- v_+ - u_+ v_-)}{\mu( \mu^2 - 1) (u_- - u_+ - v_- + v_+ )} =& \mu^2 - 1,\\
\frac{\mu v_+ - \mu u_+ + 1}{\lambda_- \mu ( \mu^2 - 1) (u_- - u_+ - v_- + v_+ )} =&  
\frac{\varphi^2 (1-\mu^2)^2 + 4\mu^2 + \varphi( 1-\mu^2) \sqrt{ \varphi^2 (1-\mu^2)^2 + 4\mu^2}}{2\mu[    \varphi^2 (1-\mu^2)^2 + 4\mu^2]},\\  
\frac{\mu v_- - \mu u_- + 1}{\lambda_+ \mu ( \mu^2 - 1) (u_- - u_+ - v_- + v_+ )} =&  \frac{-[\varphi^2 (1-\mu^2)^2 + 4\mu^2] + \varphi( 1-\mu^2) \sqrt{ \varphi^2 (1-\mu^2)^2 + 4\mu^2}}{2\mu[    \varphi^2 (1-\mu^2)^2 + 4\mu^2]},\\  
\frac{ u_- - v_- + 1}{u_- - u_+ - v_- + v_+ } =& \frac{ \varphi^2 (1-\mu^2)^2 + 4\mu^2 +[ 2\mu - \varphi( 1-\mu^2 ) ]\sqrt{ \varphi^2 (1-\mu^2)^2 + 4\mu^2} }{2[\varphi^2 (1-\mu^2)^2 + 4\mu^2]}
\end{split}
\end{equation}\fi

\subsection{The matrix exponential $Q(x)= \exp( xA )$}
 In our notations, $A = V \Lambda V^{-1}$, where $\Lambda$ is the diagonal matrix 
 \begin{equation}
\Lambda = \begin{pmatrix}
 \lambda_1 & 0 & 0\\
    0 & \lambda_+ & 0\\
    0 & 0 & \lambda_-
\end{pmatrix}.
 \end{equation}
 %\textcolor{red}{I suppose this $\Delta$ is not the discriminant (right, fixed)}. 
 In  the solution of the evolution equations, we need the following matrix exponential 
   the matrix $Q$ defined by
\begin{equation}\label{defQ}
  \exp( xA) = \exp( x(V \Lambda V^{-1})) =  V \exp( x \Lambda) V^{-1}.
\end{equation}
% The three scalar components of Eq. (\ref{solMatM}) read
%\begin{equation}\label{solCompM}
%M_i( x ) = \sum_{j} Q_{ij}(x) M_j(a) + \int_a^x \sum_j\sum_k Q_{ik}( x- y) 
%(R\mathcal{C})_{kj} e^{\lambda_j y} dy,\qquad (1\leq i \leq 3),
%\end{equation}
% with the notation
With the notations
\begin{equation}\label{lambdaNot}
\lambda_1 := 0,\qquad \lambda_2 := \lambda_+, \qquad \lambda_3:= \lambda_-. 
\end{equation}
 \begin{equation}
 \begin{split}
 Q_{ik}( x ) =& \sum_l V_{il} e^{\lambda_l x} (V^{-1})_{lk} = \sum_l q^{(l)}_{ik} e^{\lambda_l x},\\
{\mathrm{with}}\;\;\;\;q^{(l)}_{ik} =& V_{il}  (V^{-1})_{lk}.
\end{split}
\end{equation}
 The matrices $q^{(1)}$, $q^{(2)}$, $q^{(3)}$ are readily evaluated in terms of the components of the entries of the presentation matrix $V$ of eigenvectors:
\begin{equation}\label{qExpr}
\begin{split}
q^{(1)} = & 
\frac{1}{\det( V )}
\begin{pmatrix}
 v_+ - v_- & u_- - u_+ & u_+ v_- - u_- v_+\\ 
 v_+ - v_-  &  u_- - u_+ & u_+ v_- - u_- v_+\\ 
 0 & 0 & 0 
\end{pmatrix},\\ 
\qquad q^{(2)} =&
\frac{1}{\det( V )}
\begin{pmatrix}
 -u_+ & u_+ & u_+(u_- - v_- )\\ 
 -v_+ & v_+ & v_+(u_- - v_- )\\ 
 -1 & 1 &  u_- - v_- 
\end{pmatrix},\\ 
\qquad q^{(3)} =& 
\frac{1}{\det( V )}
\begin{pmatrix}
 u_- & -u_- & -u_-(u_+ - v_+ )\\ 
 v_- & -v_- & -v_-(u_+ - v_+ )\\ 
 1 & -1 & -( u_+ - v_+) 
\end{pmatrix},\\
{\mathrm{with}}\qquad &\det( V ) = u_- - u_+ - v_- + v_+.
\end{split}
\end{equation}

{\bf{Remark.}} Calculating matrix products, it is easy to check that the vector $(u_-,v_-,1)^T$ is an eigenvector of the matrices $q^{(1)}$, $q^{(2)}$ and $q^{(3)}$ (associated with the eigenvalues $0$, $0$ and $1$ respectively):\\
\begin{equation}\label{idAlgoP}
    q^{(1)} \begin{pmatrix}
      u_-\\
      v_-\\
       1\\
    \end{pmatrix} = \vec{0},\qquad
    \qquad q^{(2)} 
    \begin{pmatrix}
     u_-\\
    v_-\\
    1\\
    \end{pmatrix}
    = \vec{0},\qquad
    \qquad q^{(3)} \begin{pmatrix}
     u_- \\
     v_-\\
     1\\
    \end{pmatrix} = 
    \begin{pmatrix}
    u_-\\
    v_-\\
    1\\
    \end{pmatrix}.
\end{equation}
 These identities are useful in evaluating the exit probability in the presence of a positive drift (see Eq. (\ref{ExSymbPositiveDrift})).

\subsection{Limit of instantaneous tumble}\label{appShortTumble}

 In the limit of short tumble events, $\varphi \gg 1$, and Taylor expansion yields 
\begin{equation}\label{limsLambda}
\begin{split}
 \lambda_+ &\underset{\varphi\gg 1}{\sim} \frac{\varphi}{\mu},\\
 \lambda_- &\underset{\varphi\gg 1}{\sim} -\frac{\mu}{1-\mu^2}.
\end{split}
\end{equation}

 From the expression of the eigenvectors in Eq. (\ref{uv+-}), we obtain the following equivalents in the limit of instantaneous tumble events:
\begin{equation}\label{equivEntriesInf}
\begin{split}
u_+ \underset{\varphi\to\infty}{\sim}& -\frac{1-\mu^2}{\mu} \varphi,\\
u_- \underset{\varphi\to\infty}{\sim}& - \frac{1}{\mu},\\
v_+ \underset{\varphi\to\infty}{\sim}& \mu,\\
v_- \underset{\varphi\to\infty}{\sim}& -\frac{1}{\mu}.\\
\end{split}
\end{equation}
These equivalents induce the following limits for the matrices $q^{(1)},q^{(2)},q^{(3)}$:

\begin{equation}\label{qLimExpr}
q^{(1)} \underset{\varphi\gg 1}{\simeq} 
\qinst^{(1)}:=\begin{pmatrix}
 0 &1 & \frac{1}{\mu}\\ 
 0 &1 & \frac{1}{\mu}\\ 
 0 & 0 & 0 
\end{pmatrix}, 
\qquad q^{(2)} \underset{\varphi\gg 1}{\simeq} \qinst^{(2)}:=
\begin{pmatrix}
 1 & - 1 & 0\\ 
 0 &0 & 0\\ 
 0 & 0 & 0 
\end{pmatrix}, 
\qquad q^{(3)} \underset{\varphi\gg 1}{\simeq} \qinst^{(3)}:=
\begin{pmatrix}
 0 &0 & - \frac{1}{\mu}\\ 
 0 &0 & - \frac{1}{\mu}\\ 
 0 & 0 & 1 
\end{pmatrix}.    
\end{equation}

The eigenvalues satisfy the following:
\begin{equation}\label{lambdaLim}
\begin{split}
\lambda_+ \underset{\varphi\to +\infty}{\sim}&  \frac{\varphi}{\mu}, \\
\lambda_- \underset{\varphi\to +\infty}{\sim}& -\frac{\mu}{1 - \mu^2}.\\
\end{split}
\end{equation}
In particular:\\
$\bullet$ {\bf{If $\mu <0$,}}
\begin{equation}
\begin{split}
\lambda_+ \underset{\varphi\to +\infty}{\longrightarrow}&  -\infty. \\
%\lambda_- \underset{\varphi\to 0}{\sim}& \frac{1}{1 + \mu}.\\
\end{split}
\end{equation}
$\bullet$ {\bf{If $\mu >0$,}}
\begin{equation}
\begin{split}
\lambda_+ \underset{\varphi\to +\infty}{\longrightarrow}&  +\infty. \\
%\lambda_- \underset{\varphi\to 0}{\sim}& -\frac{1}{1-\mu}.\\
\end{split}
\end{equation}

%\subsubsection{Application to ${\mathcal{E}}_a(a,+;[a,b],\varphi)$ for a negative drift}

\section{The matrices $W^{(i,j)}$, ($1\leq i,j \leq 3$)}\label{AppW}
 Based on their definition in Eq. (\ref{defW}), the matrices $W^{(i,j)}$ have the following expressions, for $i,j$ in $[1..3]$:
{\tiny{
\begin{equation}\label{WExpl}
\begin{split}
W^{(1,1)} =& \frac{v_- -v_+ +\mu ^2\,u_--\mu ^2\,u_+-\mu ^2\,v_-+\mu ^2\,v_++\mu \,u_-\,v_+-\mu \,u_+\,v_-}{\mu(\mu^2 - 1)(u_- - u_+ - v_- + v_+)^2}\times
\begin{pmatrix} 
v_--v_+ & -(u_--u_+) & u_-\,v_+ -u_+\,v_-\\
v_--v_+ & - (u_--u_+ )& u_-\,v_+-u_+\,v_-\\
0 & 0 & 0 \end{pmatrix},\\
W^{(1,2)} =&
\frac{\mu u_+ - \mu u_- + u_+ v_- - u_+ v_+ + \mu u_- v_+^2 - \mu u_+ v_- v_+  }{\mu (\mu^2 - 1) (u_- - u_+ - v_- + v_+ )^2}
\times
\begin{pmatrix}
1 & -1 &  v_- - u_-\\
1 & -1 & v_- - u_-\\
0 &  0 &  0
\end{pmatrix},\\
W^{(1,3)} =& \frac{\mu u_- - \mu u_+ - u_- v_- + u_- v_+ + \mu u_+ v_-^2 - \mu u_- v_- v_+  }{\mu (\mu^2 - 1) (u_- - u_+ - v_- + v_+ )^2}
  \times \begin{pmatrix}
1 & -1 &  v_+ - u_+\\
1 & -1 & v_+ - u_+\\
0 &  0 &  0
\end{pmatrix},\\
W^{(2,1)} =& \frac{\mu v_- - \mu u_- + 1}{\mu (\mu^2 - 1) (u_- - u_+ - v_- + v_+ )^2}\times 
\begin{pmatrix}
u_+( v_- - v_+ ) & -u_+ ( u_- - u_+ ) & u_+ (u_- v_+ - u_+ v_-)\\
v_+( v_- - v_+ )  & -v_+ ( u_- - u_+ ) & v_+ (u_- v_+ - u_+ v_-)\\
 v_- - v_+   &   -( u_- - u_+ ) &  u_- v_+ - u_+ v_-
\end{pmatrix},\\
W^{(2,2)} =& \frac{ \mu - u_+ - \mu^2 u_- + \mu^2 u_+ + \mu^2 v_- - \mu^2 v_+ + \mu u_- v_+ - \mu v_- v_+ }{\mu (\mu^2 - 1) (u_- - u_+ - v_- + v_+ )^2}\times \begin{pmatrix}
u_+( v_- - v_+ ) & -u_+ ( u_- - u_+ ) & u_+ (u_- v_+ - u_+ v_-)\\
v_+( v_- - v_+ )  & -v_+ ( u_- - u_+ ) & v_+ (u_- v_+ - u_+ v_-)\\
 v_- - v_+   &  -( u_- - u_+ ) &  u_- v_+ - u_+ v_-
\end{pmatrix},\\
W^{(2,3)} =& \frac{ - \mu v_-^2 + \mu u_- v_- + \mu - u_-}{\mu (\mu^2 - 1) (u_- - u_+ - v_- + v_+ )^2}\times 
\begin{pmatrix}
u_+ & -u_+ & - u_+ (u_+ - v_+)\\
v_+  & -v_+& - v_+ (u_+ - v_+)\\
 1  &  -1&  - ( u_+ - v_+ ) \\
\end{pmatrix},\\
W^{(3,1)} =& \frac{\mu v_+ - \mu u_+ + 1 }{\mu (\mu^2 - 1) (u_- - u_+ - v_- + v_+ )^2}\times 
\begin{pmatrix}
- u_-( v_- - v_+ ) & u_- ( u_- - u_+ ) & -u_- (u_- v_+ - u_+ v_-)\\
- v_-( v_- - v_+ )  & v_- ( u_- - u_+ ) & -v_- (u_- v_+ - u_+ v_-)\\
 - ( v_- - v_+ )  &    u_- - u_+  &  -( u_- v_+ - u_+ v_-)
\end{pmatrix},\\
W^{(3,2)} =& \frac{ - \mu v_+^2 + \mu u_+ v_+ + \mu - u_+}{\mu (\mu^2 - 1) (u_- - u_+ - v_- + v_+ )^2}\times 
\begin{pmatrix}
u_- & -u_- & - u_- (u_- - v_-)\\
v_-  & -v_- & - v_- (u_- - v_-)\\
 1  &  -1&  - ( u_- - v_- ) \\
\end{pmatrix},\\
W^{(3,3)} =&  \frac{  \mu - u_- + \mu^2 u_- - \mu^2 u_+ - \mu^2 v_- + \mu^2 v_+ + \mu u_+ v_- - \mu v_- v_+}{\mu (\mu^2 - 1) (u_- - u_+ - v_- + v_+ )^2}\times 
\begin{pmatrix}
-u_- & u_- & u_- (u_+ - v_+)\\
-v_-  & v_- &  v_- (u_+ - v_+)\\
 -1  &  1&    u_+ - v_+ \\
\end{pmatrix}
\end{split}
\end{equation}
}}

\section{Asymptotic expansions}
\subsection{The vector of exit probabilities with a positive drift in the large-$\varphi$ limit}
The numerators and denominators in the expression of the coefficients $Z_a(a)$ and ${\mathcal{E}}_a(a,+;[a,b],\varphi)$ have the following asymptotic expansion in the limit of large $\varphi$, where $\lambda_+$ goes to infinity and $\lambda_-$ goes to $-\mu/(1-\mu^2)$:

\begin{equation}\label{denomCom}
\begin{split}
Q_{12} Q_{21}  - Q_{11}Q_{22} - Q_{11} Q_{23} +& Q_{13}Q_{21} - Q_{11} Q_{32} +  Q_{12} Q_{31} - Q_{11} Q_{33} + Q_{13}Q_{31}\\
= e^{\lambda_+(b-a)}&\left[  (\qinst^{(1)}_{12} + \qinst_{12}^{(3)} e^{-\frac{\mu(b-a)}{1-\mu^2}}) \qinst^{(2)}_{21} + \qinst^{(2)}_{12}(\qinst^{(1)}_{21} + \qinst_{21}^{(3)} e^{-\frac{\mu(b-a)}{1-\mu^2}} )     \right.\\
&-(\qinst^{(1)}_{11} + \qinst_{11}^{(3)} e^{-\frac{\mu(b-a)}{1-\mu^2}}) \qinst^{(2)}_{22} - \qinst^{(2)}_{11}(\qinst^{(1)}_{22} + \qinst_{22}^{(3)} e^{-\frac{\mu(b-a)}{1-\mu^2}})\\
&-(\qinst^{(1)}_{11} + \qinst_{11}^{(3)} e^{-\frac{\mu(b-a)}{1-\mu^2}}) \qinst^{(2)}_{23} - \qinst^{(2)}_{11}(\qinst^{(1)}_{23} + \qinst_{23}^{(3)} e^{-\frac{\mu(b-a)}{1-\mu^2}})\\
&+ (\qinst^{(1)}_{13} + \qinst_{13}^{(3)} e^{-\frac{\mu(b-a)}{1-\mu^2}}) \qinst^{(2)}_{21} + \qinst^{(2)}_{13}(\qinst^{(1)}_{21} + \qinst_{21}^{(3)} e^{-\frac{\mu(b-a)}{1-\mu^2}} )\\ 
&-(\qinst^{(1)}_{11} + \qinst_{11}^{(3)} e^{-\frac{\mu(b-a)}{1-\mu^2}}) \qinst^{(2)}_{32} - \qinst^{(2)}_{11}(\qinst^{(1)}_{32} + \qinst_{32}^{(3)} e^{-\frac{\mu(b-a)}{1-\mu^2}}\\
&+(\qinst^{(1)}_{12} + \qinst_{12}^{(3)} e^{-\frac{\mu(b-a)}{1-\mu^2}}) \qinst^{(2)}_{31} + \qinst^{(2)}_{12}(\qinst^{(1)}_{31} + \qinst_{31}^{(3)} e^{-\frac{\mu(b-a)}{1-\mu^2}} )\\    
&-(\qinst^{(1)}_{11} + \qinst_{11}^{(3)} e^{-\frac{\mu(b-a)}{1-\mu^2}}) \qinst^{(2)}_{33} - \qinst^{(2)}_{11}(\qinst^{(1)}_{33} + \qinst_{33}^{(3)} e^{-\frac{\mu(b-a)}{1-\mu^2}})\\
&+\left.  (\qinst^{(1)}_{13} + \qinst_{13}^{(3)} e^{-\frac{\mu(b-a)}{1-\mu^2}}) \qinst^{(2)}_{31} + \qinst^{(2)}_{13}(\qinst^{(1)}_{31} + \qinst_{31}^{(3)} e^{-\frac{\mu(b-a)}{1-\mu^2}} ) \right] +  o(e^{\lambda_+(b-a)})\\
= e^{\lambda_+(b-a)}&\left[  (1 + 0 e^{-\frac{\mu(b-a)}{1-\mu^2}}) \times 0  - (0 + 0 e^{-\frac{\mu(b-a)}{1-\mu^2}} )     \right.\\
&-(0 + 0 e^{-\frac{\mu(b-a)}{1-\mu^2}}) \times 0 - (1 + 0 e^{-\frac{\mu(b-a)}{1-\mu^2}})\\
&-(0 + 0 e^{-\frac{\mu(b-a)}{1-\mu^2}}) \times 0 - (\mu^{-1} - \mu^{-1} e^{-\frac{\mu(b-a)}{1-\mu^2}})\\
&+ (\mu^{-1} - \mu^{-1} e^{-\frac{\mu(b-a)}{1-\mu^2}}) \times 0 + 0\times (0 + 0 e^{-\frac{\mu(b-a)}{1-\mu^2}} )\\ 
&-(0 + 0 e^{-\frac{\mu(b-a)}{1-\mu^2}}) \times 0 - (0 + 0 e^{-\frac{\mu(b-a)}{1-\mu^2}})\\
&+(1 + 0 e^{-\frac{\mu(b-a)}{1-\mu^2}}) \times 0 - (0 + 0 e^{-\frac{\mu(b-a)}{1-\mu^2}} )\\    
&-(0 + 0 e^{-\frac{\mu(b-a)}{1-\mu^2}}) \times 0 - (0 +  e^{-\frac{\mu(b-a)}{1-\mu^2}})\\
&+\left.  (\mu^{-1} - \mu^{-1} e^{-\frac{\mu(b-a)}{1-\mu^2}}) \times 0 + 0\times (0 + 0 e^{-\frac{\mu(b-a)}{1-\mu^2}} ) \right] +  o(e^{\lambda_+(b-a)})\\
=& e^{\lambda_+(b-a)} \left[ -1 - \frac{1}{\mu} + \left(   -1 + \frac{1}{\mu} \right) e^{-\frac{\mu(b-a)}{1-\mu^2}} \right] + o( e^{\lambda_+(b-a)}).
\end{split}
\end{equation}

\begin{equation}\label{numer1}
\begin{split}
&( Q_{22} + Q_{23} + Q_{32} + Q_{33})( Q_{12} - Q_{13} ) -( Q_{12} + Q_{13}) (Q_{22} - Q_{23}+Q_{32} -Q_{33})\\
&= e^{\lambda_+(b-a)}\left[  [ \qinst^{(1)}_{22} + \qinst^{(1)}_{23} + \qinst^{(1)}_{32} + \qinst^{(1)}_{33} +  (\qinst^{(3)}_{22} + \qinst^{(3)}_{23} + \qinst^{(3)}_{32} + \qinst^{(3)}_{33} )e^{-\frac{\mu(b-a)}{1-\mu^2}}  ](  \qinst^{(2)}_{12} - \qinst^{(2)}_{13} )   \right.\\
& +( \qinst^{(2)}_{22} + \qinst^{(2)}_{23} + \qinst^{(2)}_{32} + \qinst^{(2)}_{33})
[ \qinst^{(1)}_{12} - \qinst^{(1)}_{13} + (\qinst^{(1)}_{12} - \qinst^{(1)}_{13}) e^{-\frac{\mu(b-a)}{1-\mu^2}}  ]\\ 
&-[\qinst^{(1)}_{12} + \qinst^{(1)}_{13} + ( \qinst^{(3)}_{12} + \qinst^{(3)}_{13}) e^{-\frac{\mu(b-a)}{1-\mu^2}} ] (\qinst^{(2)}_{22} - \qinst^{(2)}_{23}+\qinst^{(2)}_{32} -\qinst^{(2)}_{33})\\ 
&\left. -( \qinst^{(2)}_{12} + \qinst^{(2)}_{13}) [\qinst^{(1)}_{22} - \qinst^{(1)}_{23}+\qinst^{(1)}_{32} -\qinst^{(1)}_{33} + ( \qinst^{(3)}_{22} - \qinst^{(3)}_{23}+\qinst^{(3)}_{32} -\qinst^{(3)}_{33})  e^{-\frac{\mu(b-a)}{1-\mu^2}}  ]  \right] +  o(e^{\lambda_+(b-a)})\\
&= e^{\lambda_+(b-a)}\left[  [ \qinst^{(1)}_{22} + \qinst^{(1)}_{23} + \qinst^{(1)}_{32} + \qinst^{(1)}_{33} +  (\qinst^{(3)}_{22} + \qinst^{(3)}_{23} + \qinst^{(3)}_{32} + \qinst^{(3)}_{33} )e^{-\frac{\mu(b-a)}{1-\mu^2}}  ](  \qinst^{(2)}_{12} - \qinst^{(2)}_{13} )   \right.\\
& +( \qinst^{(2)}_{22} + \qinst^{(2)}_{23} + \qinst^{(2)}_{32} + \qinst^{(2)}_{33})
[ \qinst^{(1)}_{12} - \qinst^{(1)}_{13} + (\qinst^{(3)}_{12} - \qinst^{(3)}_{13}) e^{-\frac{\mu(b-a)}{1-\mu^2}}  ]\\ 
&-[\qinst^{(1)}_{12} + \qinst^{(1)}_{13} + ( \qinst^{(3)}_{12} + \qinst^{(3)}_{13}) e^{-\frac{\mu(b-a)}{1-\mu^2}} ] (\qinst^{(2)}_{22} - \qinst^{(2)}_{23}+\qinst^{(2)}_{32} -\qinst^{(2)}_{33})\\ 
&\left. -( \qinst^{(2)}_{12} + \qinst^{(2)}_{13}) [\qinst^{(1)}_{22} - \qinst^{(1)}_{23}+\qinst^{(1)}_{32} -\qinst^{(1)}_{33} + ( \qinst^{(3)}_{22} - \qinst^{(3)}_{23}+\qinst^{(3)}_{32} -\qinst^{(3)}_{33})  e^{-\frac{\mu(b-a)}{1-\mu^2}}  ]  \right] +  o(e^{\lambda_+(b-a)})\\
&= e^{\lambda_+(b-a)}\left[  [ 1+\mu^{-1} +  (1-\mu^{-1} )e^{-\frac{\mu(b-a)}{1-\mu^2}}  ]\times(-1)   \right.\\
& +0\times
[ \qinst^{(1)}_{12} - \qinst^{(1)}_{13} + (\qinst^{(3)}_{12} - \qinst^{(3)}_{13}) e^{-\frac{\mu(b-a)}{1-\mu^2}}  ]\\ 
&-[\qinst^{(1)}_{12} + \qinst^{(1)}_{13} + ( \qinst^{(3)}_{12} + \qinst^{(3)}_{13}) e^{-\frac{\mu(b-a)}{1-\mu^2}} ] \times 0\\ 
&\left. + 1\times [ 1- \mu^{-1}+ ( \mu^{-1} -1)  e^{-\frac{\mu(b-a)}{1-\mu^2}}  ]  \right] +  o(e^{\lambda_+(b-a)})\\
&= e^{\lambda_+(b-a)}[ -2\mu^{-1} + 2( \mu^{-1} - 1)   e^{-\frac{\mu(b-a)}{1-\mu^2}} ] + o( e^{\lambda_+(b-a)} ).
\end{split}
\end{equation}
 
\begin{equation}\label{numer2}
\begin{split}
&(-Q_{21} -Q_{31})( Q_{12} - Q_{13} ) + Q_{11} (Q_{22} - Q_{23}+Q_{32} -Q_{33})\\
=& e^{\lambda_+(b-a)}\left[  - [\qinst^{(1)}_{21} + \qinst^{(1)}_{31} + (\qinst^{(3)}_{21} + \qinst^{(3)}_{31}) e^{-\frac{\mu(b-a)}{1-\mu^2}}     ]( \qinst^{(2)}_{12} - \qinst^{(2)}_{13} ) \right.\\
&-( \qinst^{(2)}_{21} + \qinst^{(2)}_{31} )[ \qinst^{(1)}_{12} - \qinst^{(1)}_{13} + (\qinst^{(3)}_{12} - \qinst^{(3)}_{13}  )e^{-\frac{\mu(b-a)}{1-\mu^2}}   ]\\
&+ [ \qinst^{(1)}_{11} + \qinst^{(3)}_{11}e^{-\frac{\mu(b-a)}{1-\mu^2}} ](\qinst^{(2)}_{22} - \qinst^{(2)}_{23}+\qinst^{(2)}_{32} -\qinst^{(2)}_{33})\\
&+\left.  \qinst^{(2)}_{11} [ \qinst^{(1)}_{22} - \qinst^{(1)}_{23} + \qinst^{(1)}_{32} - \qinst^{(1)}_{33} +  (\qinst^{(3)}_{22} - \qinst^{(3)}_{23} + \qinst^{(3)}_{32} - \qinst^{(3)}_{33}) e^{-\frac{\mu(b-a)}{1-\mu^2}}  ]  \right] +  o(e^{\lambda_+(b-a)}\\
=& e^{\lambda_+(b-a)}\left[  - 0\times ( \qinst^{(2)}_{12} - \qinst^{(2)}_{13} ) \right.\\
&- 0\times [ \qinst^{(1)}_{12} - \qinst^{(1)}_{13} + (\qinst^{(3)}_{12} - \qinst^{(3)}_{13}  )e^{-\frac{\mu(b-a)}{1-\mu^2}}   ]\\
&+ [ \qinst^{(1)}_{11} + \qinst^{(3)}_{11}e^{-\frac{\mu(b-a)}{1-\mu^2}} ]\times 0\\
&+\left.  1\times [ 1-\mu^{-1} +  (\mu^{-1} - 1 )e^{-\frac{\mu(b-a)}{1-\mu^2}}  ]  \right] +  o(e^{\lambda_+(b-a)} ) \\
=& e^{\lambda_+(b-a)}(1-\mu^{-1})( 1 - e^{-\frac{\mu(b-a)}{1-\mu^2}}) + o( e^{\lambda_+(b-a)}) .
\end{split}
\end{equation}

\subsection{The vector $\vec{\mathcal{E}}(a)$ with a positive drift in the large-$b$ limit}
Consider a positive drift. The eigenvalue $\lambda_+$ is positive. Each of the entries of the matrix $A$  entering the expression of the determinant of the matrix $S_+$ defined in Eq. (\ref{defSPlus}) has terms of order $e^{\lambda_+(b-a)}$, $1$ and $e^{\lambda_-(b-a)}$ in the large-$b$ limit. Hence there exist coefficients $D_0$, $D_1$, $D_2$ such that
\begin{equation}\label{detSPlusAsymp}
\begin{split}
\det( S_+ ) =& Q_{12}Q_{21} - Q_{11}Q_{22} - Q_{11}Q_{23} + Q_{13}Q_{21} - Q_{11}Q_{32} + Q_{12}Q_{31} - Q_{11}Q_{33} + Q_{13}Q_{31} \\
=&  D_2 e^{2\lambda_+(b-a)} + D_1 e^{\lambda_+(b-a)} + D_0 + o(1).\\
\end{split}
\end{equation}
Calculating,
\begin{equation}
\begin{split}
D_2 =& q^{(2)}_{12} q^{(2)}_{21} - q^{(2)}_{11}q^{(2)}_{22} - q^{(2)}_{11}q^{(2)}_{23} + q^{(2)}_{13} q^{(2)}_{21} - q^{(2)}_{11}q^{(2)}_{32} + q^{(2)}_{12} q^{(2)}_{31} - q^{(2)}_{11} q^{(2)}_{33} + q^{(2)}_{13} q^{(2)}_{31}\\
=& 0,
\end{split}
\end{equation}

\begin{equation}\label{detSPlusEq}
\begin{split}
D_1 =& q^{(1)}_{12} q^{(2)}_{21} - q^{(1)}_{11}q^{(2)}_{22} - q^{(1)}_{11}q^{(2)}_{23} + q^{(1)}_{13} q^{(2)}_{21} - q^{(1)}_{11}q^{(2)}_{32} + q^{(1)}_{12} q^{(2)}_{31} - q^{(1)}_{11} q^{(2)}_{33} + q^{(1)}_{13} q^{(2)}_{31}\\
&+ q^{(2)}_{12} q^{(1)}_{21} - q^{(2)}_{11}q^{(1)}_{22} - q^{(2)}_{11}q^{(1)}_{23} + q^{(2)}_{13} q^{(1)}_{21} - q^{(2)}_{11}q^{(1)}_{32} + q^{(2)}_{12} q^{(1)}_{31} - q^{(2)}_{11} q^{(1)}_{33} + q^{(2)}_{13} q^{(1)}_{31}\\
=& \frac{(v_- - 1)(v_+ - u_+ + 1)}{u_- - u_+ - v_- + v_+}.
\end{split}
\end{equation}

\begin{equation}
\begin{split}
D_0 = q^{(1)}_{12} q^{(1)}_{21} - q^{(1)}_{11}q^{(1)}_{22} - q^{(1)}_{11}q^{(1)}_{23} + q^{(1)}_{13} q^{(1)}_{21} - q^{(1)}_{11}q^{(1)}_{32} + q^{(1)}_{12} q^{(1)}_{31} - q^{(1)}_{11} q^{(1)}_{33} + q^{(1)}_{13} q^{(1)}_{31}.
\end{split}
\end{equation}

In the numerator of the expression of $Z_a(a)$ and ${\mathcal{E}}_a(a,+;[a,b],\varphi)$ in Eq. (\ref{ZEEabxpPositiveDrift}), the terms of order $e^{2\lambda_+(b-a)}$ are zero. Moreover, the numerators in these expressions have the following asymptotic expansions in the large-$b$ limit:

%{\tiny{
\begin{equation}\label{NumEaEqPos}
\begin{split}
(-Q_{21} -Q_{31})&( Q_{12} - Q_{13} ) + Q_{11} (Q_{22} - Q_{23}+Q_{32} -Q_{33})\\
=& [(-q^{(1)}_{21} - q^{(1)}_{31} )   ( q^{(2)}_{12} - q^{(2)}_{13} ) + q^{(1)}_{11}   ( q^{(2)}_{22} - q^{(2)}_{23} + q^{(2)}_{32} - q^{(2)}_{33} )\\
          &+ (-q^{(2)}_{21} - q^{(2)}_{31} )   ( q^{(1)}_{12} - q^{(1)}_{13} ) + q^{(2)}_{11}   ( q^{(1)}_{22} - q^{(1)}_{23} + q^{(1)}_{32} - q^{(1)}_{33} )] e^{\lambda_+(b-a)}\\
          &+ (-q^{(1)}_{21} - q^{(1)}_{31} )   ( q^{(1)}_{12} - q^{(1)}_{13} ) + q^{(1)}_{11}   ( q^{(1)}_{22} - q^{(1)}_{23} + q^{(1)}_{32} - q^{(1)}_{33} )+ o( 1 )\\    
          =& \frac{(v_- + 1)(v_+ - u_+ + 1)}{u_- - u_+ - v_- + v_+}e^{\lambda_+(b-a)}  + o( 1 ).
 \end{split}         
\end{equation}
%}}

In the numerator of the expression of $Z_a(a)$, the terms of order $e^{2\lambda_+(b-a)}$ are zero. Moreover,
%{\tiny{
\begin{equation}\label{NumEaEqPos}
\begin{split}
( Q_{22} + Q_{23} + Q_{32} + Q_{33})&( Q_{12} - Q_{13} ) -( Q_{12} + Q_{13}) (Q_{22} - Q_{23}+Q_{32} -Q_{33})\\
 =& [(q^{(1)}_{22} + q^{(1)}_{23} + q^{(1)}_{32} + q^{(1)}_{33} )  \times( q^{(2)}_{12} - q^{(2)}_{13} ) - (q^{(1)}_{12} + q^{(1)}_{13}) \times ( q^{(2)}{22} - q^{(2)}_{23} + q^{(2)}_{32} - q^{(2)}_{33} )\\
          &+ (q^{(2)}_{22} + q^{(2)}_{23} + q^{(2)}_{32} + q^{(2)}_{33} ) \times ( q^{(1)}_{12} - q^{(1)}_{13} ) - (q^{(2)}_{12} + q^{(2)}_{13})   ( q^{(1)}_{22} - q^{(1)}_{23} + q^{(1)}_{32} - q^{(1)}_{33} ) ] e^{\lambda_+(b-a)} \\
          &+ (q^{(1)}_{22} + q^{(1)}_{23} + q^{(1)}_{32} + q^{(1)}_{33} ) \times ( q^{(1)}_{12} - q^{(1)}_{13} ) - (q^{(1)}_{12} + q^{(1)}_{13})   ( q^{(1)}_{22} - q^{(1)}_{23} + q^{(1)}_{3,2} - q^{(1)}_{33} ) + o( 1) \\
          &= \frac{2 u_-(v_+ - u_+ + 1)}{u_- - u_+ - v_- + v_+ }
 e^{\lambda_+(b-a)} + o( 1 ).
 \end{split}         
\end{equation}
%}}
In the large-$b$ limit, the vector of exit probabilities $\vec{\mathcal{E}}(a)$ therefore enjoys an asymptotic expansion of the form
\begin{equation}\label{noCorr}
\vec{\mathcal{E}}(a) = 
\begin{pmatrix}
\frac{u_-}{v_- - 1}\\
\frac{v_-}{v_- - 1}\\
\frac{1}{v_- - 1}\\ 
\end{pmatrix} + o( e^{-\lambda_-( b-a)}).
\end{equation}

\subsection{The vector $\vec{M}(a)$ with a positive drift in the large-$b$ limit}\label{appVecMa}

We know from Eq. (\ref{ExSymbPositiveDrift}) that $\vec{\mathcal{E}}(a)$ enjoys the following expansion when $b$ goes to infinity:
\begin{equation}
 \vec{\mathcal{E}}(a) = 
 \begin{pmatrix}
 \frac{u_-}{v_- - 1}\\
 \frac{v_-}{v_- - 1}\\
 \frac{1}{v_- - 1}    
 \end{pmatrix} + o(1).
\end{equation}

 We have shown (see Eq. (\ref{noCorr})) that there is no term of order $e^{-\lambda_+(b-a)}$ in the asymptotic expansion of $Z_a(a)$ and ${\mathcal{E}}_a(a,+;[a,b],\varphi)$ in the limit of large $b$. Hence an asymptotic expansion  of the vector $\vec{J}(b)$ (at the precision $O(1)$) is given as follows:
\begin{equation}
\begin{split}
\vec{J}(b) = & e^{\lambda_+(b-a)} 
  \left[ (b-a) W^{(2,2)} + \frac{1}{\lambda_+}( W^{(1,2)} + W^{(2,1)})
 + \frac{1}{\lambda_+ - \lambda_-}( W^{(2,3)} + W^{(3,2)}) \right] 
\begin{pmatrix}
 \frac{u_-}{v_- - 1}\\
 \frac{v_-}{v_- - 1}\\
 \frac{1}{v_- - 1}\\
\end{pmatrix}
\\
&
+\left[(b-a) W^{(1,1)}
-\frac{1}{\lambda_-}( W^{(1,3)} + W^{(3,1)})
-\frac{1}{\lambda_+}( W^{(1,2)} + W^{(2,1)})\right]
\begin{pmatrix}
 \frac{u_-}{v_- - 1}\\
 \frac{v_-}{v_- - 1}\\
 \frac{1}{v_- - 1}\\
\end{pmatrix}
 + o( 1 )\\    
 =& e^{\lambda_+(b-a)} \times
  \frac{1}{\lambda_+ - \lambda_-}( W^{(2,3)} + W^{(3,2)}) 
\begin{pmatrix}
 \frac{u_-}{v_- - 1}\\
 \frac{v_-}{v_- - 1}\\
 \frac{1}{v_- - 1}\\
\end{pmatrix}
-\frac{1}{\lambda_-}( W^{(1,3)} + W^{(3,1)})
\begin{pmatrix}
 \frac{u_-}{v_- - 1}\\
 \frac{v_-}{v_- - 1}\\
 \frac{1}{v_- - 1}\\
\end{pmatrix}
 + o( 1 )\\    
 =& \vec{K} e^{\lambda_+(b-a)} + \vec{L} + o(1), 
\end{split}    
\end{equation}
with the notations
\begin{equation}
\begin{split}
\vec{K} :=&  \frac{1}{\lambda_+ - \lambda_-}    \frac{- \mu v_-^2 + \mu u_- v_- + \mu - u_-}{\mu(1-\mu^2)(v_- - 1 )(-u_- + u_+ + v_- - v_+ )}
\begin{pmatrix}
u_+ \\
v_+ \\
1
\end{pmatrix},\\
\qquad \text{and}\qquad \vec{L} :=& -\frac{1}{\lambda_-}   \frac{\mu u_- - \mu u_+ - u_- v_- + u_- v_+ + \mu u_+ v_-^2 - \mu u_- v_- v_+}{\mu(1-\mu^2)(v_- - 1)(-u_- + u_+ + v_- - v_+ )}
\begin{pmatrix}
1\\
1\\
0
\end{pmatrix}.
\end{split}
\end{equation}

 We have used the algebraic identities

\begin{equation}\label{algId}
\begin{split}
   W^{(1,1)}  
   \begin{pmatrix}
  u_-\\
  v_-\\
  1
   \end{pmatrix} =& 0,\qquad  
   W^{(2,2)}  
   \begin{pmatrix}
  u_-\\
  v_-\\
  1
  \end{pmatrix} = 0,\qquad 
  ( W^{(1,2)} + W^{(2,1)})
  \begin{pmatrix}
  u_-\\
  v_-\\
  1
   \end{pmatrix} = 0. \\
(W^{(3,1)} + W^{(1,3)}) 
\begin{pmatrix}
  u_-\\
  v_-\\
  1
   \end{pmatrix} =&
   \frac{\mu u_- - \mu u_+ - u_- v_- + u_- v_+ + \mu u_+ v_-^2 - \mu u_- v_- v_+}{\mu(1-\mu^2)(-u_- + u_+ + v_- - v_+ )}
\begin{pmatrix}
1\\
1\\
0
\end{pmatrix},\\
(W^{(3,2)} + W^{(2,3)})  \begin{pmatrix}
  u_-\\
  v_-\\
  1
   \end{pmatrix} =&
   \frac{- \mu v_-^2 + \mu u_- v_- + \mu - u_-}{\mu(1-\mu^2)(-u_- + u_+ + v_- - v_+ )}
\begin{pmatrix}
u_+ \\
v_+ \\
1 
\end{pmatrix}.\\
   \end{split}
   \end{equation}
Gathering the terms of order $e^{\lambda_+(b-a)}$ in the numerator and using Eq. (\ref{detSPlusEq}) for the denominator yields
\begin{equation}
\begin{split}
M_1( a ) \underset{b\to\infty}{\sim}& \frac{-u_- + u_+ + v_- - v_+}{(v_- - 1)( u_+ - v_+ - 1)}\\
&\times\left[  (  q^{(2)}_{22} + q^{(2)}_{23} +  q^{(2)}_{32} + q^{(2)}_{33} )L_1  + (  q^{(1)}_{22} + q^{(1)}_{23} +  q^{(1)}_{32} + q^{(1)}_{33})K_1 \right.\\
&\left.  -(  q^{(2)}_{12} + q^{(2)}_{13})(L_2+L_3) -( q^{(1)}_{12} + q^{(1)}_{13})(K_2 + K_3)  \right]\\
=& \frac{1}{\mu(1-\mu^2)( v_- -1)^2(u_+ - v_+ - 1)}\\
&\times\left[  (  q^{(2)}_{22} + q^{(2)}_{23} +  q^{(2)}_{32} + q^{(2)}_{33} )\left(-\frac{1}{\lambda_-}   (\mu u_- - \mu u_+ - u_- v_- + u_- v_+ + \mu u_+ v_-^2 - \mu u_- v_- v_+ )\right)\right.\\
&\left.+ (  q^{(1)}_{22} + q^{(1)}_{23} +  q^{(1)}_{32} + q^{(1)}_{33}) \left( \frac{1}{\lambda_+ - \lambda_-}    (- \mu v_-^2 + \mu u_- v_- + \mu - u_-) \right) u_+ \right.\\
&\left.  -(  q^{(2)}_{12} + q^{(2)}_{13}) \left(-\frac{1}{\lambda_-}   (\mu u_- - \mu u_+ - u_- v_- + u_- v_+ + \mu u_+ v_-^2 - \mu u_- v_- v_+)\right) \right.\\
&\left.-( q^{(1)}_{12} + q^{(1)}_{13}) \left( \frac{1}{\lambda_+ - \lambda_-}    (- \mu v_-^2 + \mu u_- v_- + \mu - u_-) \right)(v_+ + 1)  \right]\\
=& \frac{1}{\mu(1-\mu^2)( v_- -1)^2(u_+ - v_+ - 1)(u_- - u_+ - v_- + v_+)}\\
&\times\left[  (  v_+ + 1 )(u_- - v_- + 1)\left(-\frac{1}{\lambda_-}   (\mu u_- - \mu u_+ - u_- v_- + u_- v_+ + \mu u_+ v_-^2 - \mu u_- v_- v_+ )\right)\right.\\
&\left.+ ( u_- - u_+ - u_- v_+ + u_+ v_-) \left( \frac{1}{\lambda_+ - \lambda_-}    (- \mu v_-^2 + \mu u_- v_- + \mu - u_-) \right) u_+ \right.\\
&\left.  - u_+  (u_- - v_- + 1) \left(-\frac{1}{\lambda_-}   (\mu u_- - \mu u_+ - u_- v_- + u_- v_+ + \mu u_+ v_-^2 - \mu u_- v_- v_+)\right) \right.\\
&\left.-( u_- - u_+ - u_-v_+ + u_+ v_- ) \left( \frac{1}{\lambda_+ - \lambda_-}    (- \mu v_-^2 + \mu u_- v_- + \mu - u_-) \right)(v_+ + 1)  \right].\\
\end{split}
\end{equation}
Hence
\begin{equation}
\begin{split}
U( a ) \underset{b\to\infty}{\sim}& \frac{1}{\mu(1-\mu^2)( v_- -1)^2(u_- - u_+ - v_- + v_+)}\\
&\times\left[  (u_- - v_- + 1)\left(\frac{1}{\lambda_-}   (\mu u_- - \mu u_+ - u_- v_- + u_- v_+ + \mu u_+ v_-^2 - \mu u_- v_- v_+ )\right)\right.\\
&\left.+ ( u_- - u_+ - u_- v_+ + u_+ v_-) \left( \frac{1}{\lambda_+ - \lambda_-}    (- \mu v_-^2 + \mu u_- v_- + \mu - u_-) \right) \right].\\
\end{split}
\end{equation}
This is the expression of $M_1(a)$ reported in Eq. (\ref{M12aExprPosDrift}).\\

Moreover,
\begin{equation}
 \begin{split}   
M_2( a ) \underset{b\to\infty}{\sim}& \frac{-u_- + u_+ + v_- - v_+}{(v_- - 1)( u_+ - v_+ - 1)}\\    
&\times\left[  -  ( q^{(2)}_{21} + q^{(2)}_{31} ) \left(-\frac{1}{\lambda_-}   (\mu u_- - \mu u_+ - u_- v_- + u_- v_+ + \mu u_+ v_-^2 - \mu u_- v_- v_+)\right)  - ( q^{(1)}_{21} + q^{(1)}_{31})K_1 \right.\\
&\left. +  q^{(2)}_{11}(L_2+L_3) + q^{(1)}_{11}(K_2 + K_3) \right]\\
=& \frac{1}{\mu(1-\mu^2)( v_- -1)^2(u_+ - v_+ - 1)}\\
& \times\left[  -  ( q^{(2)}_{21} + q^{(2)}_{31} ) \left(-\frac{1}{\lambda_-}   (\mu u_- - \mu u_+ - u_- v_- + u_- v_+ + \mu u_+ v_-^2 - \mu u_- v_- v_+)\right)  \right.\\
&\left.- ( q^{(1)}_{21} + q^{(1)}_{31}) \left( \frac{1}{\lambda_+ - \lambda_-}    (- \mu v_-^2 + \mu u_- v_- + \mu - u_-) \right) u_+  \right.\\
&\left. +  q^{(2)}_{11}\left(-\frac{1}{\lambda_-}   (\mu u_- - \mu u_+ - u_- v_- + u_- v_+ + \mu u_+ v_-^2 - \mu u_- v_- v_+)\right) \right.\\
&\left. + q^{(1)}_{11} \left( \frac{1}{\lambda_+ - \lambda_-}    (- \mu v_-^2 + \mu u_- v_- + \mu - u_-) \right)(v_+ + 1)  \right]\\
=& \frac{1}{\mu(1-\mu^2)( v_- -1)^2(u_+ - v_+ - 1)(u_- - u_+ - v_- + v_+)}\\
& \times\left[   ( v_+ + 1) \left(-\frac{1}{\lambda_-}   (\mu u_- - \mu u_+ - u_- v_- + u_- v_+ + \mu u_+ v_-^2 - \mu u_- v_- v_+)\right)  \right.\\
&\left. +  ( v_- - v_+) \left( \frac{1}{\lambda_+ - \lambda_-}    (- \mu v_-^2 + \mu u_- v_- + \mu - u_-) \right) u_+  \right.\\
&\left. - u_+\left(-\frac{1}{\lambda_-}   (\mu u_- - \mu u_+ - u_- v_- + u_- v_+ + \mu u_+ v_-^2 - \mu u_- v_- v_+)\right) \right.\\
&\left. - (v_- - v_+) \left( \frac{1}{\lambda_+ - \lambda_-}    (- \mu v_-^2 + \mu u_- v_- + \mu - u_-) \right)(v_+ + 1)  \right].\\
\end{split}
\end{equation}
 Hence,
\begin{equation}
 \begin{split}   
M_2( a )
\underset{b\to\infty}{\sim}& \frac{1}{\mu(1-\mu^2)( v_- -1)^2(u_- - u_+ - v_- + v_+)}\\
& \times\left[ \frac{1}{\lambda_-}   (\mu u_- - \mu u_+ - u_- v_- + u_- v_+ + \mu u_+ v_-^2 - \mu u_- v_- v_+)  \right.\\
&\left.+ (v_- - v_+)\left( \frac{1}{\lambda_+ - \lambda_-}    (- \mu v_-^2 + \mu u_- v_- + \mu - u_-) \right) \right].\\
\end{split}
\end{equation}
This is the expression reported in Eq. (\ref{M12aExprPosDrift}).

\end{appendices}

\bibliography{BibRef} 
\bibliographystyle{ieeetr}

\end{document}